%% file: main.tex
\documentclass[acmtog, authorversion]{acmart}
\usepackage{booktabs} 
\renewcommand\footnotetextcopyrightpermission[1]{}
\usepackage[ruled]{algorithm2e} 

\SetAlFnt{\small}
\SetAlCapFnt{\small}
\SetAlCapNameFnt{\small}
\SetAlCapHSkip{0pt}

\usepackage{pifont}
\usepackage{multirow}
\usepackage{colortbl}
\usepackage{xcolor}
\definecolor{ForestGreen}{RGB}{34,139,34}
\usepackage{subcaption}
\usepackage{amsmath}

\newcommand{\moniker}{Chord}
\newlength{\tempdima}
\newcommand{\rowname}[1]
{\rotatebox{90}{\makebox[\tempdima][c]{\textbf{#1}}}}

\DeclareMathOperator*{\argmin}{argmin}
\definecolor{commentcolor}{RGB}{50, 100, 150}  

\newenvironment{prompt}{\scriptsize\fontfamily{qag}\selectfont}{\par}

\acmJournal{TOG}

\copyrightyear{2025}
\acmYear{2025}
\setcopyright{acmlicensed}\acmConference[SA Conference Papers '25]{SIGGRAPH Asia 2025 Conference Papers}{December 15--18, 2025}{Hong Kong, Hong Kong}
\acmBooktitle{SIGGRAPH Asia 2025 Conference Papers (SA Conference Papers '25), December 15--18, 2025, Hong Kong, Hong Kong}
\acmDOI{10.1145/3757377.3763848}
\acmISBN{979-8-4007-2137-3/2025/12}

\begin{document}
\title{\moniker: \textbf{Ch}ain \textbf{o}f \textbf{R}endering \textbf{D}ecomposition for PBR Material Estimation from Generated Texture Images}

\author{Zhi Ying}
\authornote{Both authors contributed equally to the paper}
\email{zhi.ying@ubisoft.com}
\orcid{0009-0008-8390-3366}
\affiliation{%
  \institution{Ubisoft La Forge}
  \city{Shanghai}
  \country{China}
}

\author{Boxiang Rong}
\authornotemark[1]
\email{borong@student.ethz.ch}
\orcid{0000-0002-0653-4199}
\affiliation{%
  \institution{ETH Zürich}
  \city{Zürich}
  \country{Switzerland}
}

\author{Jingyu Wang}
\email{jingyuwang@student.ethz.ch}
\orcid{0009-0001-9085-7643}
\affiliation{%
  \institution{ETH Zürich}
  \city{Zürich}
  \country{Switzerland}
}

\author{Maoyuan Xu}
\email{mao-yuan.xu@ubisoft.com}
\orcid{0000-0002-9683-7002}
\affiliation{%
  \institution{Ubisoft La Forge}
  \city{Chengdu}
  \country{China}
}

\input{include/figure_teaser}
\input{secs/0_abstract}

%
%


%
%


\maketitle

\input{secs/1_introduction}
\input{secs/2_related_work}
\input{secs/3_preliminaries}
\input{secs/4_method}
\input{secs/5_experiments}
\input{secs/6_applications}
\input{secs/7_discussions}
\input{secs/8_conclusion}

\begin{acks}
We thank the anonymous reviewers for their valuable feedback. We are also grateful to Georges Nader, Arnaud Schoentgen, Alexis Rolland, and Yves Jacquier for their helpful discussions and proofreading. Zhi Ying is grateful for the unwavering family support from Xian Gong and Li Ying. This work was supported by Ubisoft.
\end{acks}

\bibliographystyle{ACM-Reference-Format}
\bibliography{main}

\include{secs/figure_only}

\end{document}

%% file: include/figure_teaser.tex
\begin{teaserfigure}
  \includegraphics[width=\textwidth]{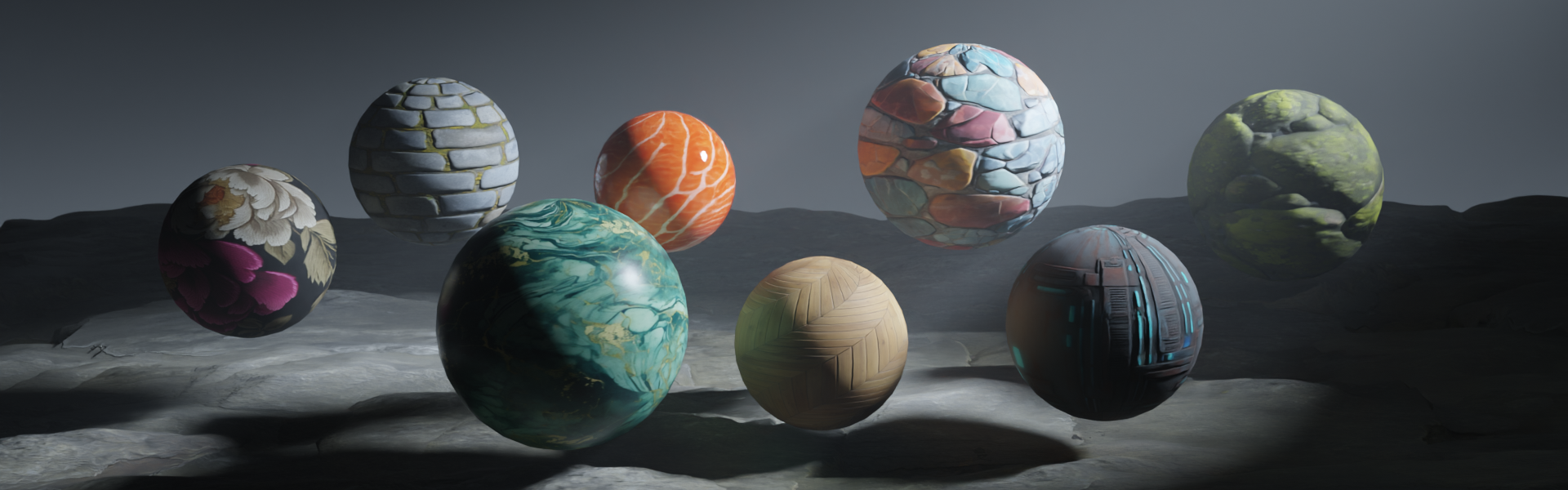}
  \caption{\textbf{PBR materials generated by our method.} Keywords in the text prompts used for generating each material (from left to right) are: \textit{fabric with flower embroidery}, \textit{cobblestone}, \textit{green marble}, \textit{salmon fish meat}, \textit{wood floor}, \textit{colorful stone wall}, \textit{cyberpunk iron wall}, and \textit{wet mossy rock}.}
  \label{fig:teaser}
\end{teaserfigure}

%% file: secs/0_abstract.tex
\begin{abstract}
Material creation and reconstruction are crucial for appearance modeling but traditionally require significant time and expertise from artists. While recent methods leverage visual foundation models to synthesize PBR materials from user-provided inputs, they often fall short in quality, flexibility, and user control.
We propose a novel two-stage generate-and-estimate framework for PBR material generation. In the generation stage, a fine-tuned diffusion model synthesizes shaded, tileable texture images aligned with user input. In the estimation stage, we introduce a chained decomposition scheme that sequentially predicts SVBRDF channels by passing previously extracted representation as input into a single-step image-conditional diffusion model. Our method is efficient, high quality, and enables flexible user control.
We evaluate our approach against existing material generation and estimation methods, demonstrating superior performance. Our material estimation method shows strong robustness on both generated textures and in-the-wild photographs. Furthermore, we highlight the flexibility of our framework across diverse applications, including text-to-material, image-to-material, structure-guided generation, and material editing.
\end{abstract}

%% file: secs/1_introduction.tex
\section{Introduction}

Generating materials that represent the appearance of objects has been a long-standing focus in the computer graphics communities. Early methods primarily synthesized RGB textures. With the rise of physically based rendering (PBR) workflows, materials are now commonly represented as spatially-varying bidirectional reflectance distribution functions (SVBRDFs). This advancement allows for more realistic representations of surface properties like reflectance and roughness but significantly increases complexity. The generation process now must handle multiple channels (e.g., basecolor, normal, roughness, and metalness maps) while ensuring precise spatial alignment. We treat these channels as distinct modalities due to their differing physical meanings. A visualization of their data distribution is shown in Fig.~\ref{fig:data_distribution}.

Recent work aims to simplify this process by compressing multiple channels into one latent space, and leverage specialized generative models~\cite{TileGen, MatFuse, ControlMat} to generate materials with certain level of user control. However, the flexibility and quality remains limited due to the constraints of their coupled architectures.

To address this challenge, we propose a two-stage \textit{generate-and-estimate} framework that enables flexible user control. The first stage leverages a generative model, guided by various control techniques, to generate texture RGB images aligned with user inputs. The second stage employs our novel \textit{chain-of-rendering-decomposition} (Chord) pipeline to predict SVBRDF channels through sequential steps.

\textit{Texture Generation} involves creating tileable texture RGB images that represent top-down views of materials. Traditional example-based methods generate textures that are visually similar to user-provided exemplars~\cite{wei2009state}. While effective, they are often limited by the quality of input examples and can produce only a narrow range of variations. More recently, text-to-image generative models have been applied to texture generation~\cite{chen2023text2tex}, showing greater flexibility and promising results. Additionally, the controllability of these models can be enhanced through additional input conditions, such as line art and depth maps~\cite{zhang2023ContrlNet}. Building on this approach, we employ a fine-tuned version of SDXL~\cite{podell2024sdxl} for generating texture RGB images. Our model is trained on texture images rendered under predefined lighting and camera conditions, with each image paired with a descriptive caption. To enable the tileability of the generated images, we use circular padding for all the convolutional layers.

\textit{Material estimation} focuses on inverse rendering, where shaded images of certain materials are used to estimate its SVBRDF channels. This task is inherently challenging due to its under-constrained nature. Recent intrinsic decomposition approaches~ \cite{kocsis24, RGBX} attempt to address this challenge by leveraging visual foundation models fine-tuned on modality-specific data, and use their strong prior knowledge to reduce the uncertainties. These data-driven methods require large amounts of training data and significant time to generalize effectively to novel materials. Moreover, they ignore the intrinsic relationships between different channels, often failing to accurately recover the expected material appearance.

Therefore, to preserve the intrinsic relationships among SVBRDF channels, we propose a novel chained scheme for image-conditional dense prediction diffusion models, where channels are predicted sequentially by conditioning on previously extracted representation. Our specific chain design named Chord is derived from the rendering equation of PBR materials, which predicts basecolor first, followed by normals with height, and finally roughness and metalness. The decomposition task is simplified by leveraging the implicitly predefined lighting and camera conditions between the two stages. To mitigate interference between different combinations of input and output modalities, we replace several U-Net blocks with a modular conditioning mechanism, termed \textit{LEGO-conditioning}.

During training, our chained scheme relies on clean intermediate representations, differing from standard diffusion training, which uses noisy samples. To address this, we adopt the single-step approach from~\cite{lotus, e2eft}, which not only ensures deterministic outputs but also improves inference efficiency by eliminating the need for multi-step denoising. Representative examples of generated PBR materials are shown in Fig.~\ref{fig:teaser}.

In summary, our main contributions are:
\begin{itemize} 
    \item A two-stage generate-and-estimate framework that enables flexible user control and simplifies material estimation via implicitly shared lighting and camera conditions.
    \item A chained scheme for multi-modal dense prediction using image-conditional diffusion model to capture inter-modal dependencies.
    \item The proposed Chord pipeline, which explicitly models the intrinsic relationships among SVBRDF modalities, combining LEGO-conditioning and single-step fine-tuning for efficient, high-quality material estimation.
\end{itemize}

%% file: secs/2_related_work.tex
\input{include/figure_data_distribution}
\section{Related Work}
\paragraph{Material Estimation}
The methods introduced by Aittala et al. \cite{Aittala15, Aittala16} aim to reduce the number of input images required for material estimation by assuming a fixed, centrally positioned point light source for illumination. These are the first optimization-based methods for few-shot material estimation. However, they rely heavily on hand-crafted heuristics and empirical priors. Deschaintre et al. \cite{Deschaintre2018} introduce the first large synthetic dataset for material estimation and use a variation of U-Net \cite{Unet} to predict SVBRDF channels. Subsequent works \cite{Li2017, Duan19, Li18, Zhou22, MaterIA, matfusion} follow similar data-driven approaches to estimate materials from single or multiple images. MATch \cite{match} makes the procedural design of materials differentiable, enabling the optimization of procedural components from images. SurfaceNet \cite{SurfaceNet} incorporates adversarial loss for unlabeled real-world images, while Material Palette \cite{MaterialPalette} adopts unsupervised domain adaptation to address the domain shift from synthetic training sets to real-world texture photos. Recently, a larger synthetic dataset, MatSynth \cite{matsynth}, was released, further improving the performance of previous data-driven methods.

\paragraph{Material Synthesis}
With the rise of generative models, they have become the most commonly used approaches for material generation. Earlier works \cite{MatFormer, PhotoMat, TileGen, Zhao20} utilize auto-regressive models or GANs to generate stationary textures with material properties. With recent advancements in diffusion models \cite{Ho20, Rombach21}, more diffusion-based methods \cite{MatFuse, ControlMat, DreamPBR} have emerged. Text2Mat \cite{Text2Mat} proposes a two-stage text-to-material pipeline, using a fine-tuned Stable Diffusion model to generate center-lit RGB exemplars, followed by an image-to-image translation model to produce SVBRDF channels. ReflectanceFusion~\cite{xue2024reflectancefusion} adopts a similar pipeline, but employs a latent code as the intermediate representation between the two stages.

Material synthesis and estimation can be unified in a single framework. MaterialGAN \cite{MaterialGAN} uses a GAN trained on a large synthetic dataset for material generation and performs material estimation by searching in the latent space. Henzler et al. \cite{Henzler20} train an image-to-image translation model with an encoder-decoder architecture, enabling both material generation and latent space sampling. MatFusion \cite{matfusion} and DiffMat \cite{DiffMat} adopt diffusion models and formulate material estimation as a conditioned generation task. MatFuse \cite{MatFuse} and ControlMat \cite{ControlMat} adapt Stable Diffusion models by modifying encoders and decoders to directly generate SVBRDF channels. MaterialPicker \cite{ma2025materialpicker} adapt a Diffusion Transformer text-to-video model for
material generation by treating material maps as video frames.

\paragraph{Intrinsic Decomposition}
Our material estimation stage draws inspiration from intrinsic decomposition methods. Most current intrinsic decomposition models are trained on synthetic indoor datasets. Kocsis et al. \cite{kocsis24} make diffusion models dependent on input images by concatenating them with noise in the latent space. RGB$\leftrightarrow$X \cite{RGBX} leverages similar idea and uses keyword text to conditionally decompose intrinsic channels. Careaga and Aksoy \cite{careagaColorful, careagaIntrinsic} achieve high-quality decompositions on in-the-wild images using a step-by-step feed forward pipeline. Our method uses RGB$\rightarrow$X model of RGB$\leftrightarrow$X as backbone and improves its performance by single-step fine-tuning, the LEGO-conditioning and the Chord pipeline.

\paragraph{Image-conditional Dense Prediction Diffusion Models}
Diffusion models \cite{Ho20, Rombach21, Song21}, as visual foundation models, have shown great promise for various downstream tasks. Marigold \cite{Marigold} is the first to propose repurposing diffusion models for dense prediction tasks. It spatially conditions the diffusion model in a lightweight manner for depth estimation. Several subsequent works \cite{GenPercept, e2eft, lotus, geowizard, lee24} have improved upon this approach, with most focusing on single-modality predictions. Notably, some methods, such as GeoWizard \cite{geowizard}, incorporate a switch to handle multiple modalities. Building on this foundation, our approach extends image-conditional dense prediction diffusion models to the multi-modal material domain, supporting four SVBRDF modalities. 

%% file: include/figure_data_distribution.tex
\begin{figure}
    \centering
    \setlength{\belowcaptionskip}{-8pt}
    \includegraphics[width=0.6\linewidth]{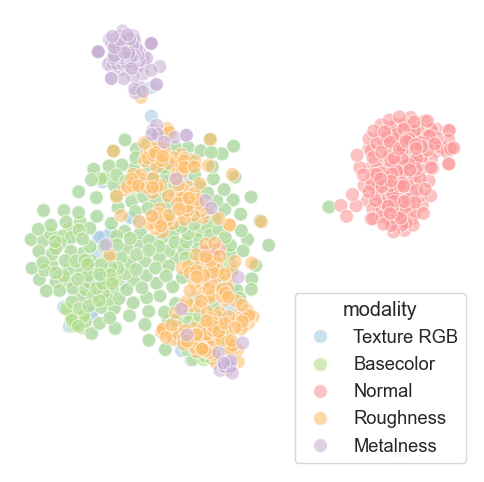}
      \caption{
        \textbf{2D t-SNE visualization of latent vectors.} Texture RGB and basecolor modalities exhibit significant overlap, suggesting a strong bijective relationship. The normal modality forms a distinct cluster. Metalness vectors are concentrated near binary values (0 and 1), while roughness vectors are more evenly distributed across the $(0, 1)$ range.
      }
      \label{fig:data_distribution}
\end{figure}

%% file: secs/3_preliminaries.tex
\section{Preliminaries}

\paragraph{Latent Diffusion Models}
LDMs such as Stable Diffusion (SD) \cite{Rombach21}, operate in lower dimensional latent space to efficiently model the high dimensional data distribution, such as high resolution images. To facilitate transformation between data space and latent space, LDMs employ a pair of auto-encoders, $\{\mathcal{E}(\cdot), \,\mathcal{D}(\cdot)\}$ to reconstruct sample $\mathbf{x} \approx \mathcal{D}(\mathcal{E}(\mathbf{x}))$. In the latent space, LDMs incorporate a forward process and a reverse process. The forward process progressively destructs the data by injecting noise. For a given time step $t \in [1, T]$, the forward process for the latent code $\mathbf{z}_t$ can be written in closed form: $\mathbf{z}_t = \sqrt{\bar{\alpha}_t}\mathbf{z}_0 + \sqrt{1-\bar{\alpha}_t}\epsilon$, where $\epsilon \sim \mathcal{N}(0,I)$, $ \bar{\alpha}_t = \prod_{i=1}^t \alpha_i$, $\alpha_t = 1-\beta_t$, and $\beta_t$ is the noise schedule. Conversely, the reverse process gradually removes noise from its inputs by predicting $\mathbf{z}_{t-1}$ given $\mathbf{z}_t$ using a trained neural network. Specifically, in the v-prediction parametrization \cite{salimans2022distillation}, the neural network $\mathbf{v}_{\theta}$ is trained to predict velocity $\hat{\mathbf{v}}_t$, and the corresponding loss at time step $t$ is: $\mathcal{L}_{\theta} = ||\mathbf{v}_t-\mathbf{v}_\theta(\mathbf{z}_t, t)||_2^2$, where $\mathbf{v}_t = \sqrt{\bar{\alpha}_t}\epsilon - \sqrt{1-\bar{\alpha}_t}\mathbf{z}_0$.

\paragraph{Image-conditional Latent Diffusion Models} In addition to the latent code of the data sample $\mathbf{z}_0 = \mathcal{E}(\mathbf{x})$, an extra conditioning image $\mathbf{I}_c$ and its latent code $\mathbf{c} = \mathcal{E}(\mathbf{I}_c)$ are provided. During training, noise is added only to $\mathbf{z}_0$. To incorporate this additional conditioning input, the parameter size of the U-Net's first convolutional layer, \textit{Conv-In} denoted as $\phi$ is doubled. The model predicts velocity as $\hat{\mathbf{v}}_t = \mathbf{v}_\theta(\mathbf{c}, \mathbf{z}_t, t)$. By treating the two halves of the doubled Conv-In layer as independent feature encoders, the predicted velocity can be reformulated as:
\begin{equation}
\label{eq:image_conditioned_ldm}
\hat{\mathbf{v}}_t = \mathbf{v}_{\theta'}(\phi_1(\mathbf{c}) + \phi_2(\mathbf{z}_t),\, t)   
\end{equation}
where $\theta' = \theta \setminus{(\phi_1 \cup \phi_2)}$, $\phi_1$ and $\phi_2$ represent the two halves of the Conv-In. This formulation is easily extensible to multiple conditioning images, as elaborated in Section~\ref{sec:lego_conditioning}.

%% file: secs/4_method.tex
\section{Method}  
\label{sec:method}
\input{include/figure_pipeline}  
Our framework follows a two-stage generation-and-estimation approach, as depicted in Fig.~\ref{fig:pipeline}.  

\subsection{Texture RGB Generation}
Our generation stage uses a fine-tuned SDXL diffusion model \cite{podell2024sdxl} to produce texture RGB images \( I_{\text{RGB}} \in \mathbb{R}^{3 \times H \times W} \) from diverse inputs including text prompts, reference images, and other control signals. The model is trained on a custom dataset of PBR textures rendered under consistent top-down view and directional lighting using our differentiable renderer \(\mathcal{R}\).

The renderer $\mathcal{R}$ implements Cook-Torrance BRDF \cite{cook1982} with: (1) Trowbridge-Reitz GGX normal distribution \cite{trowbridge1975average}, (2) Schlick-GGX geometry term \cite{karis2013real}, and (3) Schlick's Fresnel approximation \cite{schlick1994inexpensive}. This same renderer is also used for material estimation and rendering loss computation in the next stage.

\subsection{Material Estimation}
\label{sec:mat_estimation}
Given the generated image $I_\text{RGB}$, our goal is to decompose the SVBRDF channels $\text{MAT} = \{b, n, h, r, m\}$, where $b$ denotes basecolor, $n$ normal, $h$ height, $r$ roughness, and $m$ metalness. To train our material estimation model, we construct a dataset of paired samples $(\text{MAT}, I_\mathcal{R})$, where $I_{\mathcal{R}} = \mathcal{R}(\text{MAT};l)$, $l$ is a randomly rotated directional light. We assume that the rendering $I_\mathcal{R}$ should reproduce $I_{\text{RGB}}$ under the same implicitly predefined lighting and camera conditions. This assumption holds in the majority of cases, see Fig.~\ref{fig:in_the_wild} and Fig.~\ref{fig:app:editting} for examples.

The training process for our material estimation model consists of two phases:
\begin{enumerate}
    \item \textbf{Pretraining Phase} \textit{(optional)} – employs standard diffusion training to initialize and warm up the model weights.
    \item \textbf{Single-step Phase} – serves as the primary training stage, utilizing the Chord pipeline along with image-space losses.
\end{enumerate}
Below, we detail two key modifications to the image-conditional diffusion model: LEGO-conditioning and Chord pipeline.

\subsubsection{LEGO-conditioning}
\label{sec:lego_conditioning}
\input{include/figure_lego_conditioning}  
Building on the $\text{RGB}\rightarrow\text{X}$~\cite{RGBX}, we employ CLIP text embeddings $\tau$ as target channel switches when predicting different modalities. However, as shown in our ablation study (Table~\ref{tab:ablation}), directly adapting a chained pipeline reveals weight conflicts between modalities in shared layers.

We address this with LEGO-conditioning, extending Eq.\ref{eq:image_conditioned_ldm} to support multiple conditions. The v-prediction becomes:
\begin{equation}
\label{eq:lego_conditioned_ldm}
\hat{\mathbf{v}}_t = \mathbf{v}_{\theta'} \left( \frac{\sum_{i=1}^{k}\phi_i(\mathbf{c}_i) + \phi_z(\mathbf{z}_t)}{k+1},\, \tau(D_z),\, t \right),
\end{equation}
where $\mathbf{c}_i$ are conditioning latents, $z_t$ is the noisy latent of the target channel, and $D_z$ is its text description. The features are averaged to maintain consistent magnitude. Inspired by recent works~\cite{liu2023hyperhuman, hong2024supermat} and based on the SD 2.1 architecture~\cite{Rombach21}, we adopt separate first \textit{Down-Block}, last \textit{Up-Block}, and \textit{Conv-Out} modules for each target channel, while sharing intermediate U-Net blocks to preserve spatial alignment. Details of the model architecture are shown in Fig.~\ref{fig:lego_conditioning}. This design achieves effective channel separation with minimal computational and memory overhead.

During the Pretraining Phase, the training objective minimizes the loss $\mathcal{L}_{\theta} = ||\mathbf{v}_t - \hat{\mathbf{v}}_t||_2^2$, where $\mathbf{v}_t = \sqrt{\bar{\alpha}_t}\epsilon - \sqrt{1 - \bar{\alpha}_t}z_0$. In the Single-step Phase, we set $t=T$, the $\mathbf{z}_t$ term in Eq.\ref{eq:lego_conditioned_ldm} is omitted, and the loss is computed directly in the image space using Eq.\ref{eq:total_loss}. For each prediction step, active LEGO-conditioning blocks are trained jointly with the shared blocks in both phases.

\subsubsection{Chain of Rendering Decomposition (Chord)}
\label{sec:chord_pipeline}
Multi-modal dense prediction is challenging \cite{yang2025multitask}. To simplify the problem, we  follow Simchony et al. \cite{simchony1990direct} to integrate height $h$ from the normal map $n$, reducing the task to four channels: $\{b, n, r, m\}$. With the implicitly shared light assumption, while a naive gradient-based optimization of the loss $\mathcal{L}(\mathcal{R}(\hat{\text{MAT}}; l), I_{\text{RGB}})$ is possible, the problem is severely under-constrained because multiple parameter combinations can explain the same rendering. This ambiguity persists when training a model to learn an inverse rendering function $\hat{\text{MAT}} = \mathcal{R}^{-1}(I_{\text{RGB}}; l)$.

To address these challenges, our Chord pipeline divides the problem into three sequential steps:

\begin{enumerate}
\item \textbf{Basecolor Prediction:} Conditioning on $I_{\text{RGB}}$, the first step predicts $\hat{b}$.
\item \textbf{Normal Prediction:} From $\hat{b}$, we compute an approximate irradiance map $I_{\text{IRR}}$. The normal map $\hat{n}$ is then predicted by conditioning on both $I_{\text{RGB}}$ and $I_{\text{IRR}}$.
\item \textbf{Roughness \& Metalness Prediction:} Given the estimated $\hat{b}$, $\hat{n}$, and $I_\text{IRR}$, we approximate the optimal roughness and metalness combination image, denoted as $I_{\text{RM}}$. The roughness and metalness$\{\hat{r}, \hat{m}\}$ are predicted by conditioning on both $I_{\text{RGB}}$ and $I_{\text{RM}}$.
\end{enumerate}

\paragraph{Basecolor Prediction}
We begin by predicting the basecolor to minimize accumulated errors along the chain. As shown in Fig.~\ref{fig:data_distribution}, the data distribution of 
$b$ is the closest to that of the input RGB images. This makes it easier for the model to learn the transition between the two modalities.

\paragraph{Normal Prediction}
In modern PBR workflows, normal maps encode geometric details that are independent of base color variations. Directly conditioning normal prediction on $I_{\text{RGB}}$ can introduce noise since color information is largely irrelevant to geometry. Instead, under the assumption of single directional lighting, we compute an approximate irradiance image $I_{\text{IRR}}$ by:
\begin{equation}
\label{eq:approx_Irr}
    I_{\text{IRR}} = I_{\text{RGB}} / b
\end{equation}
This removes color dependence from diffuse shading and preserves lighting and geometric cues in the irradiance.
We empirically verify that the error introduced from specular terms is negligible. The resulting $I_{\text{IRR}}$ serves as a cleaner conditioning representation for normal prediction, particularly by decoupling diffuse albedo effects.

\paragraph{Roughness \& Metalness Prediction}
Similar to the motivation of previous steps, we aim to isolate and extract representation closer to the target channels $\{r, m\}$. We first estimate lighting direction $\mathbf{l}^*$ using an energy-decay heuristic on the approximate irradiance image (see Section A.2 of supplementary material). We then perform a grid search to minimize:
\begin{equation}
\text{MSE}(x) = \|\hat{I}_{\mathcal{R}}(x) - I_{\text{RGB}}(x)\|_2^2
\end{equation}
where $\hat{I}_{\mathcal{R}}(x) = \mathcal{R}(\hat{b}(x), \hat{n}(x), r(x), m(x); \mathbf{l}^*)$. The discrete search space $\mathcal{S}$ is defined as:
\begin{equation}
\mathcal{S} = \left\{\left(\tfrac{25+5i}{255}, j\right) \mid i \in \mathbb{Z}, 0 \leq i \leq 40,\; j \in \{0,1\}\right\}
\end{equation}
yielding 41 roughness values and binary metalness. The optimal per-pixel solution is:
\begin{equation}
I_{\text{RM}}(x) = \begin{bmatrix}
r^*(x) \\ m^*(x)
\end{bmatrix} = \underset{(r,m) \in \mathcal{S}}{\argmin}\, \text{MSE}(x)
\end{equation}

\paragraph{Single-step Phase Loss}
We compute loss in image space, by setting $t=T$ and decode latent using VAE decoder, and our loss combines:
\begin{equation}
\label{eq:total_loss}
\begin{aligned}
\mathcal{L} = &\underbrace{\|\hat{\text{MAT}} - \text{MAT}\|_1}_{\text{Pixel}} + 
\underbrace{\|\Phi(\hat{\text{MAT}}) - \Phi(\text{MAT})\|_1}_{\text{Perceptual}} \\
&+ \underbrace{\|\mathcal{R}(\hat{\text{MAT}}; l) - \mathcal{R}(\text{MAT}; l)\|_1}_{\text{Render}}
\end{aligned}
\end{equation}
where $\text{MAT} = \{b,n,r,m\}$, $\hat{\text{MAT}}$ are predictions, $\Phi$ is pre-trained VGG-16 \cite{johnson2016perceptual} and $l$ is randomly rotated directional light per iteration. This simplified formulation is for clarity, in the complete loss function, we use cosine similarity for the normal channel $\hat{n}$ instead of the $\ell_1$ loss used for other channels:
\begin{equation}
\mathcal{L}_n = 1 - \hat{n} \cdot n
\end{equation}
This penalizes angular discrepancies and encourages alignment between predicted and ground truth normals.
For the render loss, we combine $\ell_1$ and VGG perceptual losses. At each iteration, the lighting direction $\mathbf{l}$ is randomly sampled 8 times to generate 8 rendered image pairs. A coefficient $\lambda = 0.005$ is applied to the perceptual loss terms during combination. The complete loss is:
\begin{equation}
\label{eq:complete_total_loss}
\begin{aligned}
\mathcal{L}_{\text{complete}} =\;& \|\hat{\text{MAT}} \setminus{\hat{n}} - \text{MAT} \setminus{n}\|_1 + \mathcal{L}_n \\
&+ \lambda\|\Phi(\hat{\text{MAT}}) - \Phi(\text{MAT})\|_1 \\
&+ \|\mathcal{R}(\hat{\text{MAT}}; l) - \mathcal{R}(\text{MAT}; l)\|_1 \\
&+ \lambda\|\Phi(\mathcal{R}(\hat{\text{MAT}}; l)) - \Phi(\mathcal{R}(\text{MAT}; l))\|_1.
\end{aligned}
\end{equation}

%% file: include/figure_pipeline.tex
\begin{figure*}
    \centering
    \includegraphics[width=\linewidth]{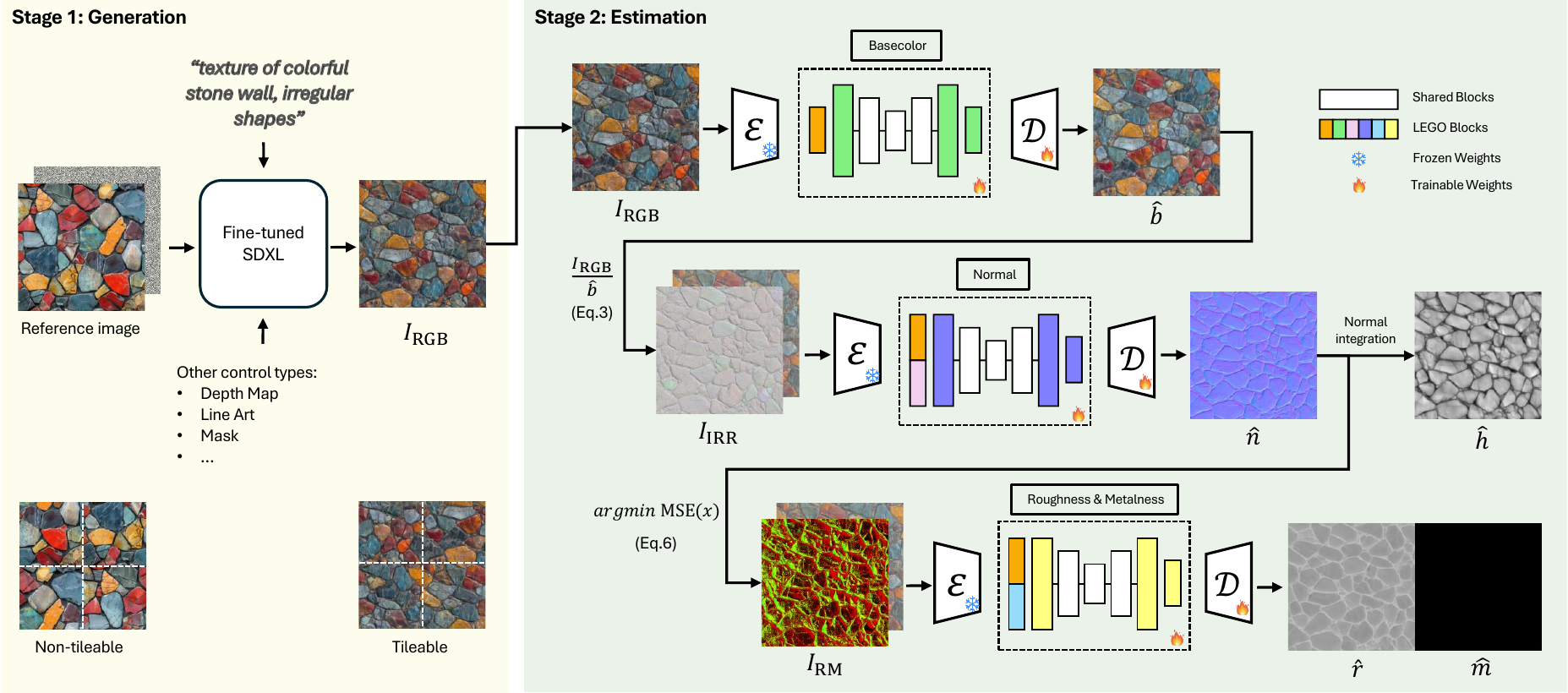}
    \caption{
        \textbf{Method Overview.} 
        \textbf{Stage 1:} Tileable texture image ($I_\text{RGB}$) generation using a fine-tuned diffusion model, controllable via user guidance (text prompts, reference images, or other control types). 
        \textbf{Stage 2:} Material estimation predicts SVBRDF channels sequentially: (1) basecolor $\hat{b}$, (2) normal $\hat{n}$ (with height $\hat{h}$ derived via normal integration), and (3) roughness $\hat{r}$ and metalness $\hat{m}$. Each step's input is computed from previous outputs. LEGO-conditioning provides modality-specific weights while maintaining shared backbone weights for alignment. See Section~\ref{sec:mat_estimation} for details. Reference image from Vecteezy.
      }
    \label{fig:pipeline}
\end{figure*}

%% file: include/figure_lego_conditioning.tex
\begin{figure}
    \centering
    \includegraphics[width=\linewidth]{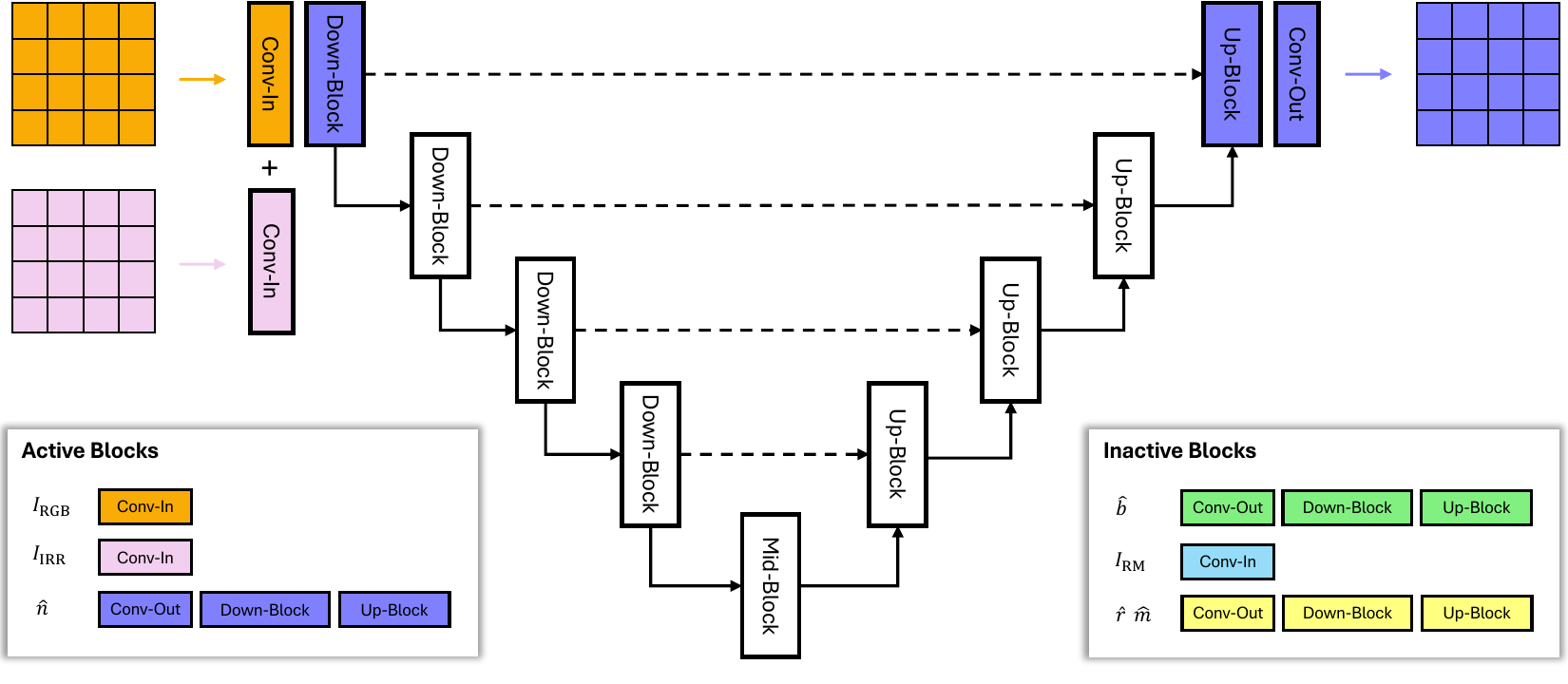}
    \caption{
        \textbf{Detailed illustration of LEGO-conditioning.}
        Using the normal prediction step as an example, we illustrate the block-level architecture of the U-Net. We also highlight the corresponding active and inactive LEGO-conditioning blocks in this step.
      }
    \label{fig:lego_conditioning}
\end{figure}

%% file: secs/5_experiments.tex
\section{Experiments}
\subsection{Implementation Details}
Our texture RGB generation model is based on SDXL \cite{podell2024sdxl}, fine-tuning it on ~1,000 high-quality texture rendering images rendered by $\mathcal{R}$ with descriptive captions.

To train our material estimation model, we constructed a dataset by combining materials from MatSynth and private data, resulting in a total of 28,344 materials after augmentation. During training, each material was resized and randomly cropped to a resolution of $512 \times 512$. 
For a fair comparison, we retrained several baselines, including SurfaceNet, MatFusion, RGB$\rightarrow$X, Lotus, and E2E-FT on our dataset. Training details for our method are provided in the supplementary material Section A.1.

\subsection{Material Generation}
\input{figures/generation_comparison/generation_comparison}
We compare our method with MatFuse~\cite{MatFuse} for conditioned material generation (Fig.~\ref{fig:generation_comparison}). While MatFuse directly predicts all four material channels, our two-stage approach generates a texture RGB image first, then decomposes it into channels. Our method significantly outperforms those of MatFuse in terms of both condition alignment and visual quality. See Section~\ref{sec:application} for additional material generation applications.

\subsection{Material Estimation}
We evaluate our method on two datasets: the test split of MatSynth~\cite{matsynth}, containing 89 unique materials, and a curated test set of 250 unique materials from the Substance Asset Library~\cite{substance_lib}, ensuring no overlap with the training data. The detailed material category distributions are provided in the supplementary material Section A.3. Each material was rendered under four directional lights positioned at the image corners, resulting in 356 and 1,000 evaluation images at a resolution of 1k. Our evaluation encompasses both single-modality estimation, comparing with other dense prediction methods, and full-modalities estimation, compared with other material estimation approaches. We use PSNR and LPIPS to evaluate the similarity between predictions and ground truth (GT) channels.

\subsubsection{Single-modality Estimation} 
\input{figures/normal_comparison/normal_comparison.tex}
\input{tables/normal_comparison.tex}
Different tasks have varying criteria for basecolor, with some PBR materials baking ambient occlusion into the basecolor and others not. Since a certain proportion of materials in our dataset include baked AO, direct comparison with intrinsic decomposition methods~\cite{careagaIntrinsic} is challenging. Therefore, we focus on normal channel estimation. Qualitative results in Fig.~\ref{fig:normal_comparison} and numerical results in Table \ref{tab:normal_comparison} show that our model remains robust and competitive with state-of-the-art methods, even when predicting four modalities simultaneously.

\subsubsection{Full-Modalities Estimation}
\input{figures/pbr_comparison/pbr_comparison}
\input{tables/pbr_comparison}
\input{tables/inference_time}
We compare our method with RGB$\rightarrow$X, SurfaceNet, MatFusion, and Material Palette for material estimation. The \textit{Relit} measurements are obtained by rendering the material under nine predefined point lights and directional lights. As shown in Fig.~\ref{fig:pbr_comparison}, Table~\ref{tab:pbr_comparison_matsynth}, Table~\ref{tab:pbr_comparison_substance} and Table~\ref{tab:inference_time}. Our model delivers superior quality while achieving an 11$\times$ speedup compared to RGB$\rightarrow$X. Additionally, our approach remains robust on unseen generated texture images and in-the-wild photographs, as illustrated in Fig.\ref{fig:art_gallery} and Fig.\ref{fig:in_the_wild}. For further visual comparisons with ControlMat~\cite{ControlMat}, SurfaceNet~\cite{SurfaceNet}, and MaterIA~\cite{MaterIA}, please refer to Section A.4.1 of supplementary material.

\subsection{Ablation Study}
\input{tables/ablation}
We validate the components of our pipeline through comprehensive ablation studies. All experiments are initialized using SD 2.1 and fine-tuned on our dataset for 20 epochs, except for the pretrained version of our method, which includes an additional 5 epochs of diffusion pretraining. As presented in Table~\ref{tab:ablation}, compared to the baseline (RGB$\rightarrow$X), our approach demonstrates that using single-step diffusion with the loss computed directly in image space (+Single-step) improves PSNR but yields worse LPIPS due to blurrier outputs. This issue can be mitigated by incorporating a combined loss (+Combined Loss) described in Eq.\ref{eq:complete_total_loss} (excluding the render loss terms), which improves the LPIPS score.

Directly applying a naive chained scheme (+Chain), which uses previous-step predictions as conditioning inputs for the current step, yields only marginal performance improvements. This is primarily due to interference caused by shared model weights across different modalities. By incorporating LEGO-conditioning into the chained pipeline (+LEGO-conditioning), this issue is mitigated through the use of modality-specific weights, leading to more effective multi-channel prediction. 

We further evaluate the effect of computing $I_\text{IRR}$ and $I_\text{RM}$, and show that both contribute to improved estimation. Finally, including render loss and Pretraining Phase yields additional performance gains.

%% file: figures/generation_comparison/generation_comparison.tex
\begin{figure}[ht]
    \tiny
    \centering
    \setlength\tabcolsep{1pt}
    \settoheight{\tempdima}{\includegraphics[width=0.18\linewidth]{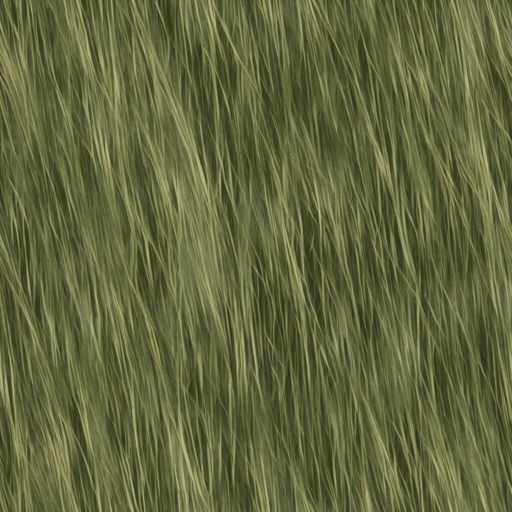}} 
    \begin{tabular}{@{}c@{\hskip 0.03in}|@{\hskip 0.03in}cccccc@{}}
    Condition(s) & & \shortstack{Diffuse (MatFuse) \\ Basecolor (Ours)} & Normal  & Roughness & \shortstack{Specular (MatFuse) \\ Metalness (Ours)} \\
    \cmidrule(r){1-1}\cmidrule(lr){2-6}
    \begin{minipage}[b]{0.18\linewidth}
    \centering
        \begin{prompt}
            \textbf{"wispy grass"}
        \end{prompt}
    \end{minipage} &
    \rowname{MatFuse}&
    \includegraphics[height=0.18\linewidth]{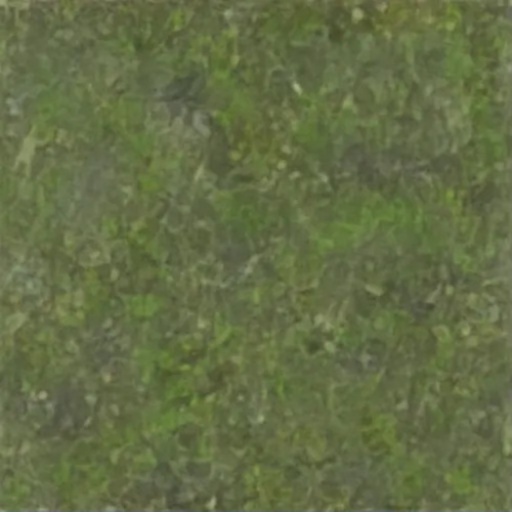} &
    \includegraphics[height=0.18\linewidth]{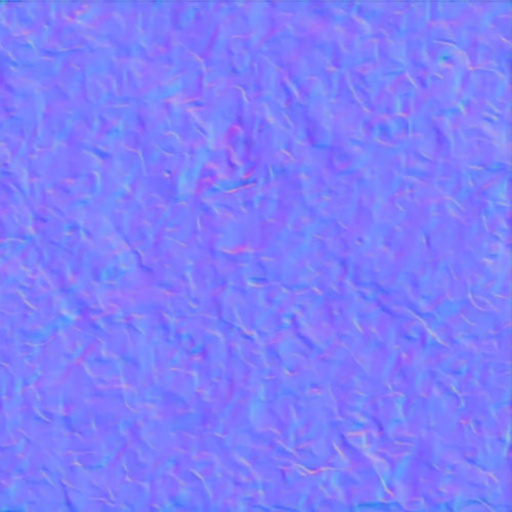} &
    \includegraphics[height=0.18\linewidth]{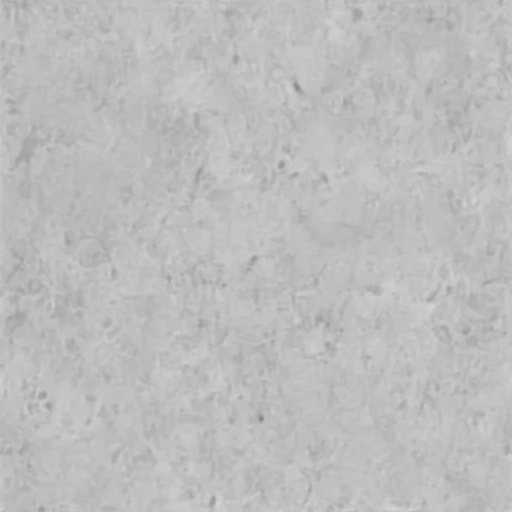} &
    \includegraphics[height=0.18\linewidth]{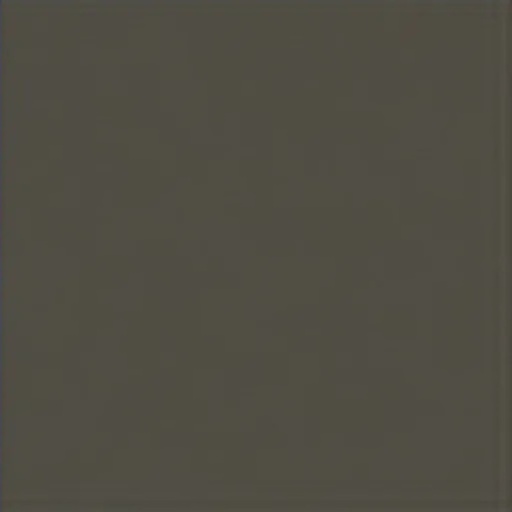} \\
    \includegraphics[height=0.18\linewidth]{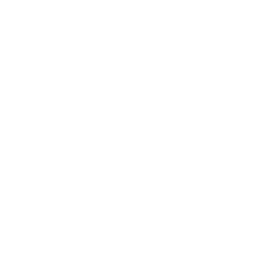} &
    \rowname{Ours}&
    \includegraphics[height=0.18\linewidth]{figures/generation_comparison/chord/grass/basecolor_512.jpg} &
    \includegraphics[height=0.18\linewidth]{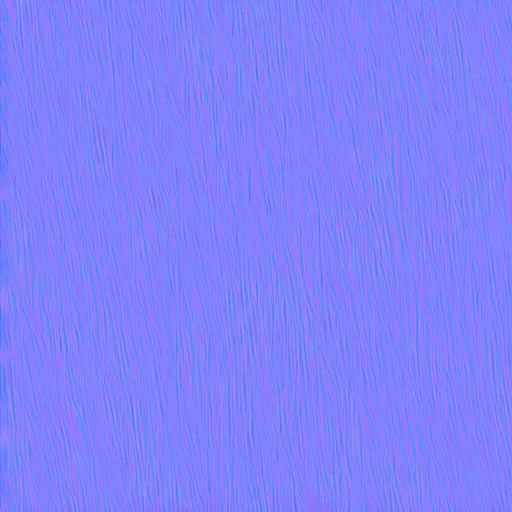} &
    \includegraphics[height=0.18\linewidth]{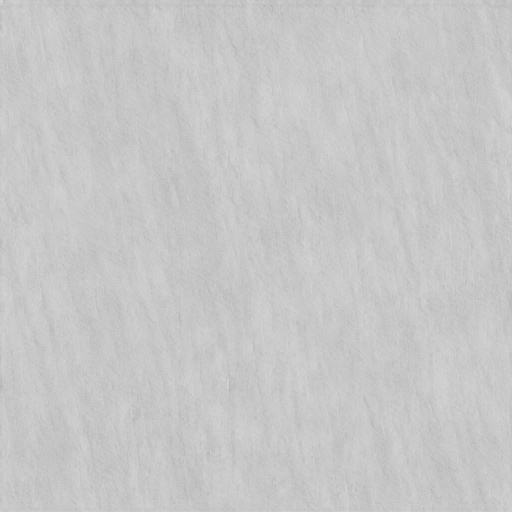} &
    \includegraphics[height=0.18\linewidth]{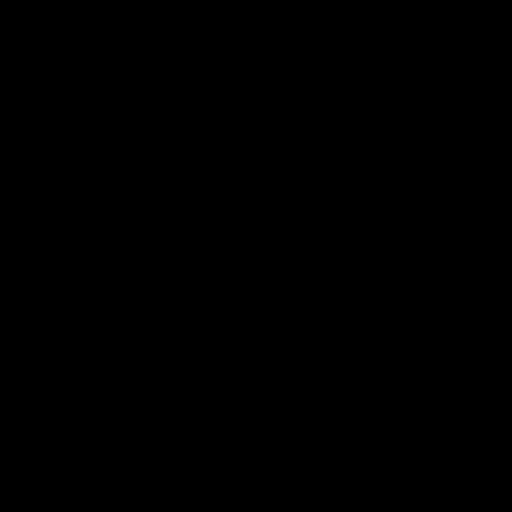} \\
    \cmidrule(r){1-1}\cmidrule(lr){2-6}
    \begin{minipage}[b]{0.18\linewidth}
    \centering
        \begin{prompt}
            \textbf{"light brown wood planks, aged appearance, rough, weathered"}
        \end{prompt}
    \vspace{1\baselineskip}
    \end{minipage} &
    \rowname{MatFuse}&
    \includegraphics[height=0.18\linewidth]{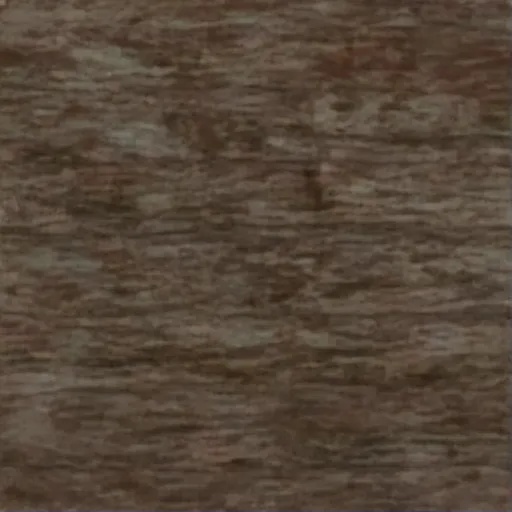} &
    \includegraphics[height=0.18\linewidth]{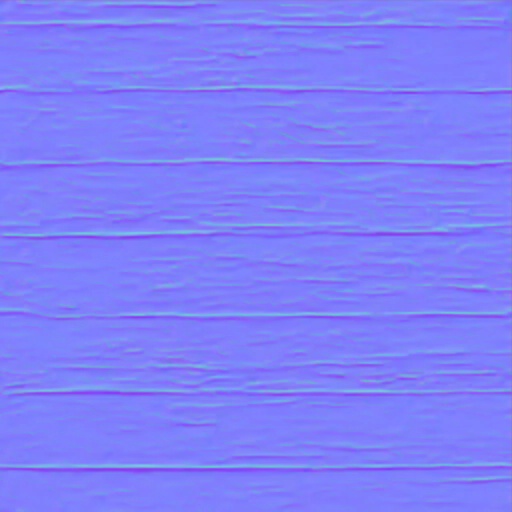} &
    \includegraphics[height=0.18\linewidth]{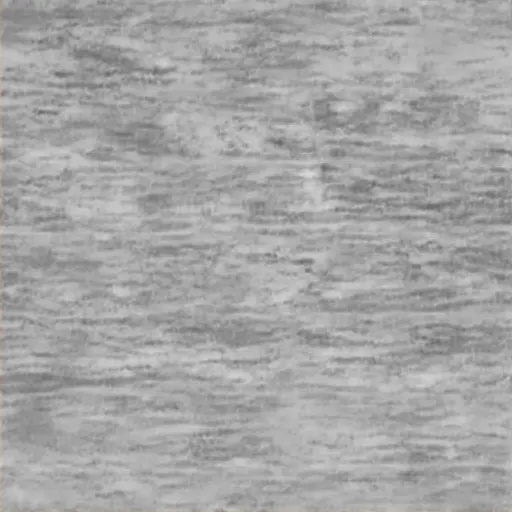} &
    \includegraphics[height=0.18\linewidth]{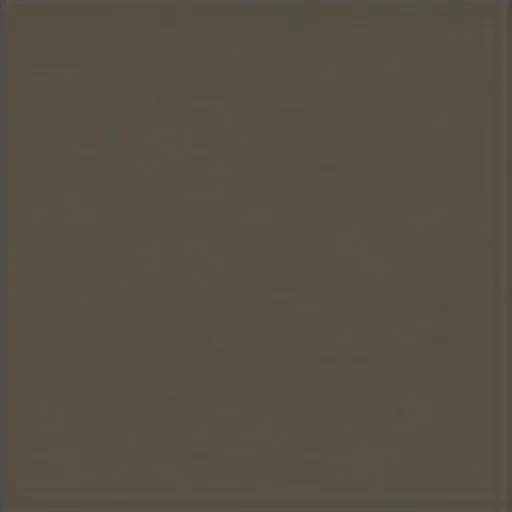} \\
    \includegraphics[height=0.18\linewidth]{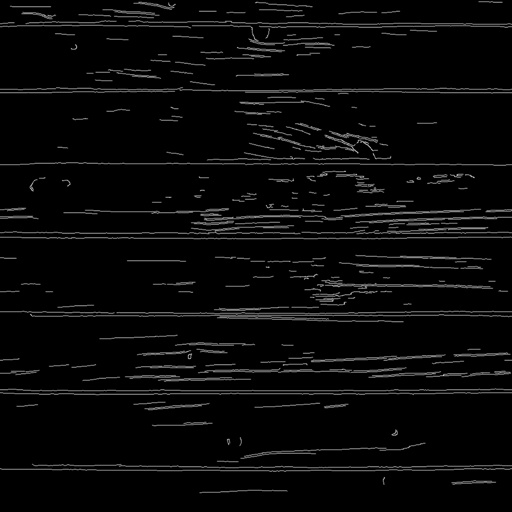} &
    \rowname{Ours}&
    \includegraphics[height=0.18\linewidth]{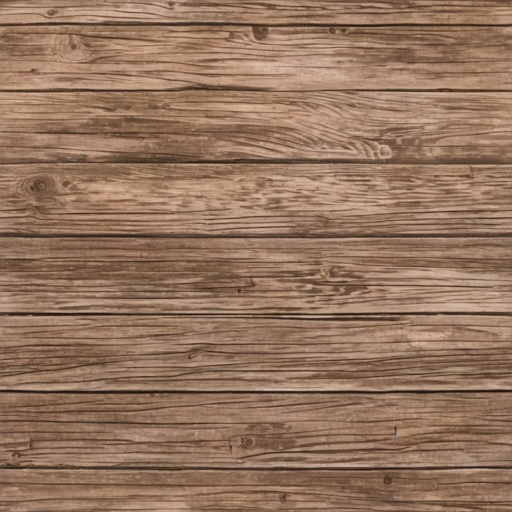} &
    \includegraphics[height=0.18\linewidth]{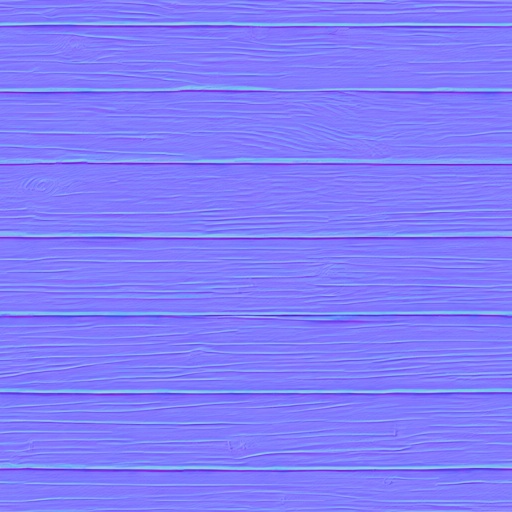} &
    \includegraphics[height=0.18\linewidth]{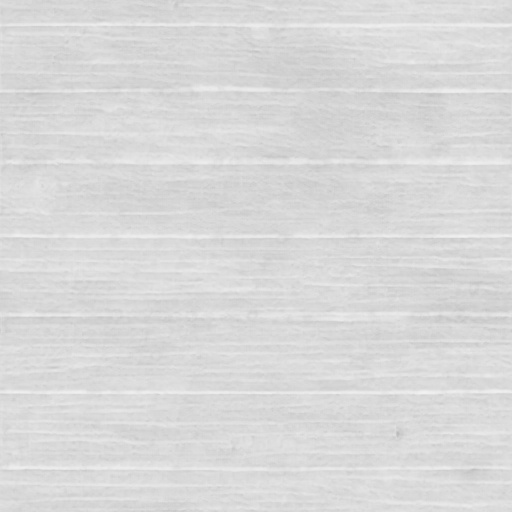} &
    \includegraphics[height=0.18\linewidth]{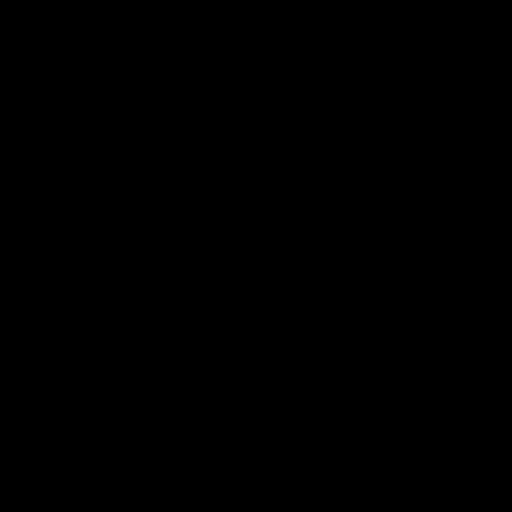} \\
    \cmidrule(r){1-1}\cmidrule(lr){2-6}
    \begin{minipage}[b]{0.18\linewidth}
    \centering
        \begin{prompt}
            \textbf{"rough cliff surface, irregular, red"}
        \end{prompt}
    \vspace{2\baselineskip}
    \end{minipage} &
    \rowname{MatFuse}&
    \includegraphics[height=0.18\linewidth]{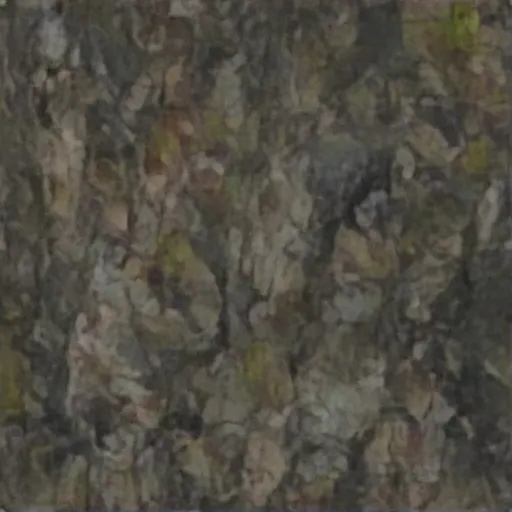} &
    \includegraphics[height=0.18\linewidth]{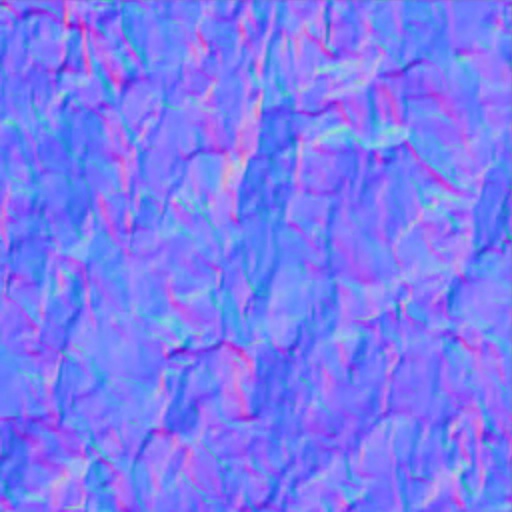} &
    \includegraphics[height=0.18\linewidth]{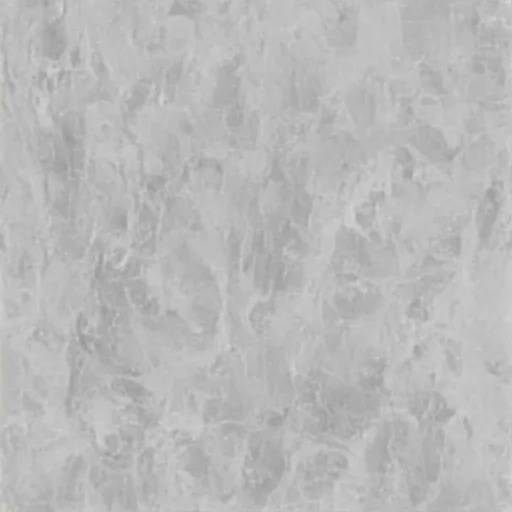} &
    \includegraphics[height=0.18\linewidth]{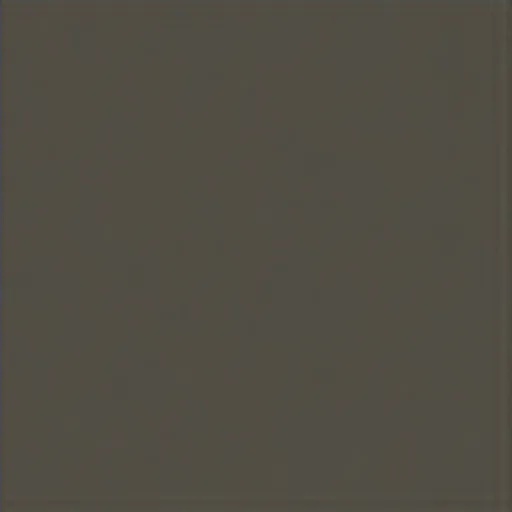} \\
    \includegraphics[height=0.18\linewidth]{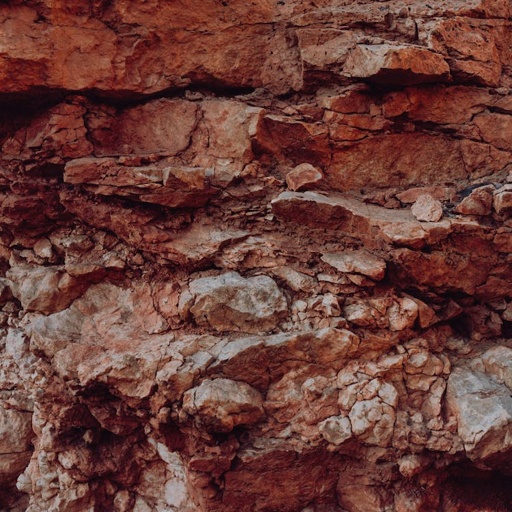} &
    \rowname{Ours}&
    \includegraphics[height=0.18\linewidth]{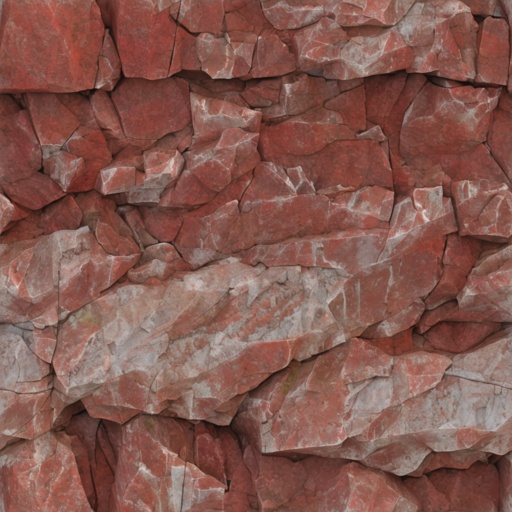} &
    \includegraphics[height=0.18\linewidth]{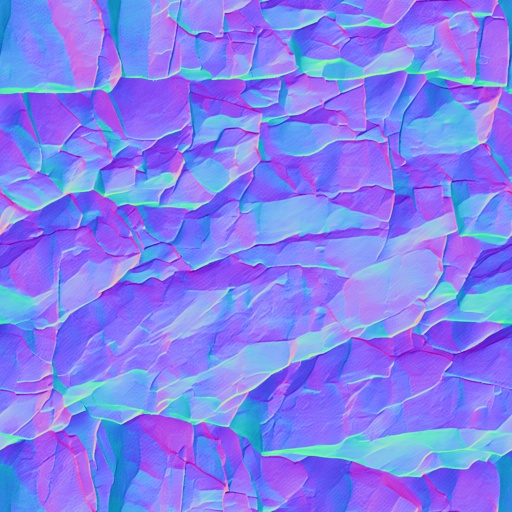} &
    \includegraphics[height=0.18\linewidth]{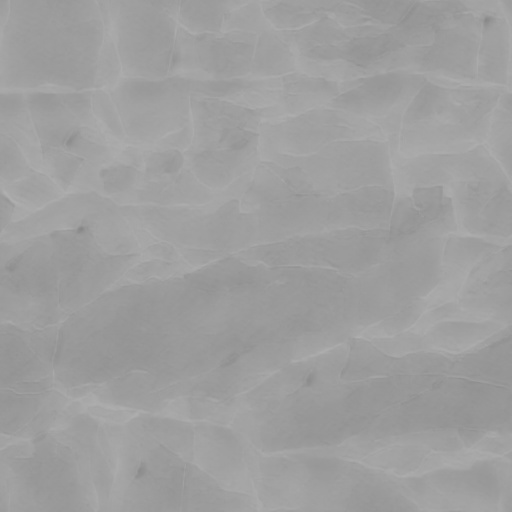} &
    \includegraphics[height=0.18\linewidth]{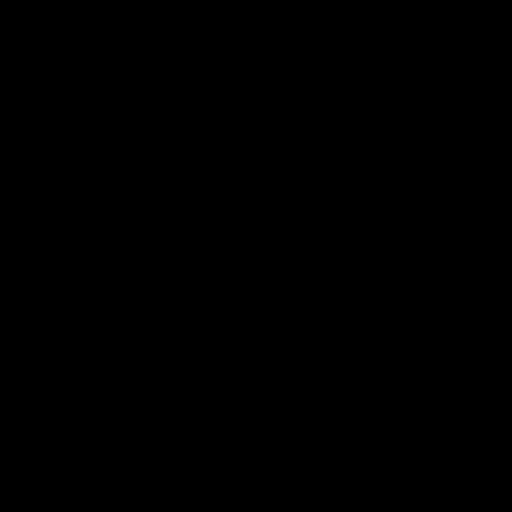} \\
    \cmidrule(r){1-1}\cmidrule(lr){2-6}
    \begin{minipage}[b]{0.18\linewidth}
    \centering
        \begin{prompt}
            \textbf{"mud ground, cracks, wasteland"}
        \end{prompt}
        \vspace{2\baselineskip}
    \end{minipage} &
    \rowname{MatFuse}&
    \includegraphics[height=0.18\linewidth]{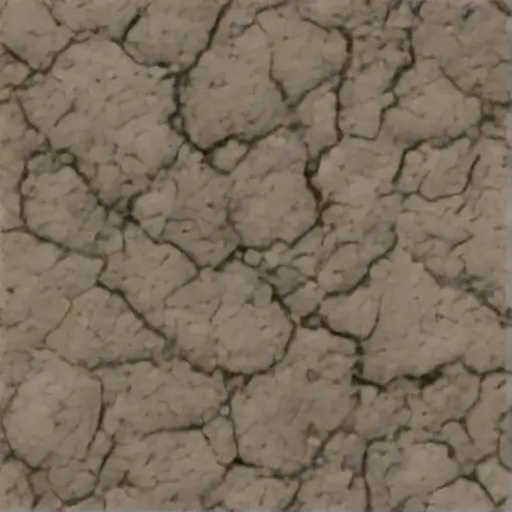} &
    \includegraphics[height=0.18\linewidth]{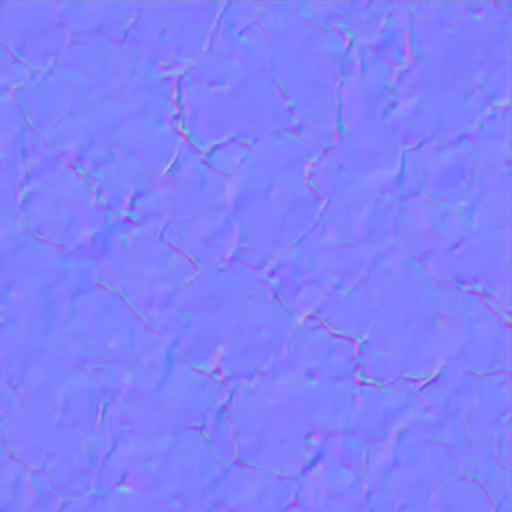} &
    \includegraphics[height=0.18\linewidth]{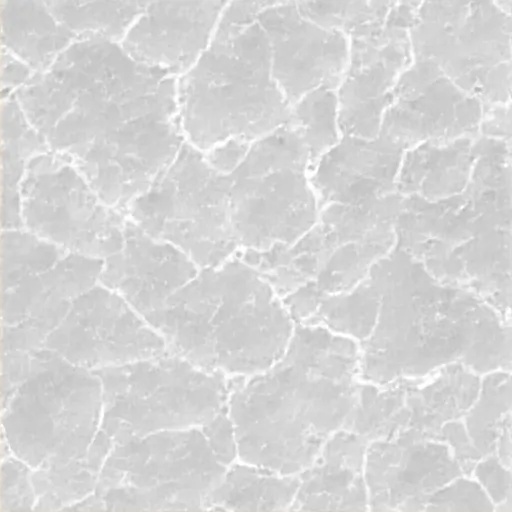} &
    \includegraphics[height=0.18\linewidth]{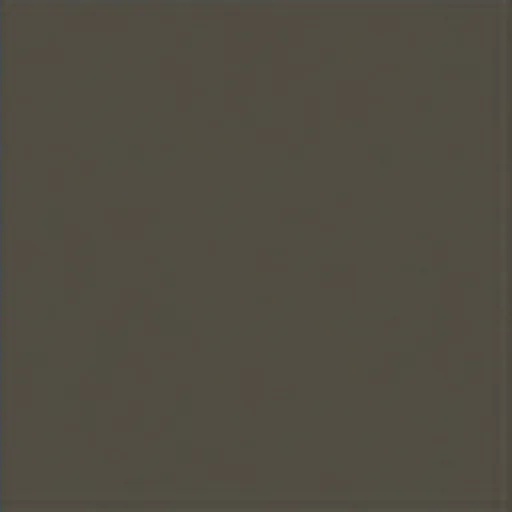} \\
    \includegraphics[height=0.18\linewidth]{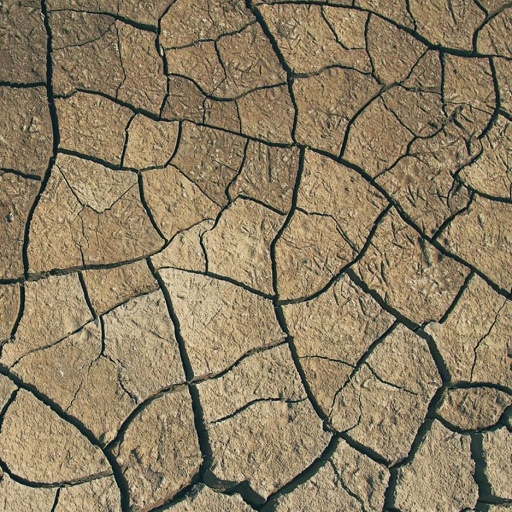} &
    \rowname{Ours}&
    \includegraphics[height=0.18\linewidth]{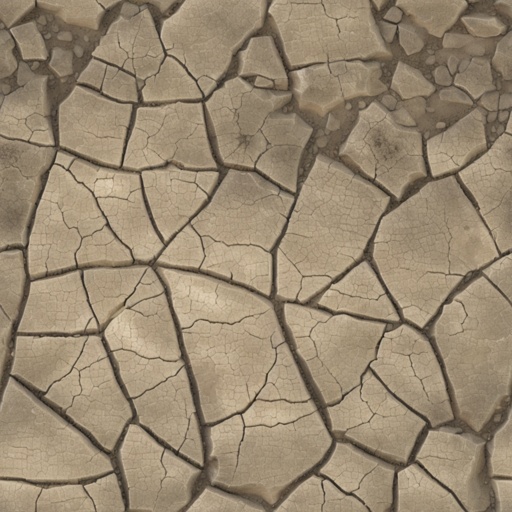} &
    \includegraphics[height=0.18\linewidth]{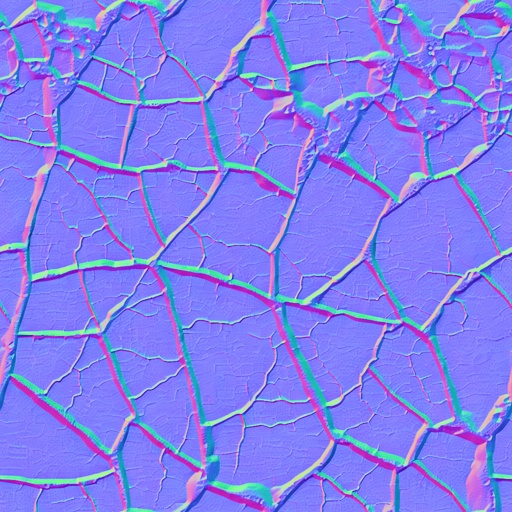} &
    \includegraphics[height=0.18\linewidth]{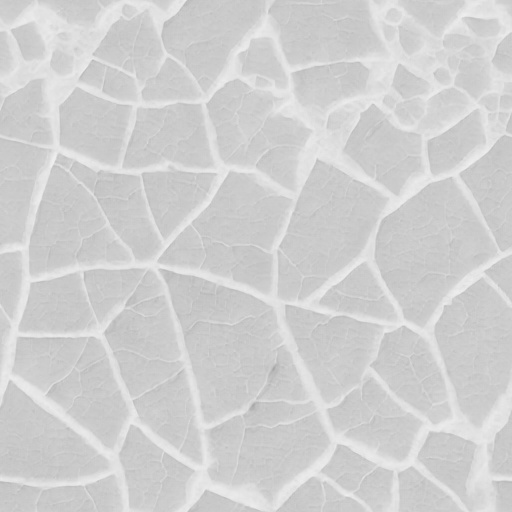} &
    \includegraphics[height=0.18\linewidth]{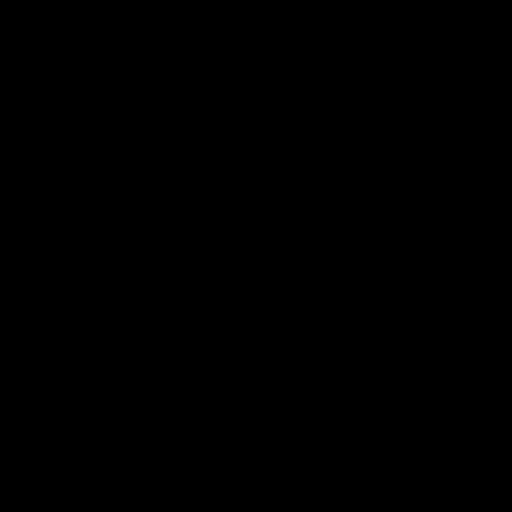} \\

    \end{tabular}
    \caption{\textbf{Material Generation Comparison with MatFuse \cite{MatFuse}.} Prompts are prefixed with \textit{"A material of"} for MatFuse and \textit{"texture of"} for our method. We compare three conditioning scenarios: text only, text with sketch, and text with reference image. Reference image from Pexels and Pixnio.}
\label{fig:generation_comparison}
\end{figure}

%% file: figures/normal_comparison/normal_comparison.tex
\begin{figure}[ht]
    \tiny
    \centering
    \setlength\tabcolsep{1pt}
    \settoheight{\tempdima}{\includegraphics[width=0.158\linewidth]{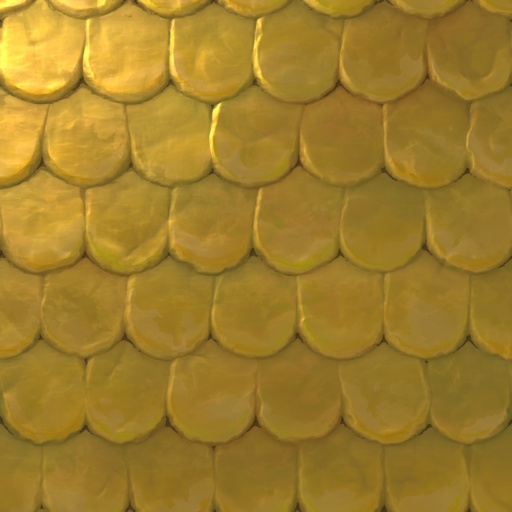}} 
    \begin{tabular}{@{} cccccc @{}}
    Input RGB & StableNormal\textsuperscript{$\ast$} & Lotus-D\textsuperscript{$\dagger$} & E2E-FT\textsuperscript{$\dagger$} & Ours & GT  \\
    \includegraphics[height=0.158\linewidth]{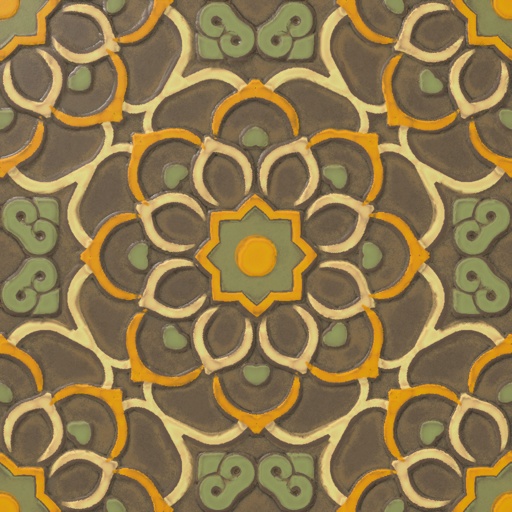} &
    \includegraphics[height=0.158\linewidth]{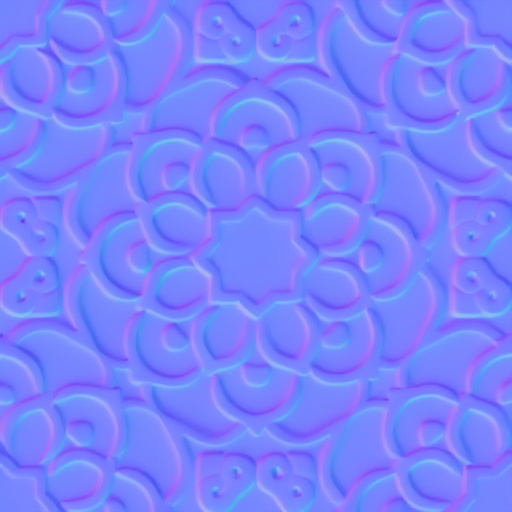} &
    \includegraphics[height=0.158\linewidth]{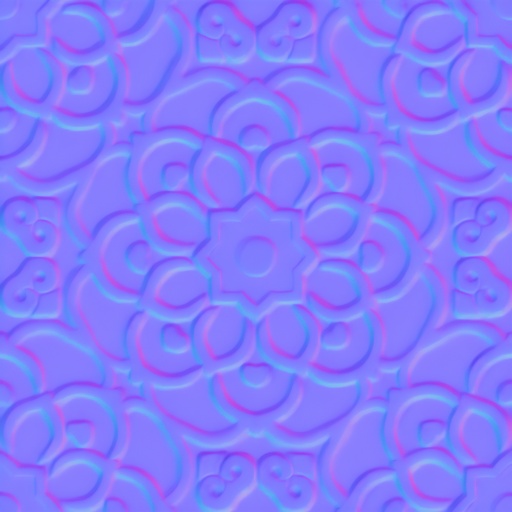} &
    \includegraphics[height=0.158\linewidth]{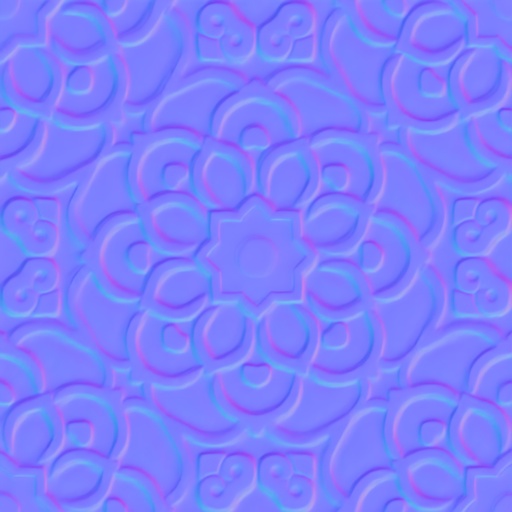} &
    \includegraphics[height=0.158\linewidth]{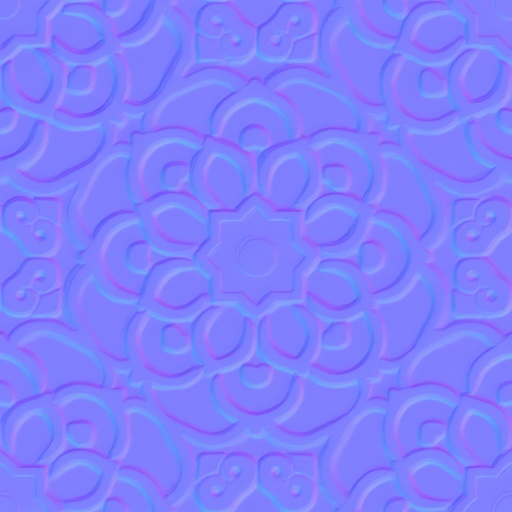} &
    \includegraphics[height=0.158\linewidth]{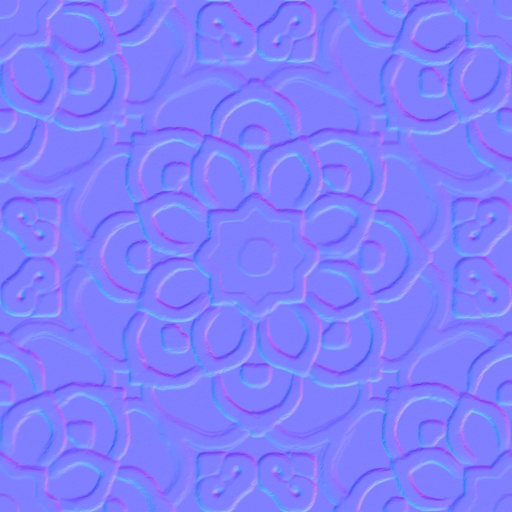} \\
    \includegraphics[height=0.158\linewidth]{figures/normal_comparison/Gold_RoofTiles_01/input_512.jpg} &
    \includegraphics[height=0.158\linewidth]{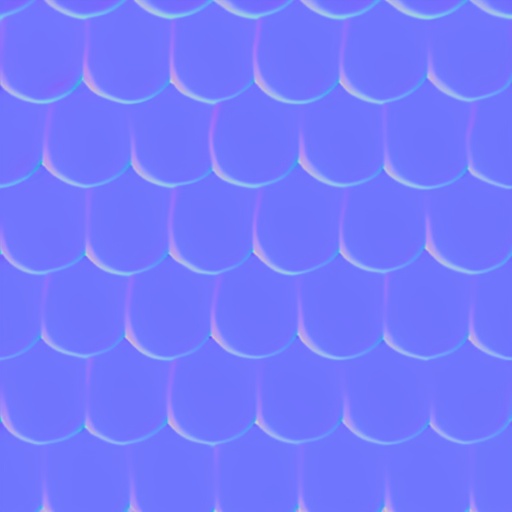} &
    \includegraphics[height=0.158\linewidth]{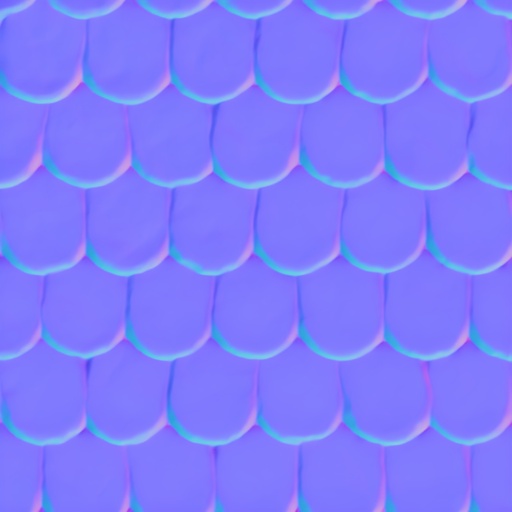} &
    \includegraphics[height=0.158\linewidth]{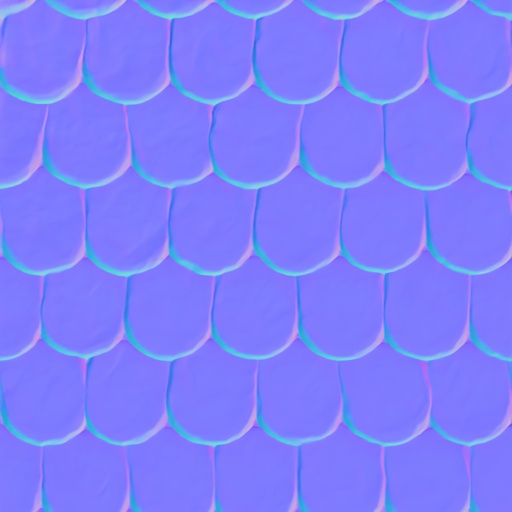} &
    \includegraphics[height=0.158\linewidth]{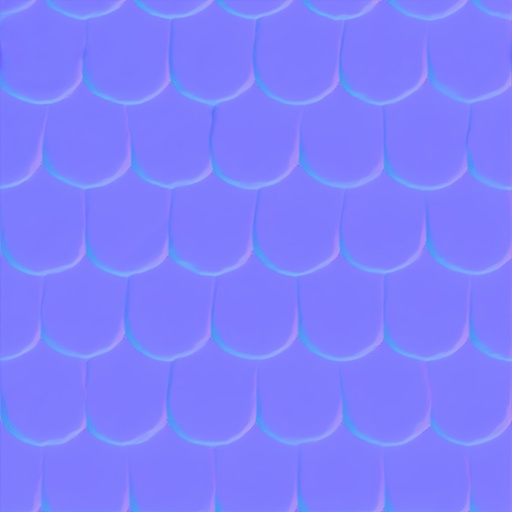} &
    \includegraphics[height=0.158\linewidth]{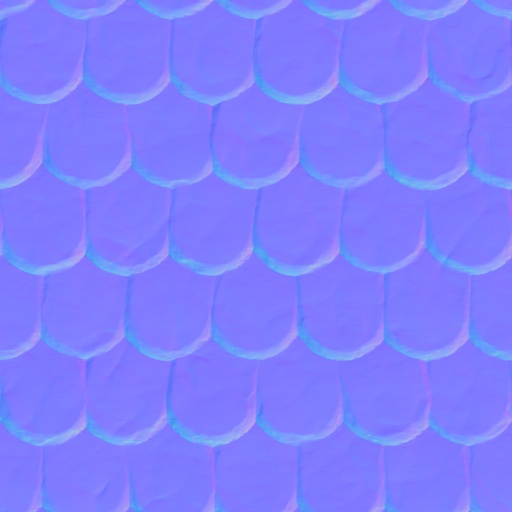} \\
    \end{tabular}
\caption{\textbf{Qualitative normal estimation comparisons.}}
\label{fig:normal_comparison}
\end{figure}

%% file: tables/normal_comparison.tex
\begin{table}[t]
    \centering
    \caption{
        \textbf{Normal estimation compared to single-modality methods on Substance test set.}  $\dagger$: trained on our dataset, $\ast$: author-provided weights.
    }
    \scalebox{0.82}{%
        \setlength{\tabcolsep}{2pt}%
        \begin{tabular}{cccc}
            \toprule
              \textbf{Method} & \textbf{PSNR}$\uparrow$ & \textbf{LPIPS}$\downarrow$ & \textbf{Number of Modalities}\\
            \midrule
              StableNormal\textsuperscript{$\ast$}  & 20.03 & 0.502 & 1\\
              Lotus-D\textsuperscript{$\dagger$}  & 22.76 & 0.371 & 1\\
              Lotus-G\textsuperscript{$\dagger$}  & 22.06 & 0.416 & 1\\
              E2E-FT\textsuperscript{$\dagger$}  & \textbf{24.07} & 0.379 & 1\\
              Ours & 23.32 & \textbf{0.334} & 4\\
            \bottomrule
        \end{tabular}
    }
    \label{tab:normal_comparison}
\end{table}

%% file: figures/pbr_comparison/pbr_comparison.tex
\begin{figure}[ht]
    \centering
    \tiny
    \setlength\tabcolsep{1pt}
    \setlength{\belowcaptionskip}{-10pt}
    \settoheight{\tempdima}{\includegraphics[width=0.183\linewidth]{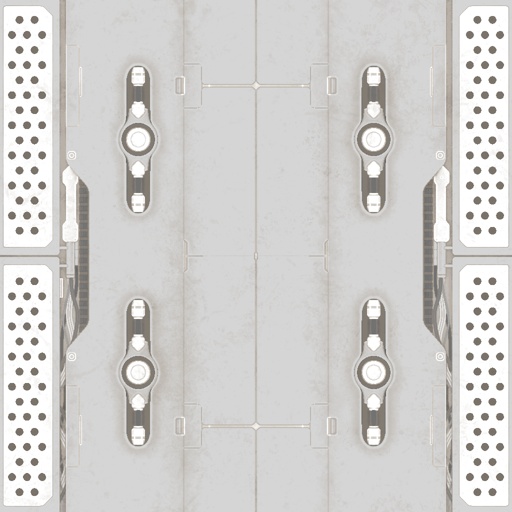}} 
    \begin{tabular}{@{}cccccc@{}}
    & Relit & Basecolor & Normal  & Roughness & Metalness \\
    \rowname{GT}&
    \includegraphics[height=0.183\linewidth]{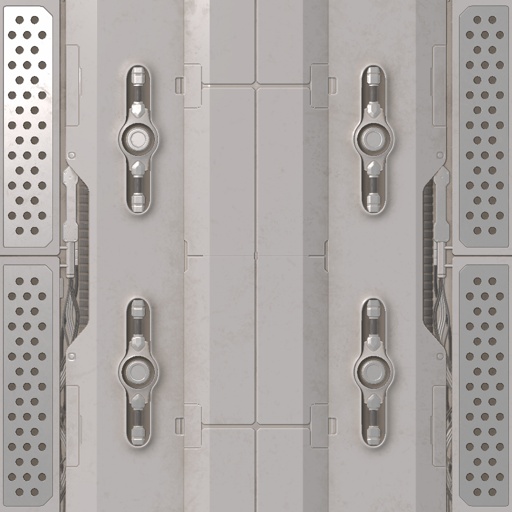} &
    \includegraphics[height=0.183\linewidth]{figures/pbr_comparison/Steel_SciFiCeiling_01/GT/basecolor_512.jpg} &
    \includegraphics[height=0.183\linewidth]{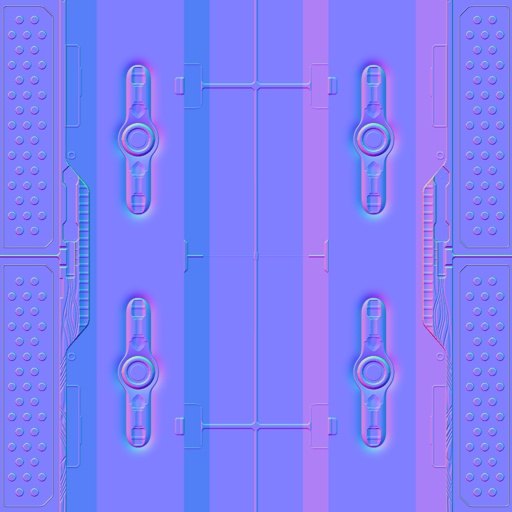} &
    \includegraphics[height=0.183\linewidth]{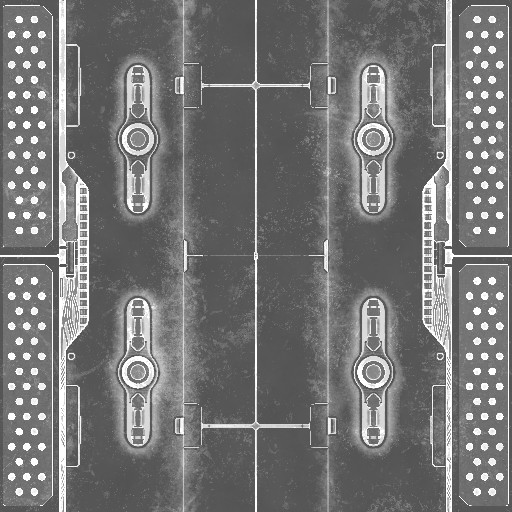} &
    \includegraphics[height=0.183\linewidth]{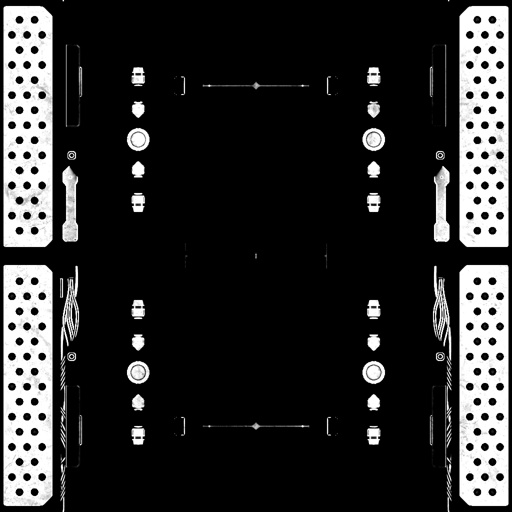} \\
    \rowname{Material Palette\textsuperscript{$\ast$}}&
    \includegraphics[height=0.183\linewidth]{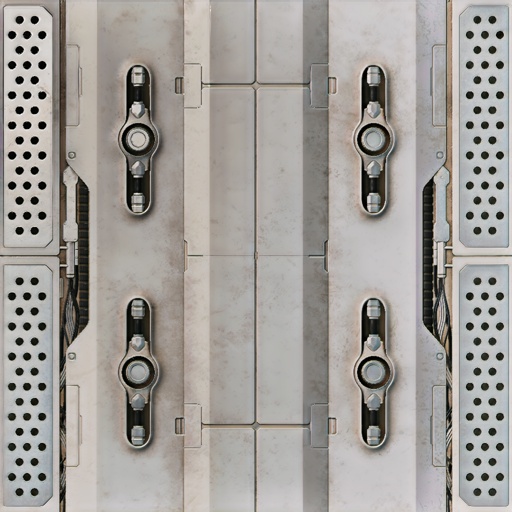} &
    \includegraphics[height=0.183\linewidth]{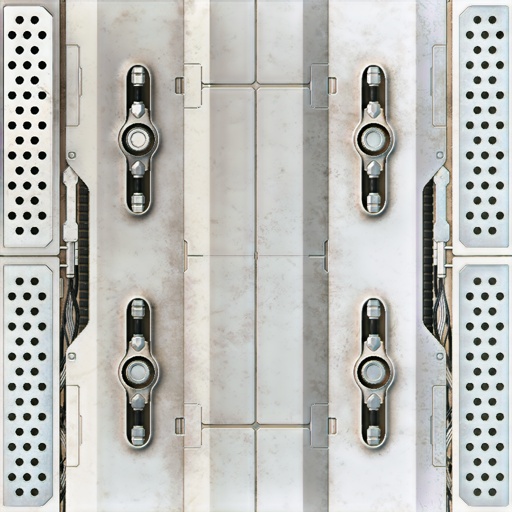} &
    \includegraphics[height=0.183\linewidth]{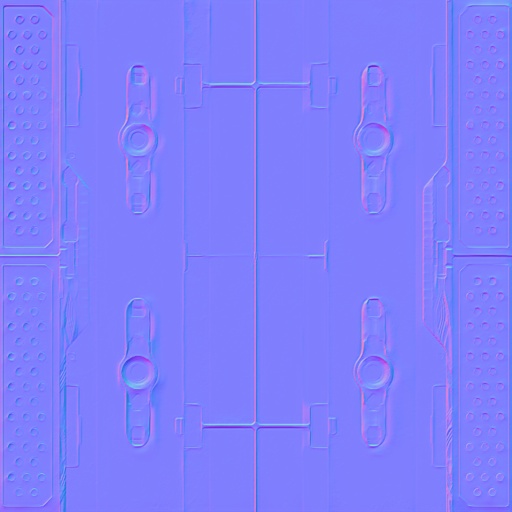} &
    \includegraphics[height=0.183\linewidth]{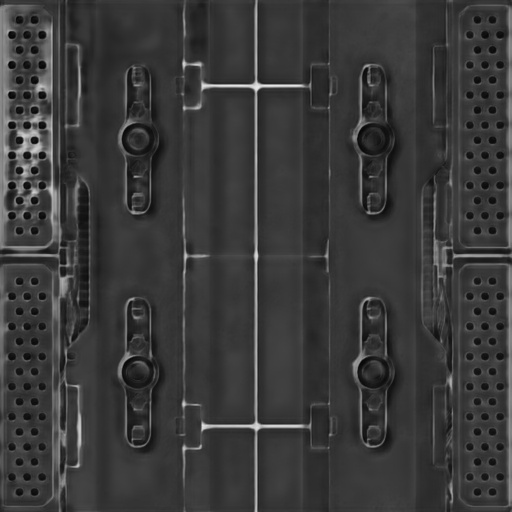} &
    \begin{minipage}[b]{0.183\linewidth}
        \centering
        \textbf{N/A}
        \vspace{3\baselineskip}
    \end{minipage}  \\
    \rowname{Sampler}&
    \includegraphics[height=0.183\linewidth]{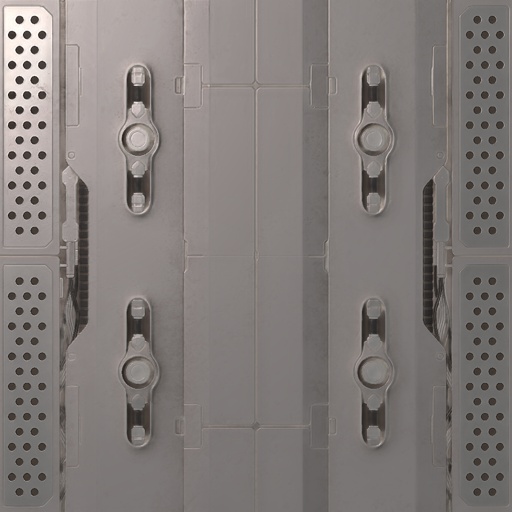} &
    \includegraphics[height=0.183\linewidth]{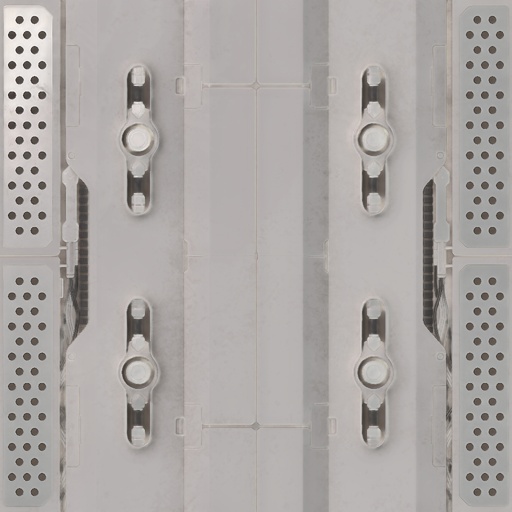} &
    \includegraphics[height=0.183\linewidth]{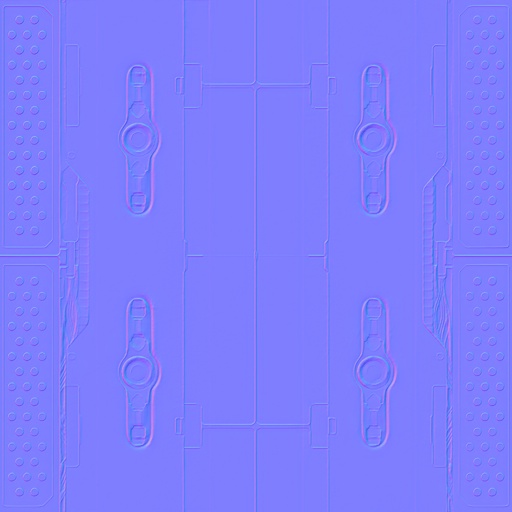} &
    \includegraphics[height=0.183\linewidth]{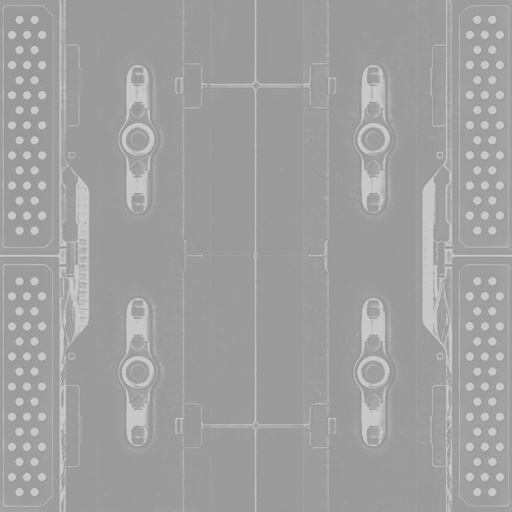} &
    \includegraphics[height=0.183\linewidth]{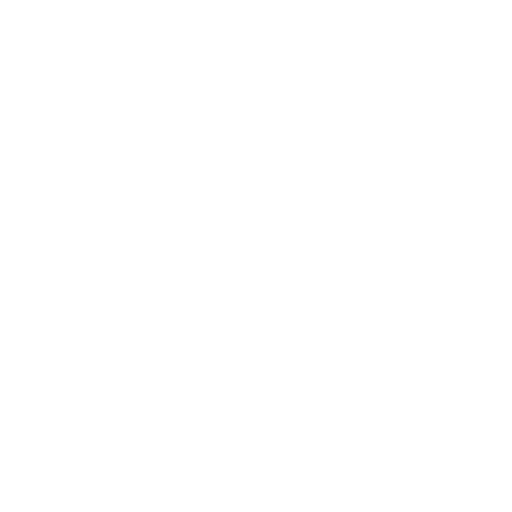} \\
    \rowname{SurfaceNet{$\dagger$}}&
    \includegraphics[height=0.183\linewidth]{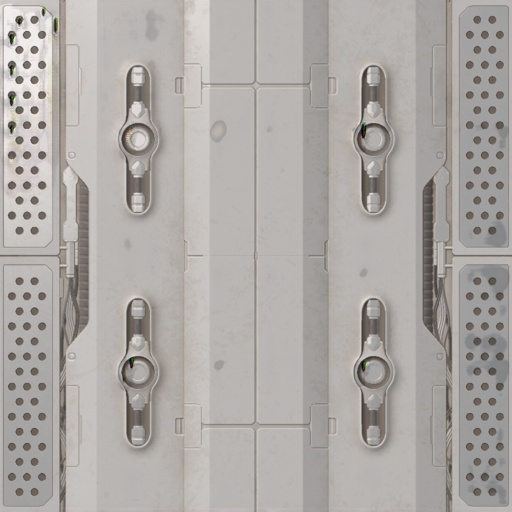} &
    \includegraphics[height=0.183\linewidth]{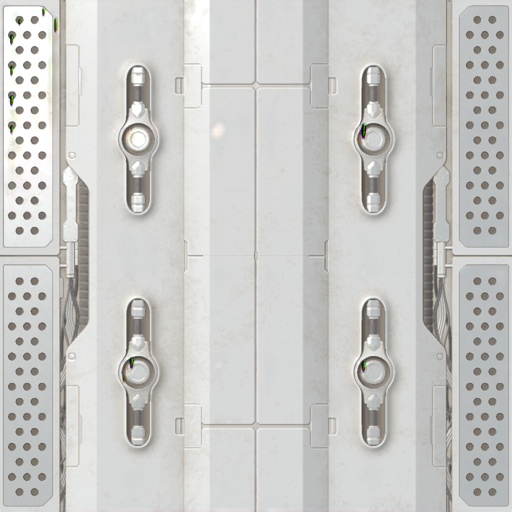} &
    \includegraphics[height=0.183\linewidth]{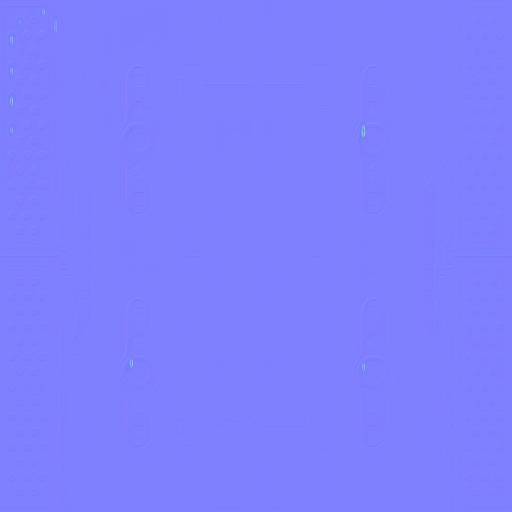} &
    \includegraphics[height=0.183\linewidth]{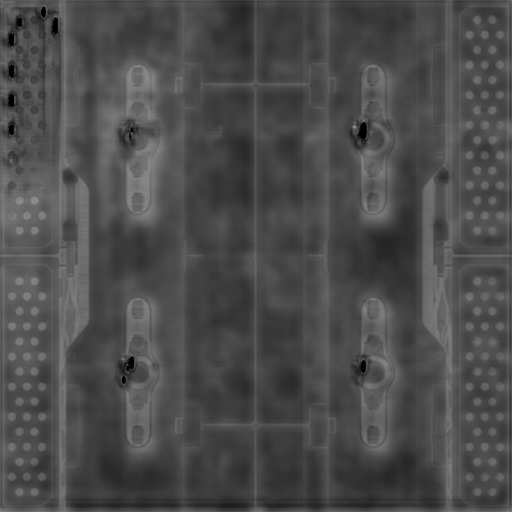} &
    \includegraphics[height=0.183\linewidth]{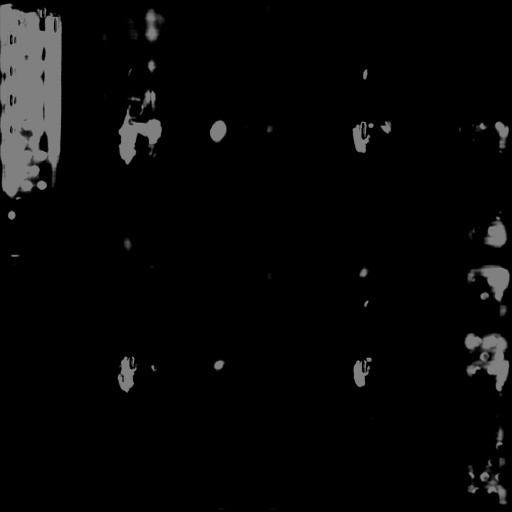}  \\
    \rowname{MatFusion\textsuperscript{$\dagger$}}&
    \includegraphics[height=0.183\linewidth]{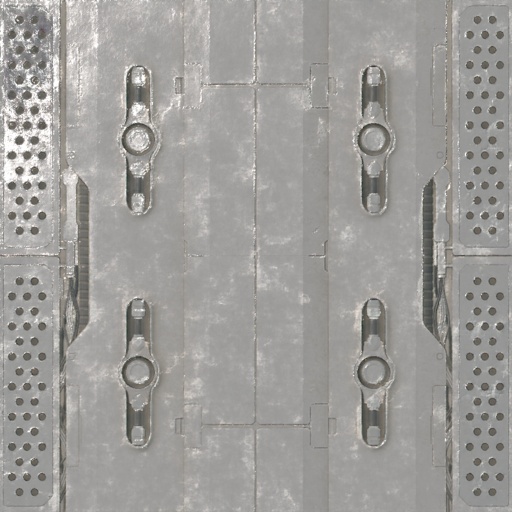} &
    \includegraphics[height=0.183\linewidth]{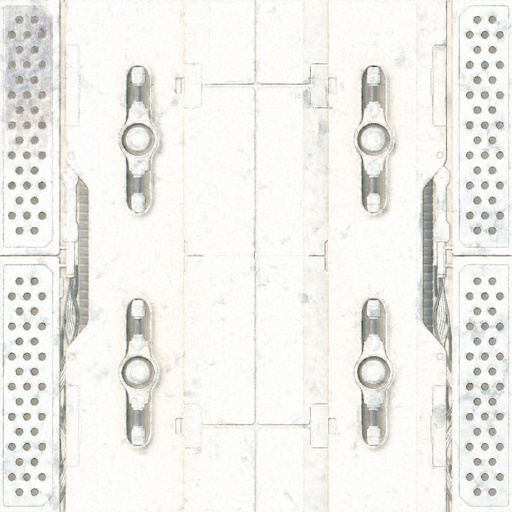} &
    \includegraphics[height=0.183\linewidth]{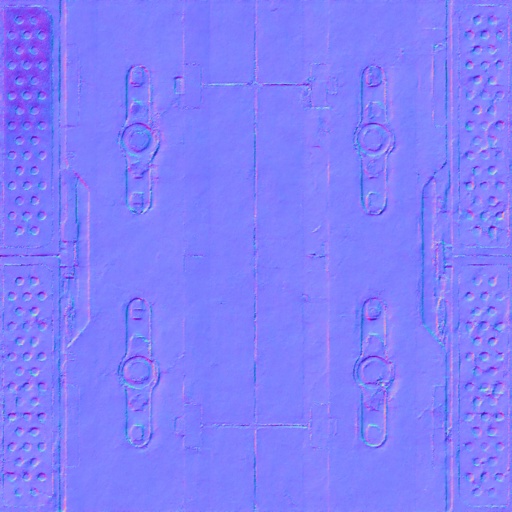} &
    \includegraphics[height=0.183\linewidth]{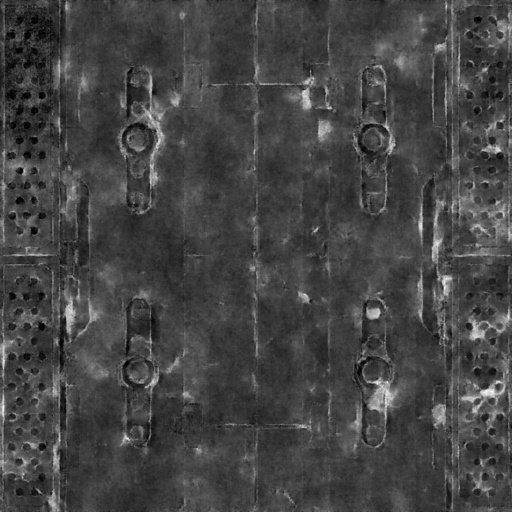} &
    \includegraphics[height=0.183\linewidth]{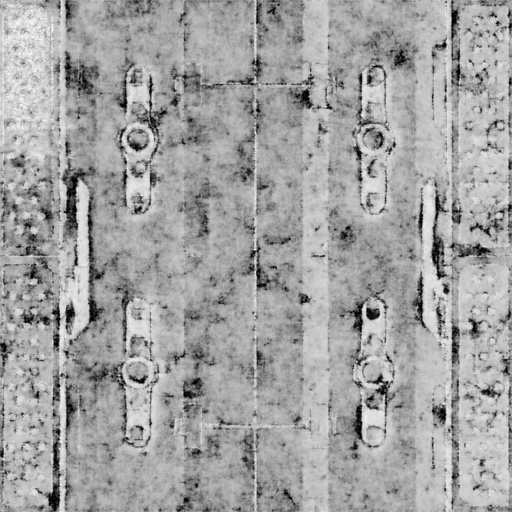} \\
    \rowname{RGB$\rightarrow$X\textsuperscript{$\dagger$}}&
    \includegraphics[height=0.183\linewidth]{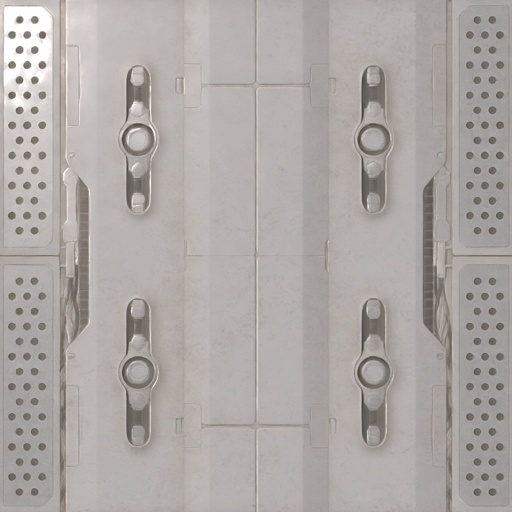} &
    \includegraphics[height=0.183\linewidth]{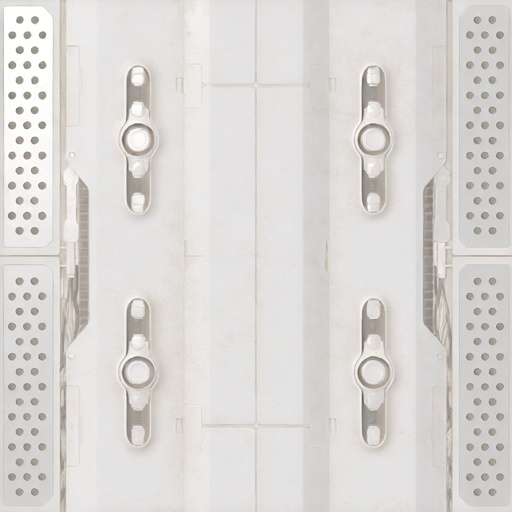} &
    \includegraphics[height=0.183\linewidth]{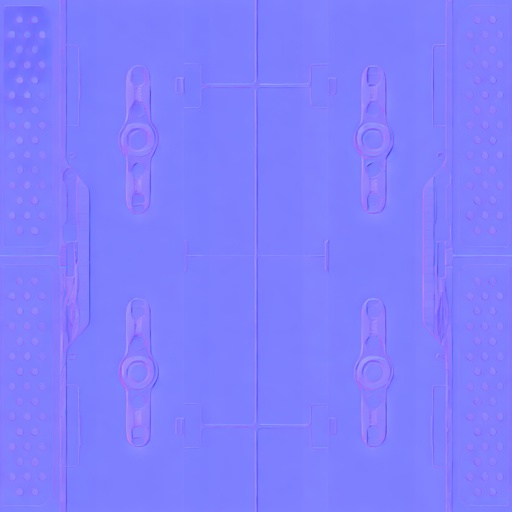} &
    \includegraphics[height=0.183\linewidth]{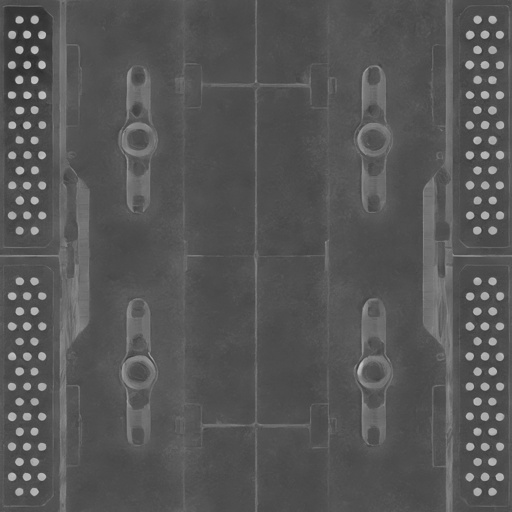} &
    \includegraphics[height=0.183\linewidth]{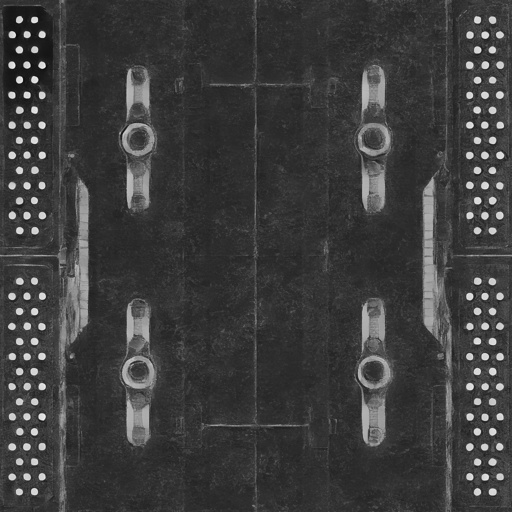}  \\
    \rowname{Ours}&
    \includegraphics[height=0.183\linewidth]{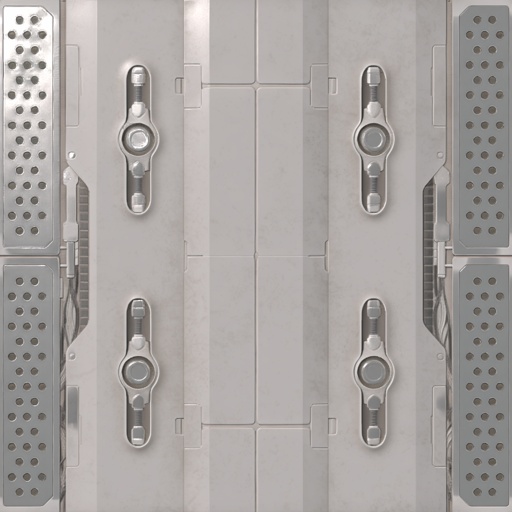} &
    \includegraphics[height=0.183\linewidth]{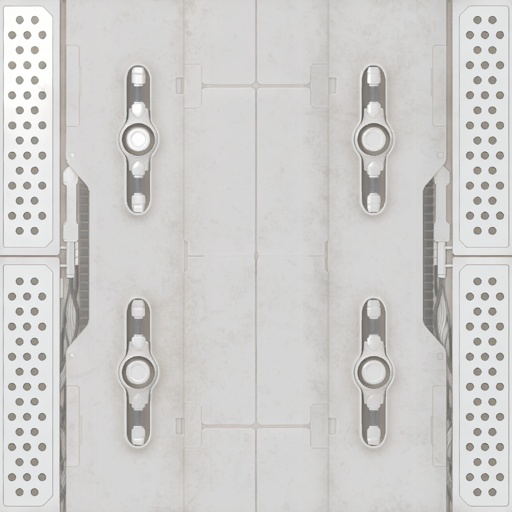} &
    \includegraphics[height=0.183\linewidth]{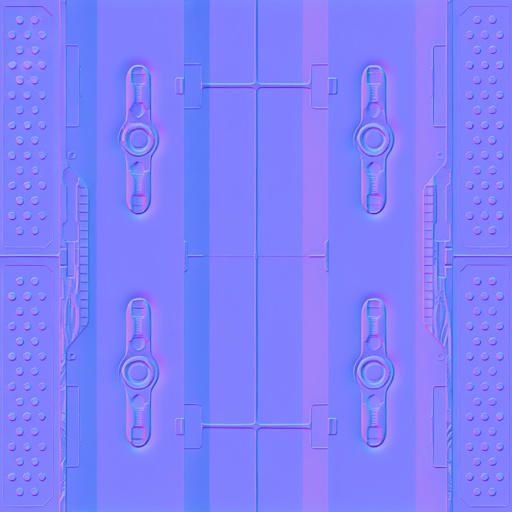} &
    \includegraphics[height=0.183\linewidth]{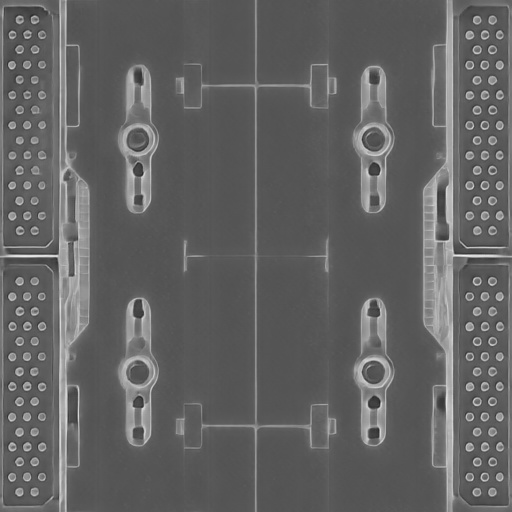} &
    \includegraphics[height=0.183\linewidth]{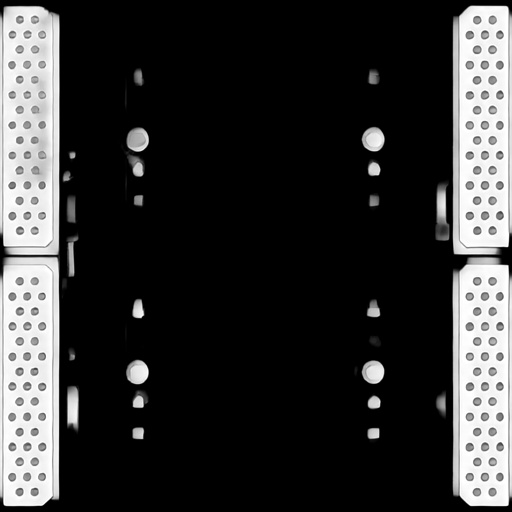}  \\
    \end{tabular}
    \caption{\textbf{Qualitative PBR estimation comparisons.} We also include results from Substance 3D Sampler's feature called "AI-Powered Image to Material".}
\label{fig:pbr_comparison}
\end{figure}

%% file: tables/pbr_comparison.tex
\begin{table}[t]
    \centering
    \caption{
        \textbf{Material estimation on MatSynth test split.} $\dagger$: trained on our dataset, $\ast$: author-provided weights.
    }    
    \scalebox{0.82}{%
        \setlength{\tabcolsep}{2pt}%
        \begin{tabular}{ccccccc}
            \toprule
            {\textbf{Method}} & {\textbf{Basecolor}} & & {\textbf{Normal}} & & {\textbf{Roughness}} & \\
            \cmidrule(lr){2-3} 
            \cmidrule(lr){4-5}
            \cmidrule(lr){6-7}
            & \textbf{PSNR}$\uparrow$ & {\textbf{LPIPS}$\downarrow$}  & \textbf{PSNR}$\uparrow$ & {\textbf{LPIPS}$\downarrow$} & \textbf{PSNR}$\uparrow$  & {\textbf{LPIPS}$\downarrow$}  \\
            \cmidrule(lr){1-7} 
            {Material Palette\textsuperscript{$\ast$}} & {17.09}  & {0.381}   &  {22.78} & {0.414} &  {15.00} &  {0.593}  \\
            {SurfaceNet\textsuperscript{$\dagger$}} & {24.93}  & {0.345}  & {25.45}  & {0.489} & {16.09} & {0.605} \\
            {MatFusion\textsuperscript{$\dagger$}}& {24.39}  & {0.377}    & {23.00} & {0.486} & {14.59} & {0.624} \\         
            {RGB$\rightarrow$X\textsuperscript{$\dagger$}} & {25.79}  & {0.313}  & {25.67}  & {0.361} & {18.11} & {\textbf{0.512}} \\
            {Ours}  & {\textbf{28.78}} & {\textbf{0.269}} & {\textbf{26.65}}  &  {\textbf{0.353}}  &  {\textbf{19.31}} &  {0.532} \\
            \midrule
            {\textbf{Method}}  & {\textbf{Metalness}} & & {\textbf{Height}} & & {\textbf{Relit}} & \\
            \cmidrule(lr){2-3} 
            \cmidrule(lr){4-5}
            \cmidrule(lr){6-7}
            {}  & \textbf{PSNR}$\uparrow$ & {\textbf{LPIPS}$\downarrow$} & \textbf{PSNR}$\uparrow$ & {\textbf{LPIPS}$\downarrow$} & \textbf{PSNR}$\uparrow$ & {\textbf{LPIPS}$\downarrow$} \\
            \cmidrule(lr){1-7}
            {SurfaceNet\textsuperscript{$\dagger$}}  & {61.16}  & {0.142}  & {17.72} & {0.534} & {22.55} & {0.363}\\
            {MatFusion\textsuperscript{$\dagger$}}  & {29.91}  & {0.436}  & {17.52} & {0.580} & {22.29} & {0.382}\\
            {RGB$\rightarrow$X\textsuperscript{$\dagger$}}  & {61.40}  & {0.102}  & {18.80} & {0.508} & {23.24} & {0.344}\\
            {Ours} & {\textbf{71.93}} & {\textbf{0.089}}  & {\textbf{19.36}}  &  {\textbf{0.467}}  &  {\textbf{24.40}} &  {\textbf{0.318}}\\
            \bottomrule
        \end{tabular}
    }%
    \label{tab:pbr_comparison_matsynth}
\end{table}

\begin{table}[t]
    \centering
    \caption{
        \textbf{Material estimation on Substance test set.} $\dagger$: trained on our dataset, $\ast$: author-provided weights.
    }    
    \scalebox{0.82}{%
        \setlength{\tabcolsep}{2pt}%
        \begin{tabular}{ccccccc}
            \toprule
            {\textbf{Method}} & {\textbf{Basecolor}} & & {\textbf{Normal}} & & {\textbf{Roughness}} & \\
            \cmidrule(lr){2-3} 
            \cmidrule(lr){4-5}
            \cmidrule(lr){6-7}
            & \textbf{PSNR}$\uparrow$ & {\textbf{LPIPS}$\downarrow$}  & \textbf{PSNR}$\uparrow$ & {\textbf{LPIPS}$\downarrow$} & \textbf{PSNR}$\uparrow$  & {\textbf{LPIPS}$\downarrow$}  \\
            \cmidrule(lr){1-7} 
            {Material Palette\textsuperscript{$\ast$}} & {16.04}  & {0.384}   &  {19.97} & {0.417} &  {14.32} &  {0.590}  \\
            {SurfaceNet\textsuperscript{$\dagger$}} & {24.60}  & {0.326}  & {21.91}  & {0.508} & {15.02} & {0.607} \\
            {MatFusion\textsuperscript{$\dagger$}} & {25.35}  & {0.347}    & {20.69} & {0.503} & {14.08} & {0.622} \\         
            {RGB$\rightarrow$X\textsuperscript{$\dagger$}} & {25.72}  & {0.317}  & {22.56}  & {0.364} & {16.72} & {\textbf{0.536}} \\
            {Ours} & {\textbf{29.05}} & {\textbf{0.269}}  & {\textbf{23.32}}  &  {\textbf{0.334}}  &  {\textbf{17.48}} &  {0.550} \\
            \midrule
            {\textbf{Method}}  & {\textbf{Metalness}} & & {\textbf{Height}} & & {\textbf{Relit}} & \\
            \cmidrule(lr){2-3} 
            \cmidrule(lr){4-5}
            \cmidrule(lr){6-7}
            {}  & \textbf{PSNR}$\uparrow$ & {\textbf{LPIPS}$\downarrow$} & \textbf{PSNR}$\uparrow$ & {\textbf{LPIPS}$\downarrow$} & \textbf{PSNR}$\uparrow$ & {\textbf{LPIPS}$\downarrow$} \\
            \cmidrule(lr){1-7}
            {SurfaceNet\textsuperscript{$\dagger$}} & {64.45}  & {0.118}  & {21.09} & {0.512} & {21.54} & {0.344}\\
            {MatFusion\textsuperscript{$\dagger$}} & {32.66}  & {0.441}  & {20.12} & {0.526} & {21.75} & {0.362}\\
            {RGB$\rightarrow$X\textsuperscript{$\dagger$}} & {64.96}  & {0.088}  & {22.64} & {0.489} & {21.91} & {0.340}\\
            {Ours}  & {\textbf{77.67}} & {\textbf{0.064}}  & {\textbf{23.36}}  &  {\textbf{0.470}}  &  {\textbf{22.94}} &  {\textbf{0.309}}\\
            \bottomrule
        \end{tabular}
    }%
    \label{tab:pbr_comparison_substance}
\end{table}

%% file: tables/inference_time.tex
\begin{table}[t]
    \centering
    \caption{
        \textbf{Inference time for full-modality methods.} Times are measured on an RTX A6000 GPU.
    }
    \scalebox{0.82}{%
        \setlength{\tabcolsep}{2pt}%
        \begin{tabular}{ccccc}
            \toprule
              \textbf{} & \textbf{SurfaceNet} & \textbf{MatFusion} & \textbf{RGB$\rightarrow$X} & \textbf{Ours}\\
            \midrule
              Time (Seconds)  & 0.6 & 46.4 & 23.6 & 2.1\\
            \bottomrule
        \end{tabular}
    }
    \label{tab:inference_time}
\end{table}

%% file: tables/ablation.tex
\begin{table*}[ht]
    \footnotesize
    \centering
    \caption{
        \textbf{Material estimation ablation study on Substance test set.} 
    }
    \scalebox{1.0}{
        \setlength{\tabcolsep}{2pt}%
        \begin{tabular}{lcccccccccccccc}
            \toprule
            {\textbf{Method}} & {\textbf{Basecolor}} & & {\textbf{Normal}} & & {\textbf{Roughness}} & & {\textbf{Metalness}} & & {\textbf{Height}} & & {\textbf{Relit}} & & {\textbf{Inference}} & {\textbf{Parameters}}\\
            \cmidrule(lr){2-3} 
            \cmidrule(lr){4-5}
            \cmidrule(lr){6-7}
            \cmidrule(lr){8-9} 
            \cmidrule(lr){10-11}
            \cmidrule(lr){12-13}
            \cmidrule(lr){14-14}
            \cmidrule(lr){15-15}
            {} & \textbf{PSNR}$\uparrow$ & {\textbf{LPIPS}$\downarrow$}  & \textbf{PSNR}$\uparrow$ & {\textbf{LPIPS}$\downarrow$} & \textbf{PSNR}$\uparrow$  & {\textbf{LPIPS}$\downarrow$} & \textbf{PSNR}$\uparrow$ & {\textbf{LPIPS}$\downarrow$}  & \textbf{PSNR}$\uparrow$ & {\textbf{LPIPS}$\downarrow$} & \textbf{PSNR}$\uparrow$  & {\textbf{LPIPS}$\downarrow$} & {(Seconds)} & {(B)}\\
            \cmidrule(lr){1-15} 
            {RGB$\rightarrow$X}  & {25.72}  & {0.317} & {22.56}  & {0.364} & {16.72} & {\textbf{0.536}} & {64.96} & {0.088}  & {22.64}  &  {0.489}  &  {21.91} &  {0.340} & {23.6} & {1.3}
            \\
            \cmidrule(lr){1-15} 
            {\quad+Single-step}  & {27.95} & {0.354}  & {23.06}  &  {0.400}  &  {18.15} &  {0.640} & {58.86} & {0.120}  & {23.33}  &  {0.479}  &  {22.55} &  {0.373} & {1.9} & {1.3}
            \\
            {\quad+Combined Loss}  & {28.12} & {0.285}  & {22.96}  &  {0.345}  &  {18.39} &  {0.566} & {73.58} & {0.071}  & {22.95}  &  {0.472}  &  {22.42} &  {0.315} & {1.9} & {1.3}
            \\
            {\quad+Chain}  & {28.29} & {0.284}  & {23.04}  &  {0.346}  &  {18.24} &  {0.571} & {73.68} & {0.083}  & {23.36}  &  {0.475}  &  {22.50} &  {0.310} & {1.9} & {1.3}
            \\
            {\quad+LEGO-conditioning}  & {28.45} & {0.283}  & {23.16}  &  {0.343}  &  {\textbf{18.59}} &  {0.559} & {72.08} & {0.085}  & {22.92}  &  {0.474}  &  {22.52} &  {0.315} & {1.9} & {1.4}
            \\
            {\quad+Approx. Irradiance}  & {28.53} & {0.291}  & {23.25}  &  {0.339}  &  {18.50} &  {0.576} & {\textbf{80.14}} & {\textbf{0.061}}  & {\textbf{23.54}}  &  {0.476}  &  {22.72} &  {0.316} & {1.9} & {1.4}
            \\
            {\quad+RM Grid Search}  & {28.39} & {0.286}  & {23.26}  &  {0.340}  &  {17.91} &  {0.594} & {70.02} & {0.076}  & {23.53}  &  {0.471}  &  {22.76} &  {0.314} & {2.1} & {1.4}
            \\
            \cmidrule(lr){1-15} 
            {\quad+Render Loss (Ours)}  & {28.59} & {0.283}  & {23.31}  &  {0.344}  &  {16.97} &  {0.633} & {78.61} & {0.077}  & {23.34}  &  {\textbf{0.469}}  &  {22.88} &  {0.311} & {2.1} & {1.4}
            \\
            {\quad+Pretraining (Ours)}  & {\textbf{29.05}} & {\textbf{0.269}}  & {\textbf{23.32}}  &  {\textbf{0.334}}  &  {17.48} &  {0.550} & {77.67} & {0.064}  & {23.36}  &  {0.470}  &  {\textbf{22.94}} &  {\textbf{0.309}} & {2.1} & {1.4}
            \\
            
            \bottomrule
        \end{tabular}
    }
    \label{tab:ablation}
\end{table*}

%% file: secs/6_applications.tex
\section{Applications}
\label{sec:application}
In this section, we explore the potential applications of our method. Due to the generate-and-estimate framework, our method inherits the versatility and controllability of the large text-to-image models. Specifically, we leverage tools such as ControlNet \cite{zhang2023ContrlNet}, IP-Adapter \cite{ye2023ip-adapter} and RePaint \cite{lugmayr2022repaint} to guide the texture RGB generation process and subsequently estimate SVBRDF channels. Examples can be found in Fig. \ref{fig:art_gallery} and Fig. \ref{fig:more_applications}.

\paragraph{Text to Material} 
In this application, users provide a text prompt to guide the generation process. This operates similarly to standard text-to-image methods, with the added stage of material estimation.

\paragraph{Image to Material} 
Leveraging the image-to-image functionality of the generation stage, users can provide a reference image to guide the generation process. This facilitates applications such as reference-based material synthesis and variation generation.

\paragraph{Structure-controlled Generation} 
With additional pre-trained ControlNet weights, such as those for line art or depth maps, users can control the generation process to preserve structural layouts specified by the input conditions, enhancing user controllability.

\paragraph{Material Editing} 
By incorporating an additional in-painting model trained using methods such as RePaint \cite{lugmayr2022repaint}, users can edit generated materials. For example, they can mask specific areas of the texture RGB and regenerate to modify the material.

%% file: secs/7_discussions.tex
\section{Discussion and Future Work}


\paragraph{Lighting Assumption} 
Our generate-and-estimate framework assumes generated RGB textures ($I_{\text{RGB}}$) have consistent directional lighting. This usually works but fails for glossy surfaces, leading to less accurate material estimation. The main problem is a conflict: $I_{\text{RGB}}$ must be tileable and lit by directional light. Our circular padding makes generated images tileable, but causes edge gradients and highlights of glossy surface to wrap around and bleed across the entire image. This spreads incorrect lighting across the image, creating inconsistencies expected by the material estimation pipeline. In the future, we could generate two images per material: one without tile constraints to capture specular effects, and one tileable for non-specular details, then process them through separate branches.

\paragraph{Baked-in Shadows in Basecolor}
Our material estimation results occasionally show baked in shadows in the predicted basecolor channel. We believe this issue arises from the presence of baked ambient occlusion (AO) in many materials within our training dataset. This problem can be reduced by using higher quality material data or by applying shadow removal techniques, such as the method proposed in~\cite{zhu2022bijective, zhao2025hunyuan3d}, to preprocess our dataset.

\paragraph{Limited Generalization Ability}
We employ single-step fine-tuning to train our material estimation model. While this approach yields stronger performance on PBR material test datasets, it can compromise generalization to in-the-wild images (see Fig.\ref{fig:in_the_wild}, row 4, basecolor). This limitation could be mitigated in future work by adopting alternative single-step or few-step generative modeling techniques, such as rectified flow\cite{liu2022flow} or mean flow~\cite{geng2025mean}.

%% file: secs/8_conclusion.tex
\section{Conclusion}
In this paper, we introduce a generate-and-estimate framework for material generation. The texture RGB generation stage leverages the capabilities of large text-to-image models to ensure prompt alignment and user control, while maintaining consistent lighting across generated images. In the material estimation stage, we introduce a novel chain-of-rendering-decomposition pipeline that leverages the known lighting assumptions to sequentially estimate SVBRDF modalities. This step-by-step approach integrates PBR cues directly into the conditioning inputs of the estimation model. Our method significantly outperforms baseline approaches in both user control alignment and output quality, and demonstrates strong versatility across a range of material generation applications.

To the best of our knowledge, this is the first work to introduce chained image-conditional diffusion models with PBR-aware rendering decomposition, offering a practical and effective solution to the inherently under-constrained problem of inverse rendering in material estimation.

%% file: secs/figure_only.tex
\input{figures/art_gallery/art_gallery}
\input{figures/in-the-wild/in-the-wild}
\input{figures/applications/applications}

%% file: figures/art_gallery/art_gallery.tex
\begin{figure*}[ht]
    \centering
    \setlength\tabcolsep{1pt}
    \settoheight{\tempdima}{\includegraphics[width=0.15\textwidth]{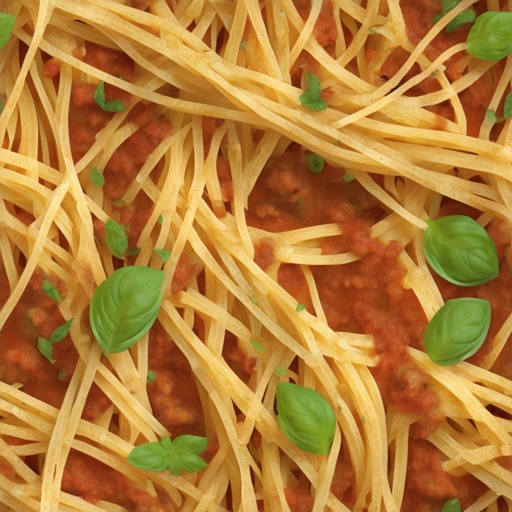}} 
    \begin{tabular}{@{} ccccccc @{}}
    Prompt & Texture RGB & Basecolor & Normal & Roughness & Metalness & Render \\
    \begin{minipage}[b]{0.12\linewidth}
    \centering
        \begin{prompt}
            \textbf{"texture of delicious spaghetti with sauce, basil leaves"}
        \end{prompt}
    \vspace{2\baselineskip}
    \end{minipage} &
    \includegraphics[height=0.12\textwidth]{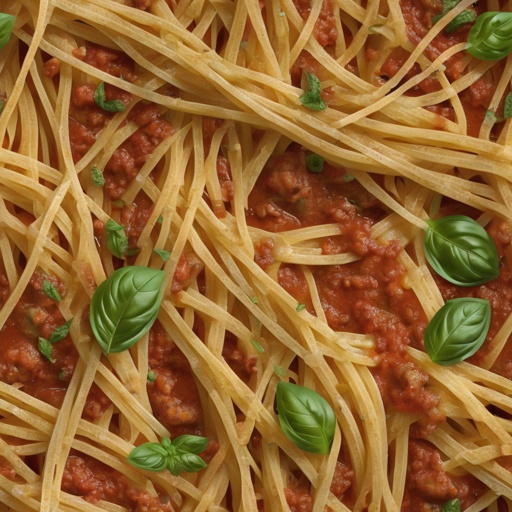} &
    \includegraphics[height=0.12\textwidth]{figures/art_gallery/pasta/basecolor_512.jpg} &
    \includegraphics[height=0.12\textwidth]{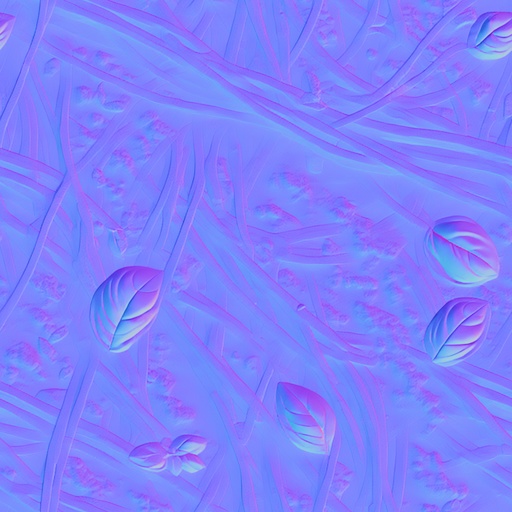} &
    \includegraphics[height=0.12\textwidth]{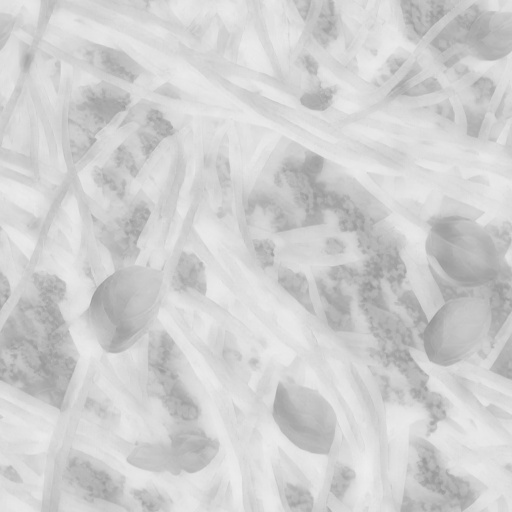} &
    \includegraphics[height=0.12\textwidth]{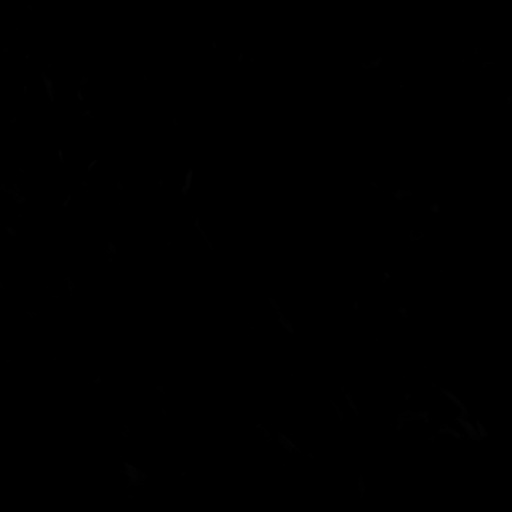} &
    \includegraphics[height=0.12\textwidth]{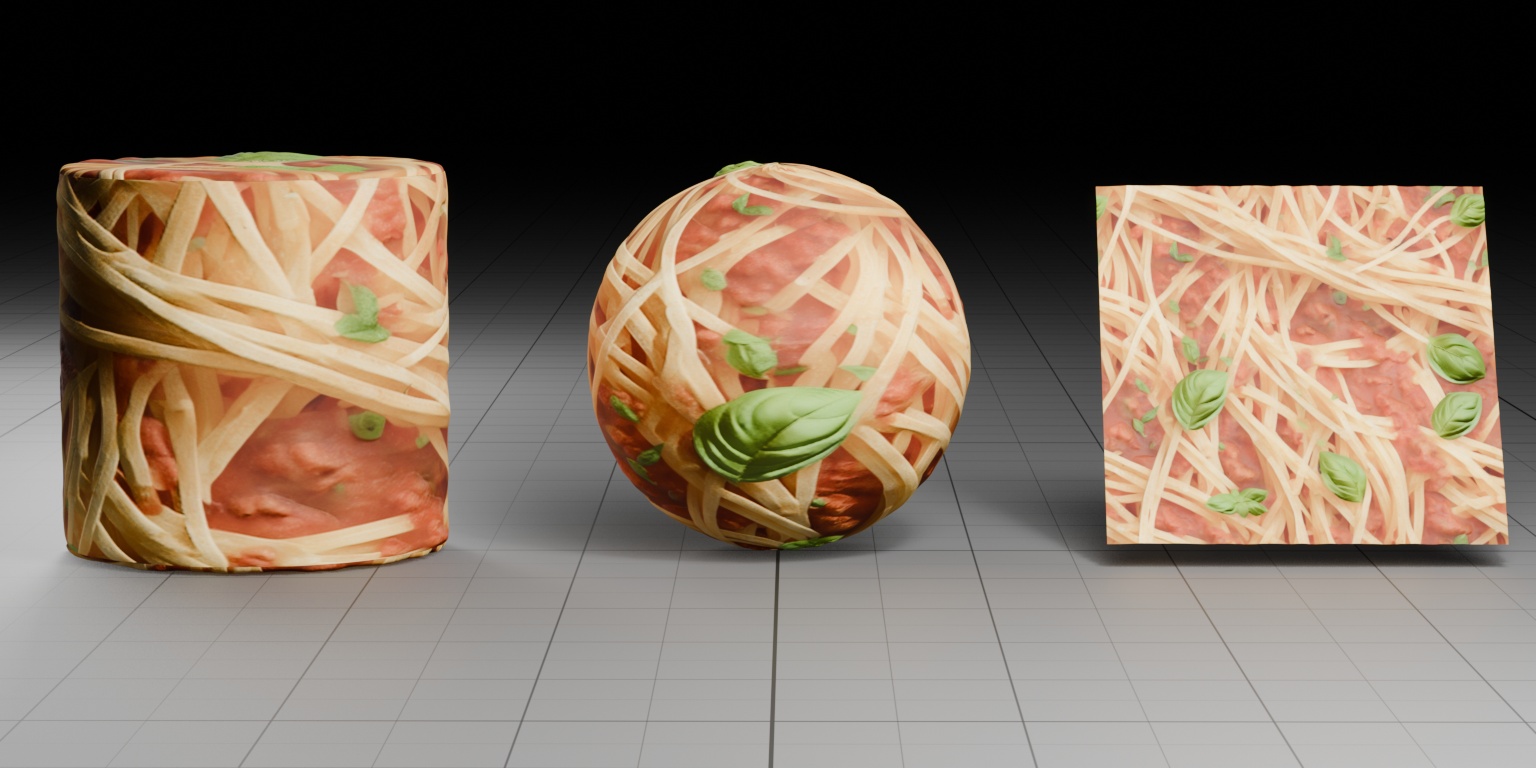}\\
    \begin{minipage}[b]{0.12\linewidth}
    \centering
        \begin{prompt}
            \textbf{"texture of gray brown rough stone wall, moss patches, lichen, organic"}
        \end{prompt}
    \vspace{2\baselineskip}
    \end{minipage} &
    \includegraphics[height=0.12\textwidth]{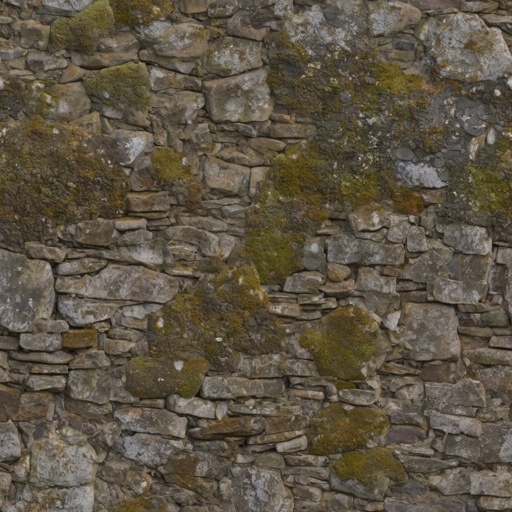} &
    \includegraphics[height=0.12\textwidth]{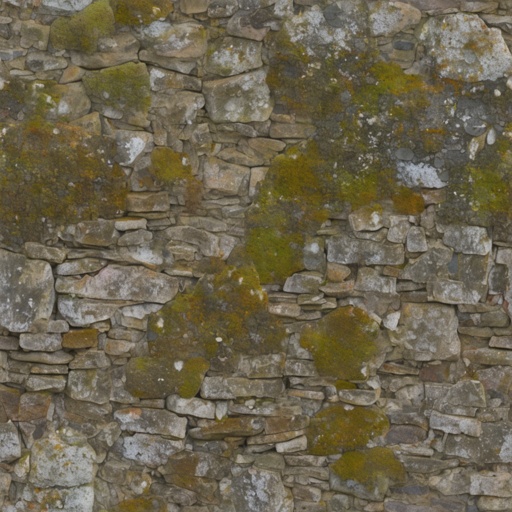} &
    \includegraphics[height=0.12\textwidth]{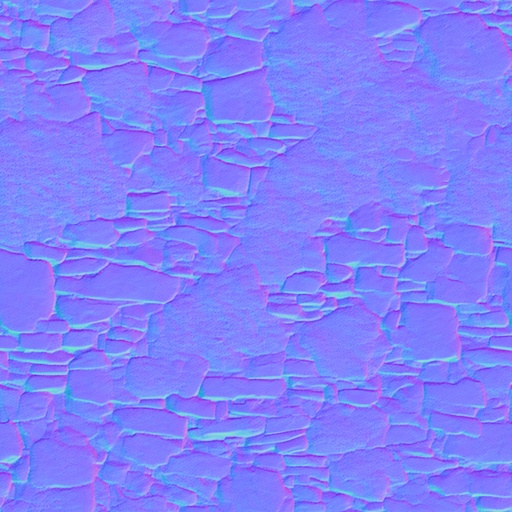} &
    \includegraphics[height=0.12\textwidth]{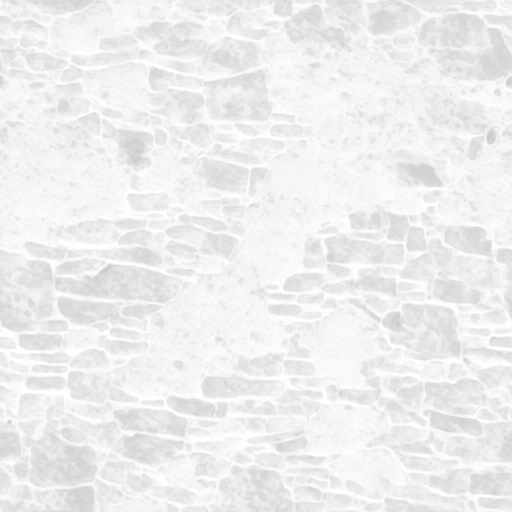} &
    \includegraphics[height=0.12\textwidth]{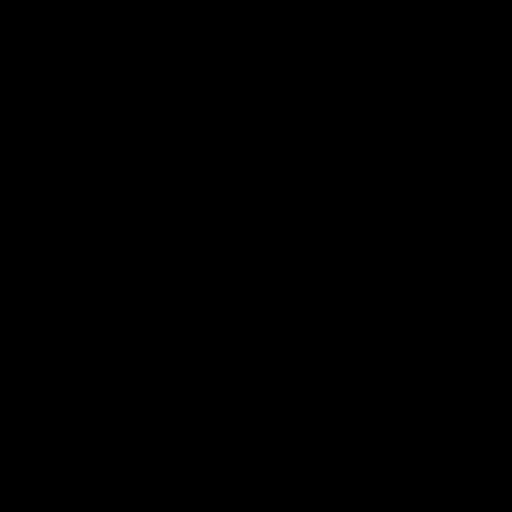} &
    \includegraphics[height=0.12\textwidth]{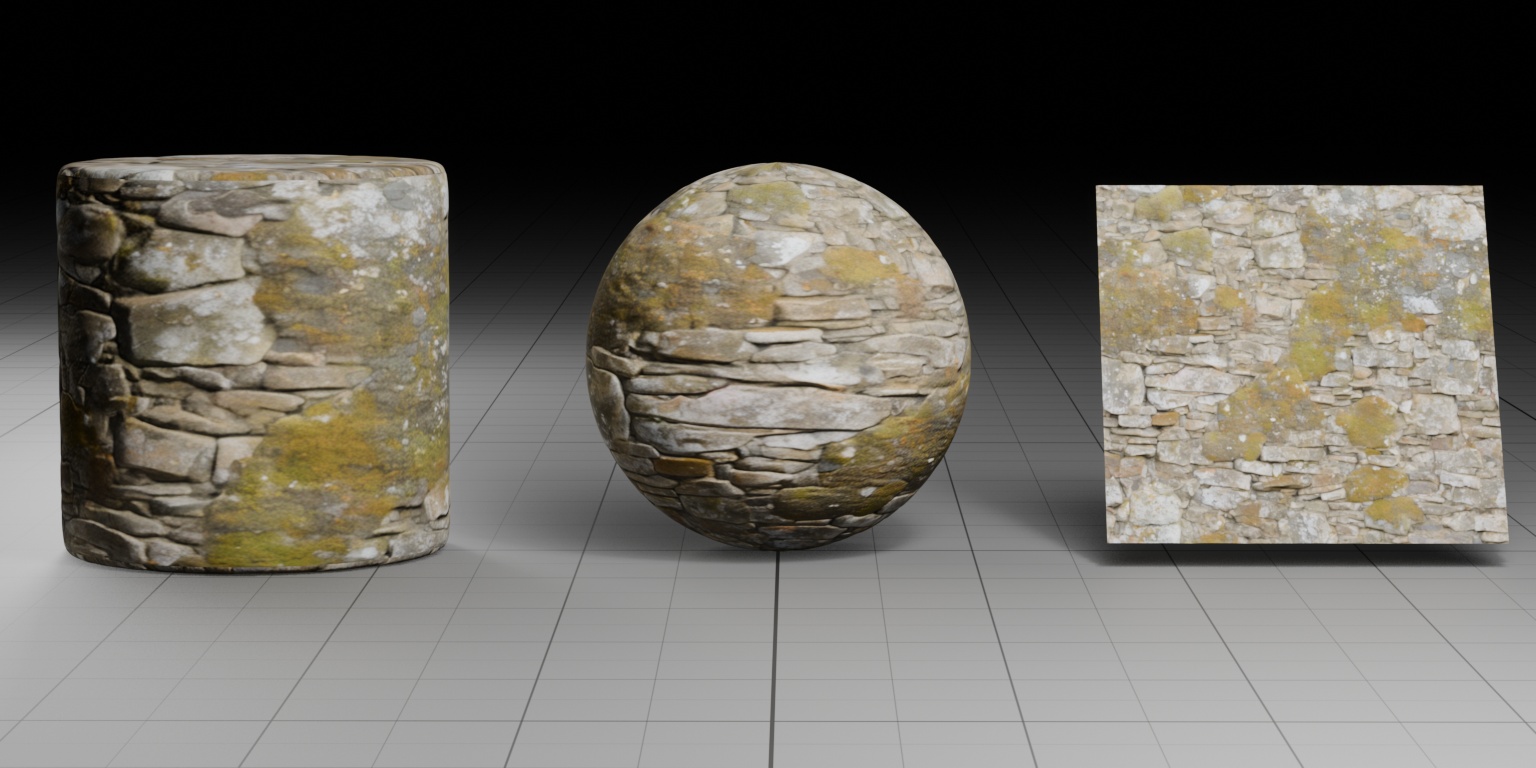} \\
    \begin{minipage}[b]{0.12\linewidth}
    \centering
        \begin{prompt}
            \textbf{"texture of alien carnivorous"}
        \end{prompt}
    \vspace{2\baselineskip}
    \end{minipage} &
    \includegraphics[height=0.12\textwidth]{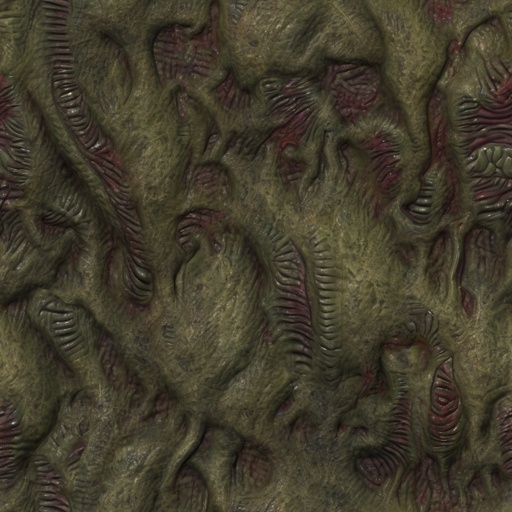} &
    \includegraphics[height=0.12\textwidth]{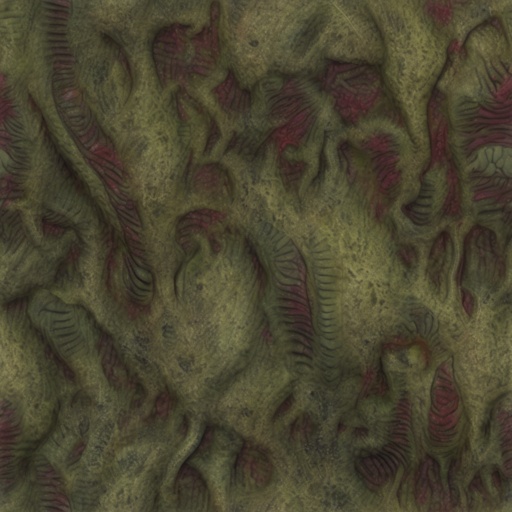} &
    \includegraphics[height=0.12\textwidth]{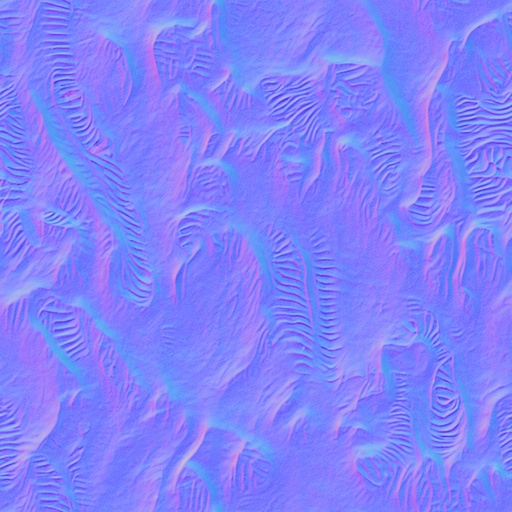} &
    \includegraphics[height=0.12\textwidth]{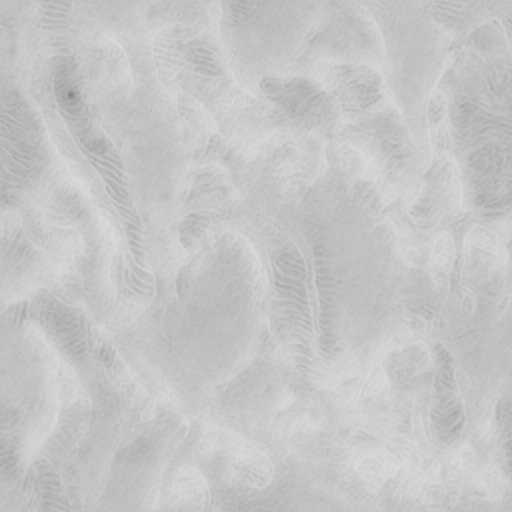} &
    \includegraphics[height=0.12\textwidth]{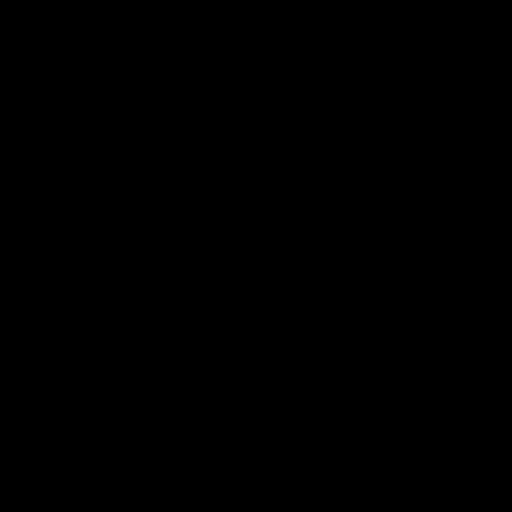} &
    \includegraphics[height=0.12\textwidth]{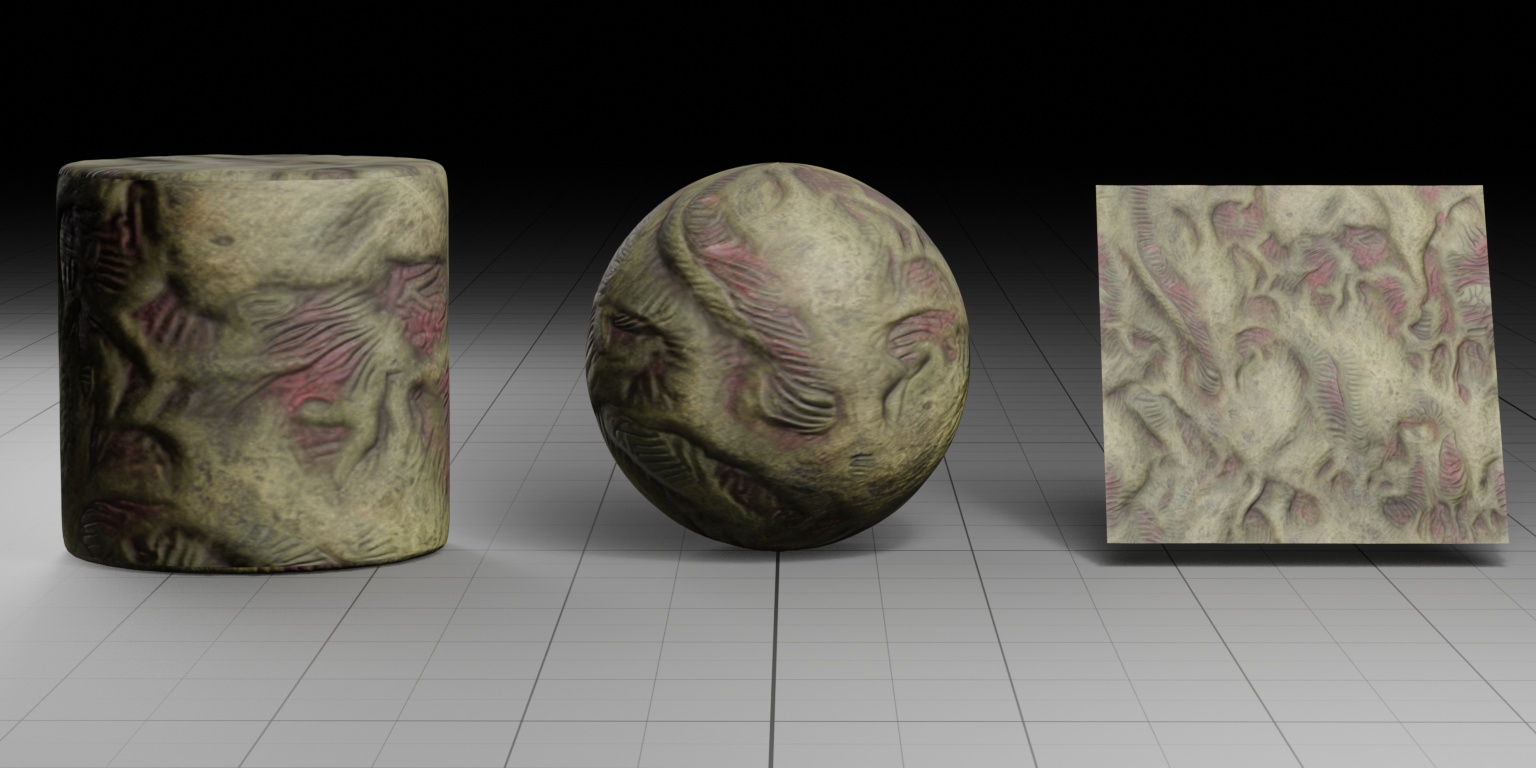} \\
    \begin{minipage}[b]{0.12\linewidth}
    \centering
        \begin{prompt}
            \textbf{"texture of dented metal surface"}
        \end{prompt}
    \vspace{2\baselineskip}
    \end{minipage} &
    \includegraphics[height=0.12\textwidth]{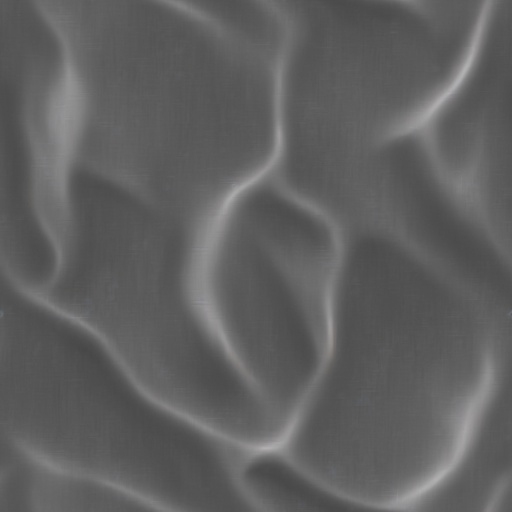} &
    \includegraphics[height=0.12\textwidth]{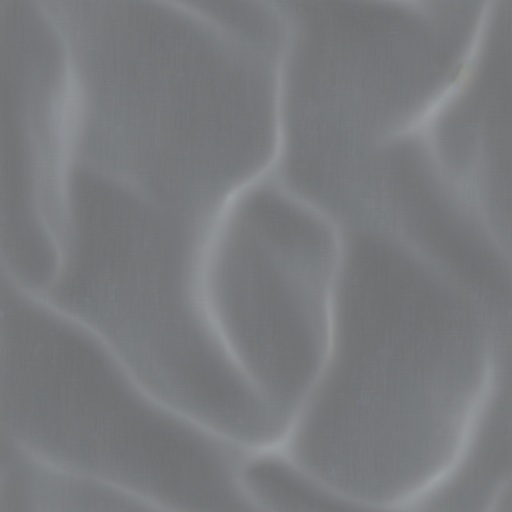} &
    \includegraphics[height=0.12\textwidth]{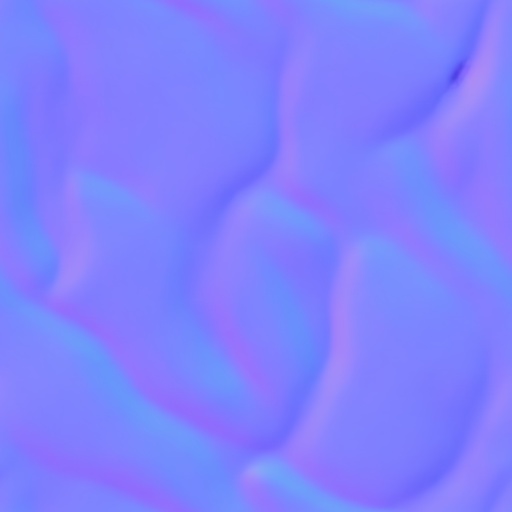} &
    \includegraphics[height=0.12\textwidth]{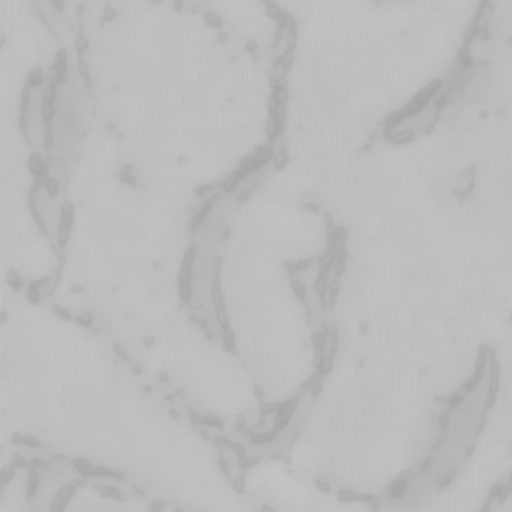} &
    \includegraphics[height=0.12\textwidth]{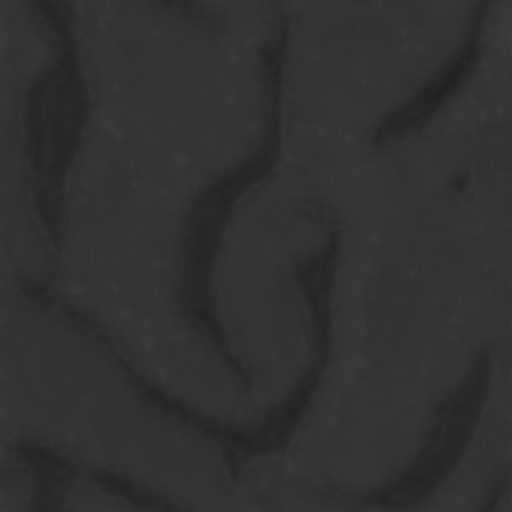} &
    \includegraphics[height=0.12\textwidth]{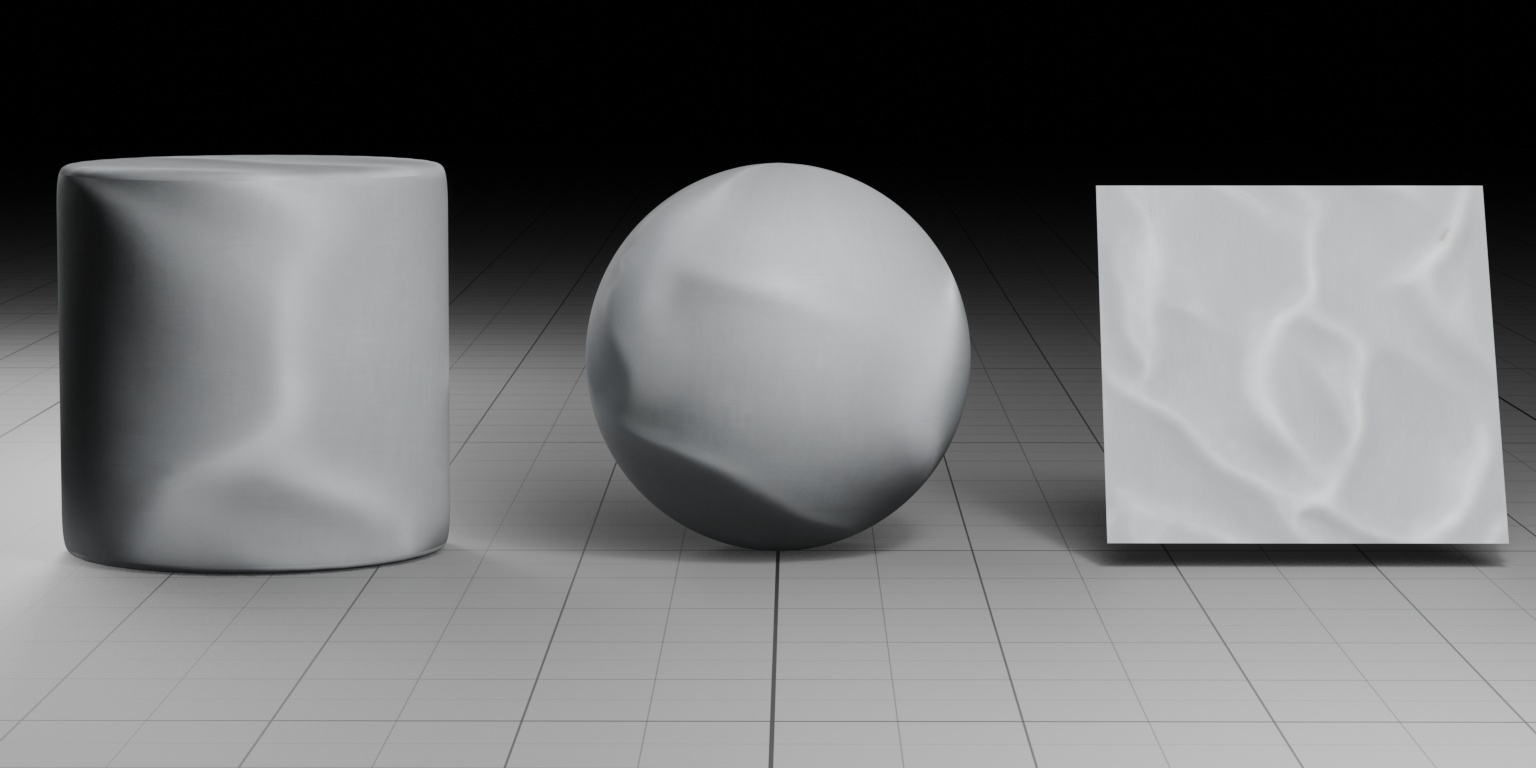} \\
    \end{tabular}
    \caption{\textbf{PBR material generated by our method.} \textit{Texture RGB} represents $I_\text{RGB}$ generated by texture generation stage, while \textit{Render} represents re-rendered image from estimated material.}
    \label{fig:art_gallery} 
\end{figure*}

%% file: figures/in-the-wild/in-the-wild.tex
\begin{figure*}[ht]
    \centering
    \setlength\tabcolsep{1pt}
    \settoheight{\tempdima}{\includegraphics[width=0.15\textwidth]{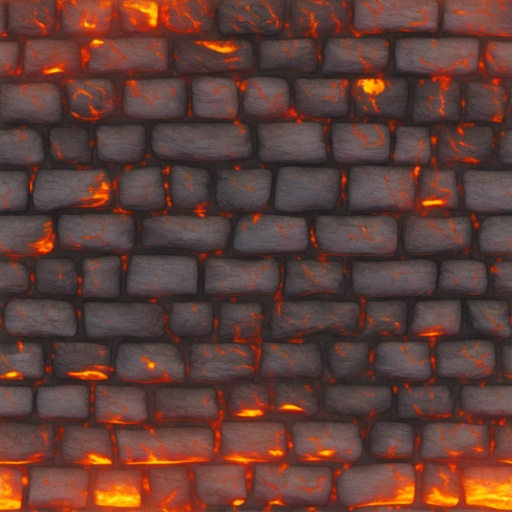}} 
    \begin{tabular}{@{} ccccccc @{}}
    Input & Relit & Basecolor & Normal & Roughness & Metalness & Render \\
    \includegraphics[height=0.12\textwidth]{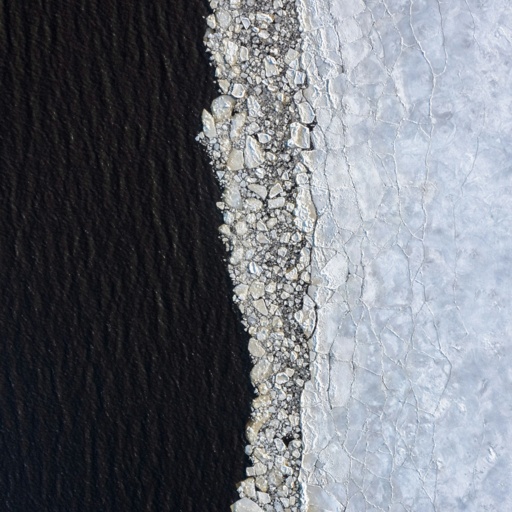}&
    \includegraphics[height=0.12\textwidth]{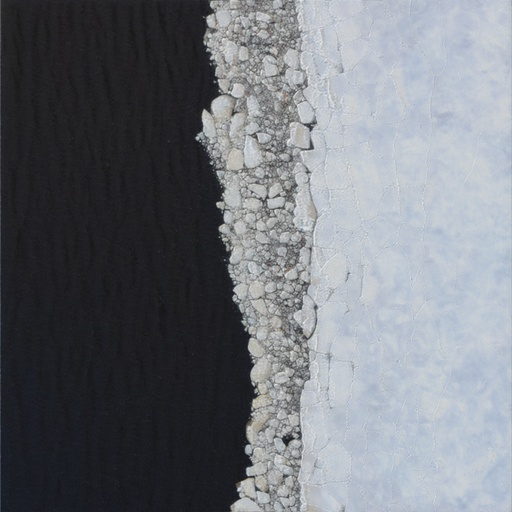} &
    \includegraphics[height=0.12\textwidth]{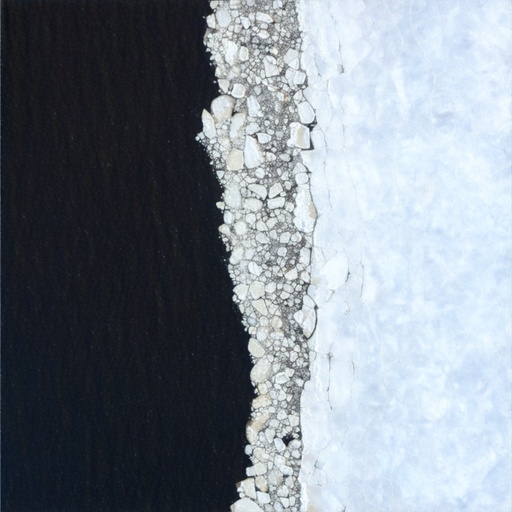} &
    \includegraphics[height=0.12\textwidth]{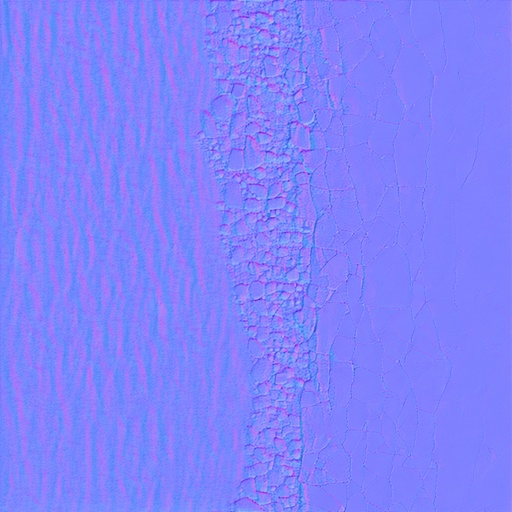} &
    \includegraphics[height=0.12\textwidth]{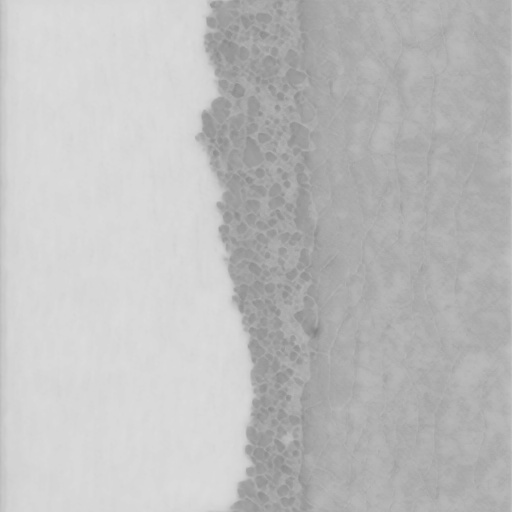} &
    \includegraphics[height=0.12\textwidth]{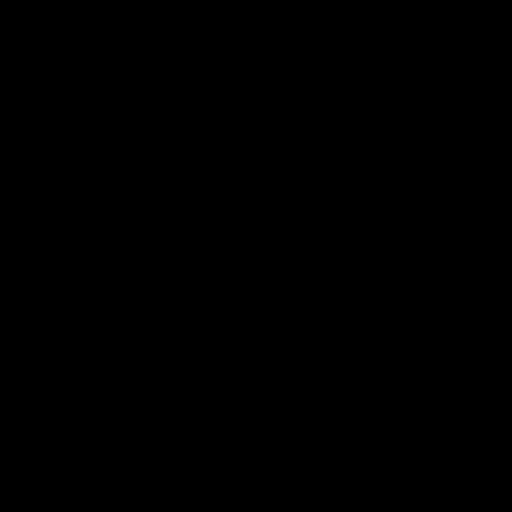} &
    \includegraphics[height=0.12\textwidth]{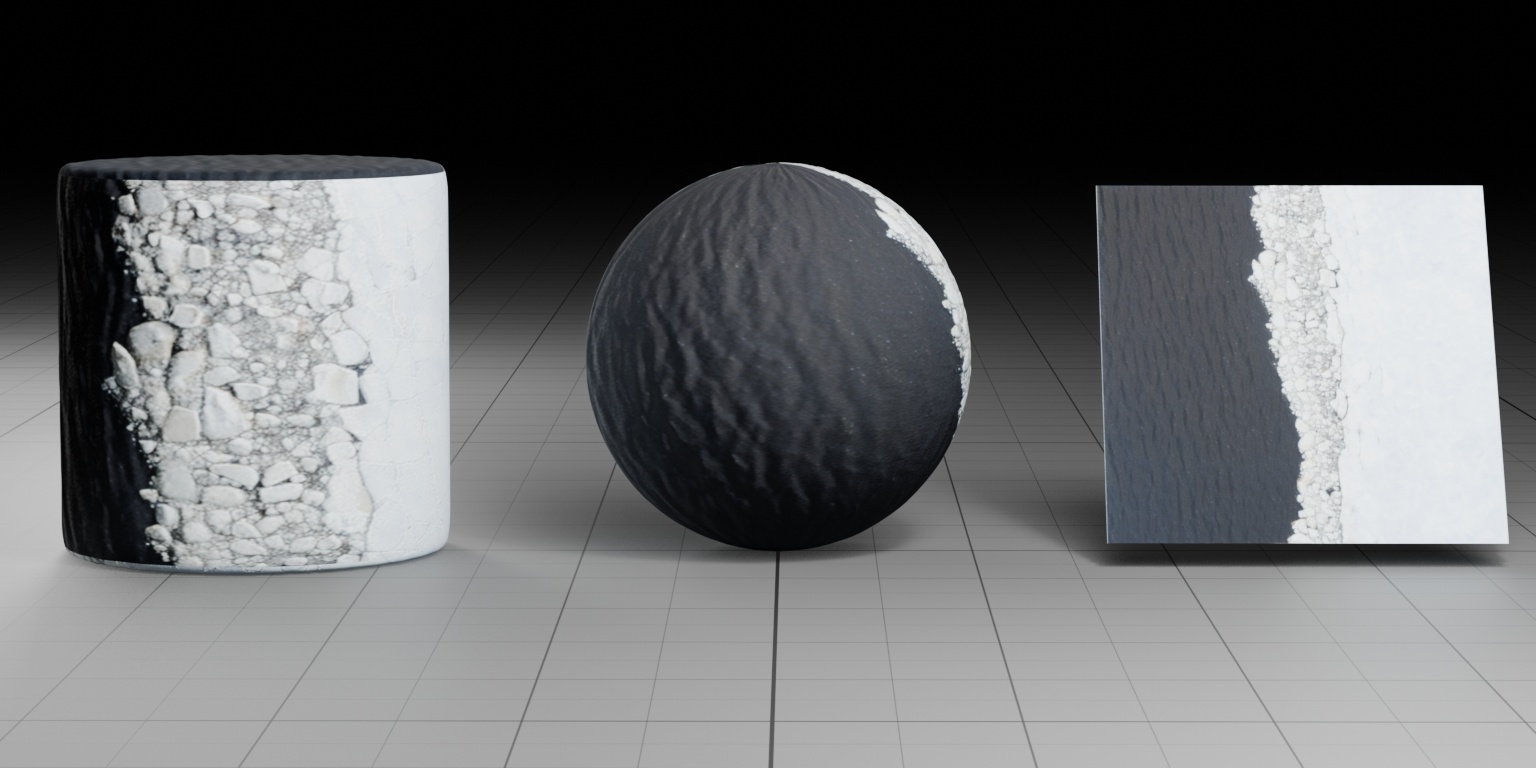}\\
    \includegraphics[height=0.12\textwidth]{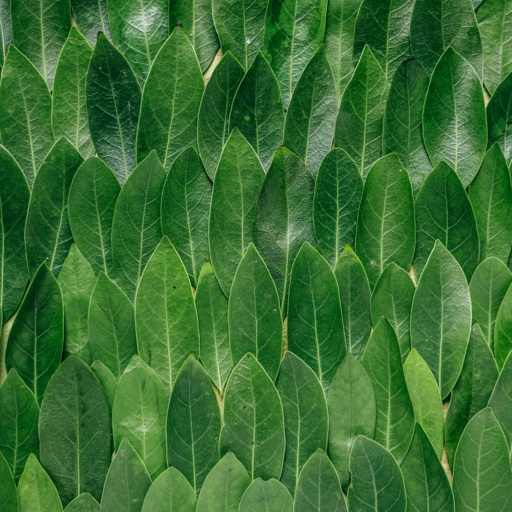}&
    \includegraphics[height=0.12\textwidth]{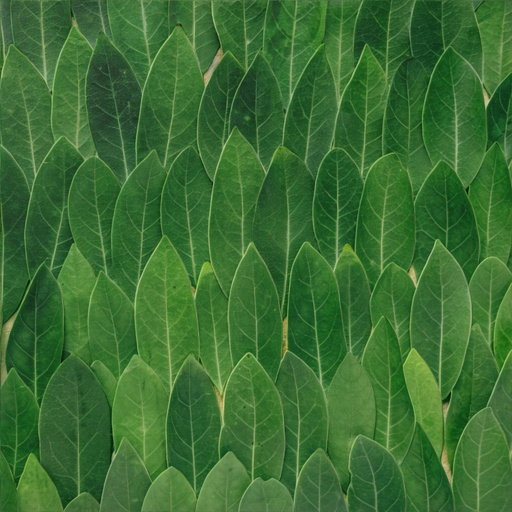} &
    \includegraphics[height=0.12\textwidth]{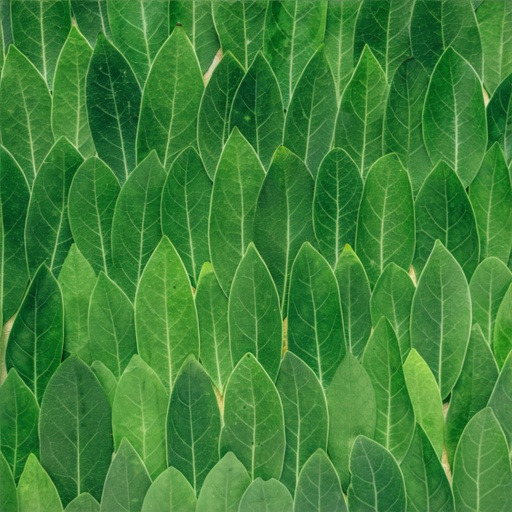} &
    \includegraphics[height=0.12\textwidth]{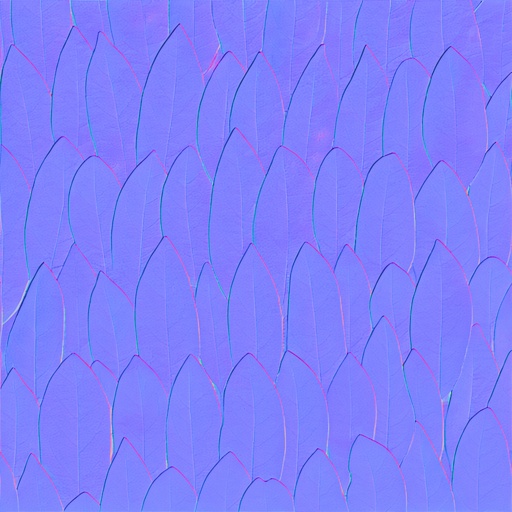} &
    \includegraphics[height=0.12\textwidth]{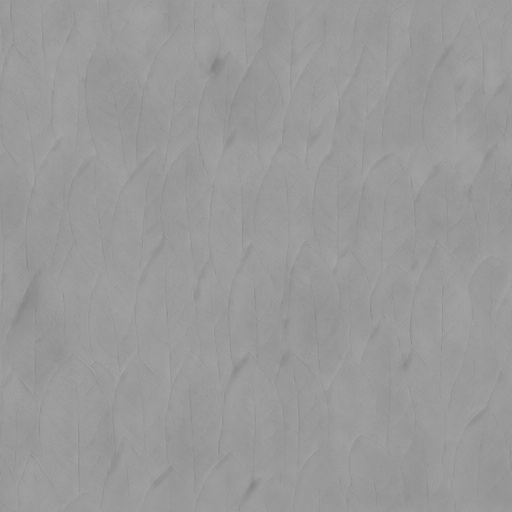} &
    \includegraphics[height=0.12\textwidth]{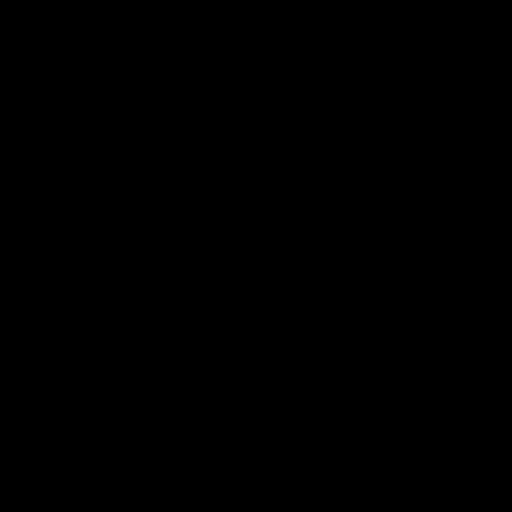} &
    \includegraphics[height=0.12\textwidth]{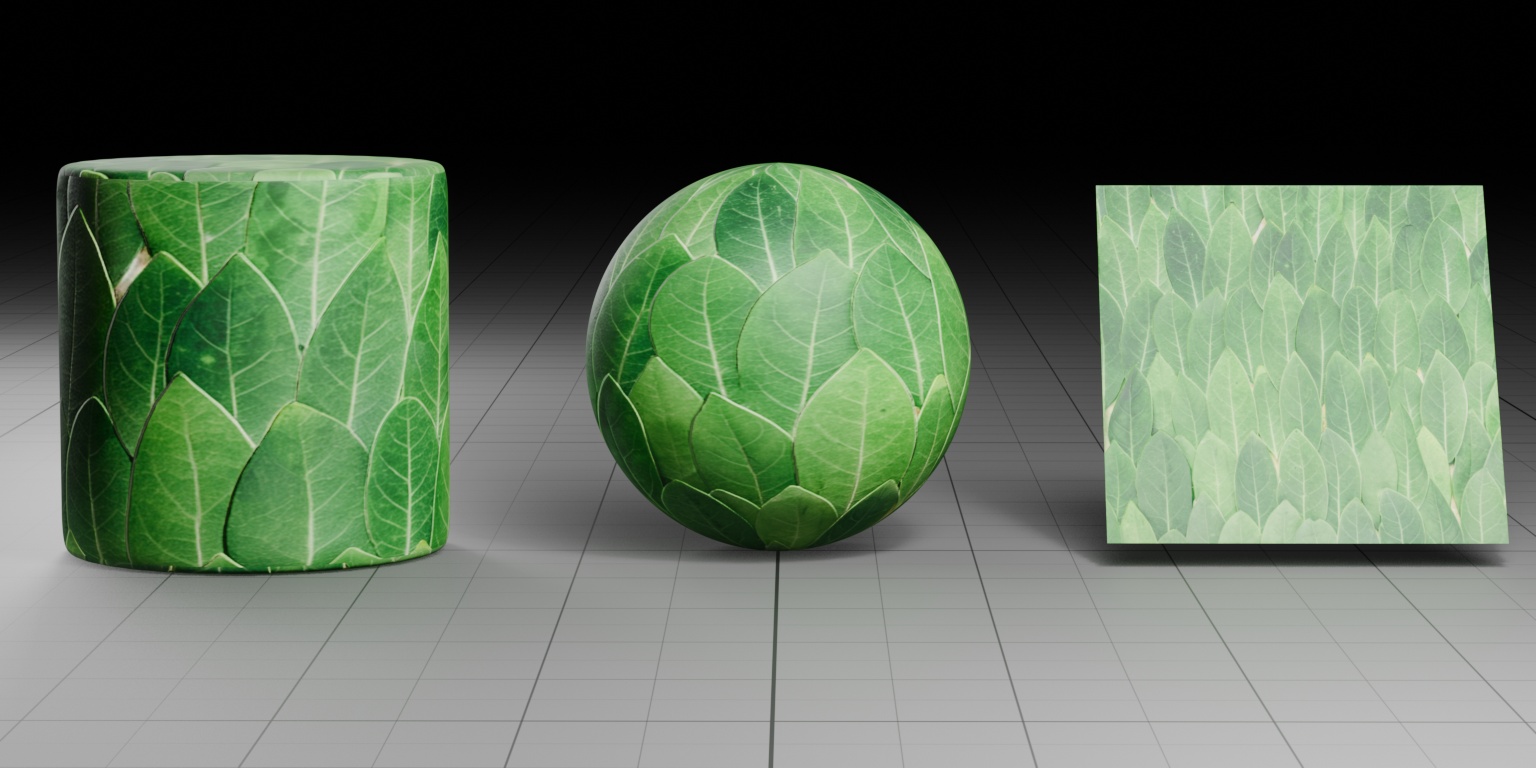}\\
    \includegraphics[height=0.12\textwidth]{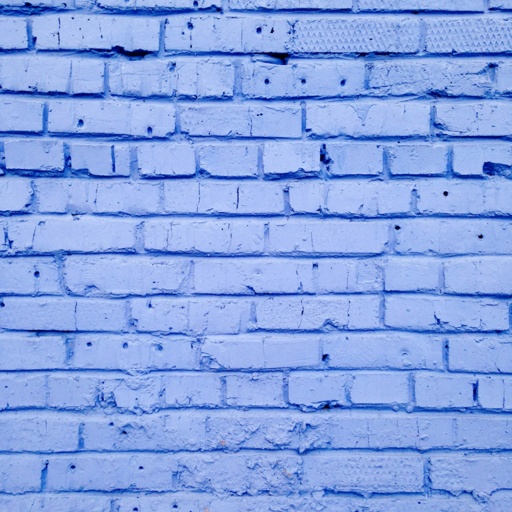}&
    \includegraphics[height=0.12\textwidth]{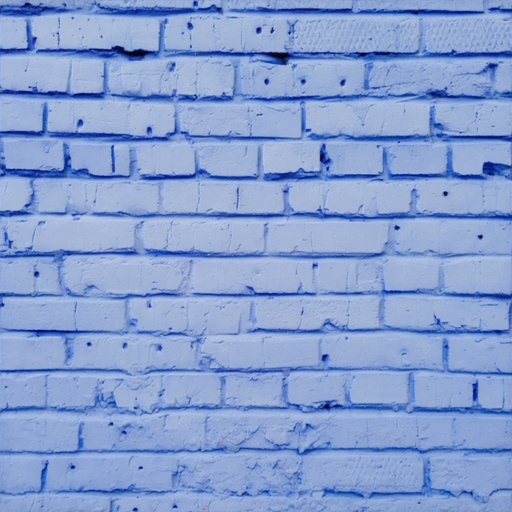} &
    \includegraphics[height=0.12\textwidth]{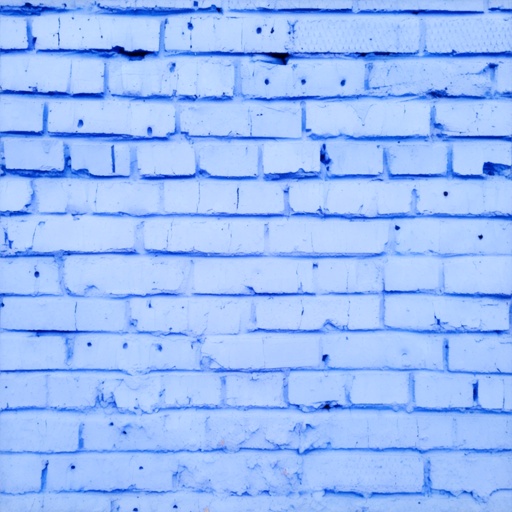} &
    \includegraphics[height=0.12\textwidth]{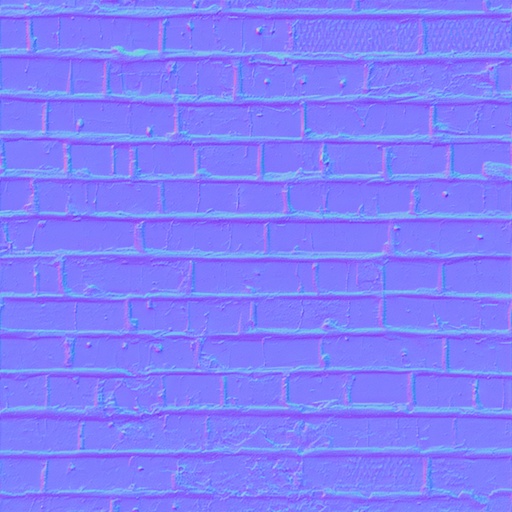} &
    \includegraphics[height=0.12\textwidth]{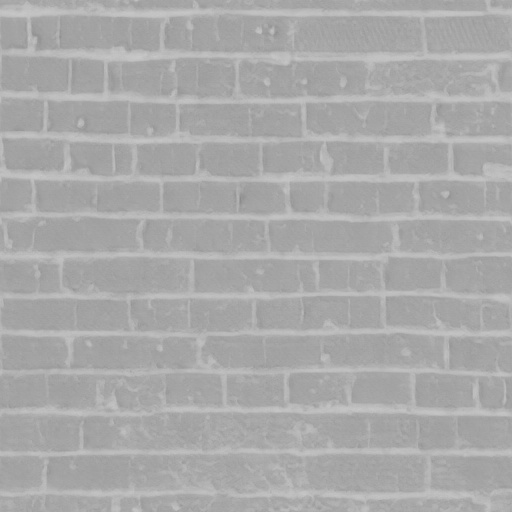} &
    \includegraphics[height=0.12\textwidth]{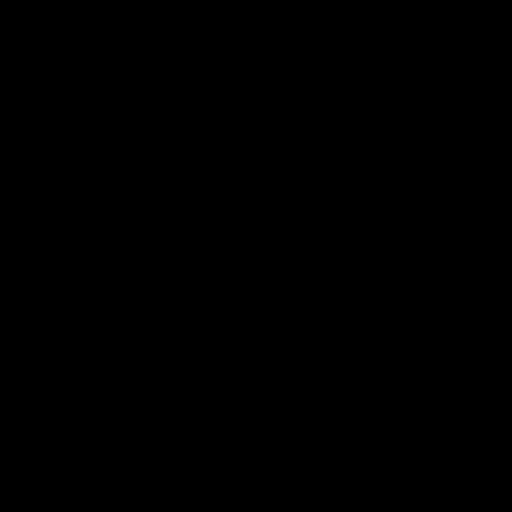} &
    \includegraphics[height=0.12\textwidth]{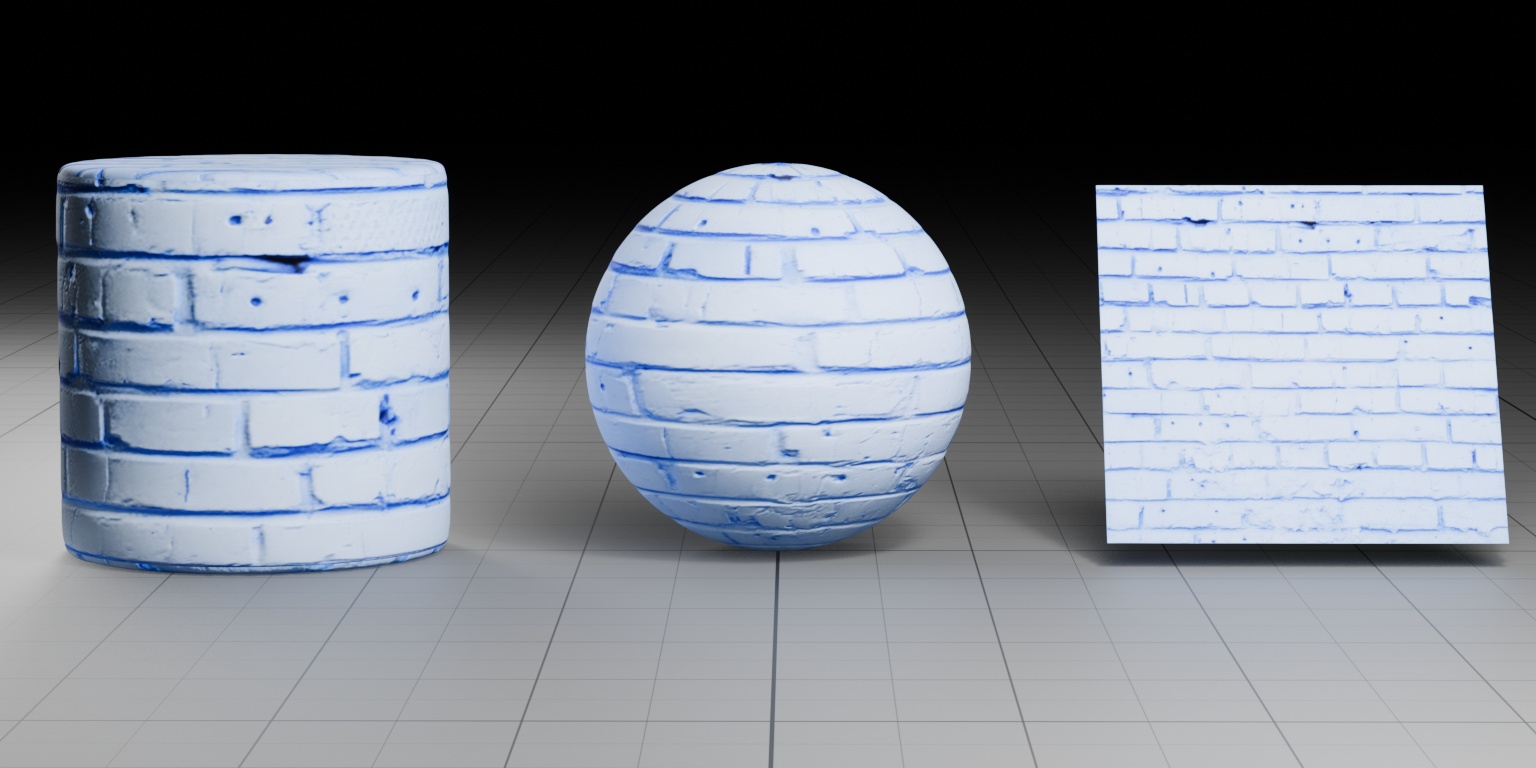}\\
    \includegraphics[height=0.12\textwidth]{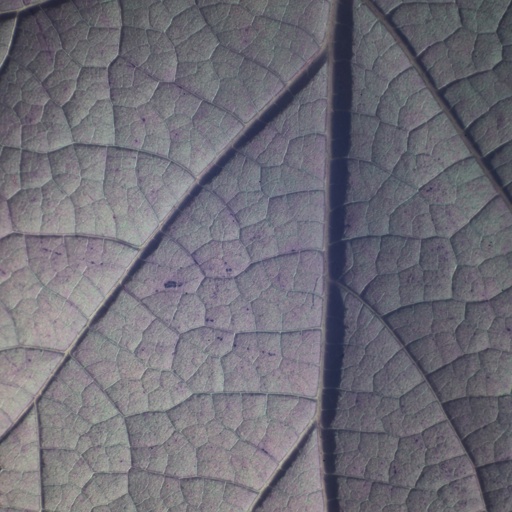}&
    \includegraphics[height=0.12\textwidth]{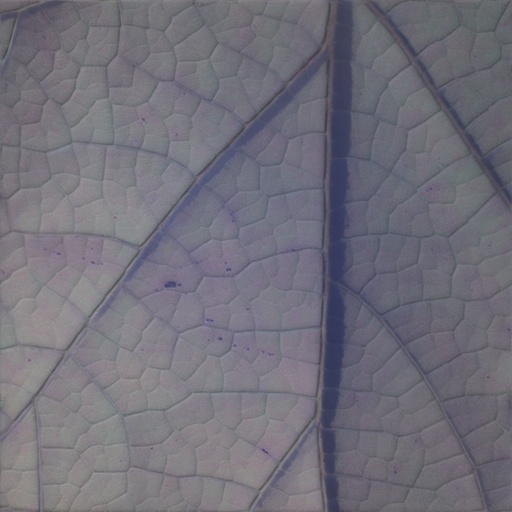} &
    \includegraphics[height=0.12\textwidth]{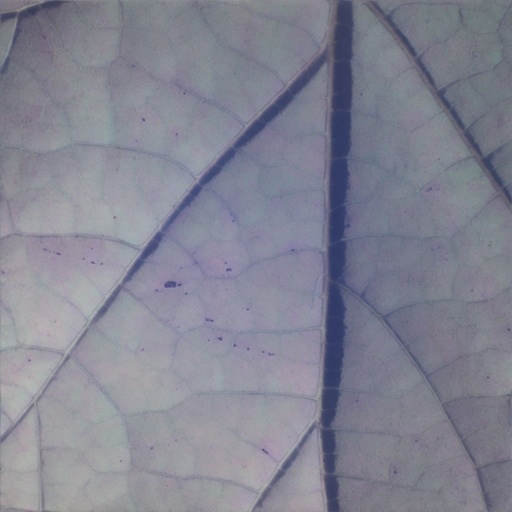} &
    \includegraphics[height=0.12\textwidth]{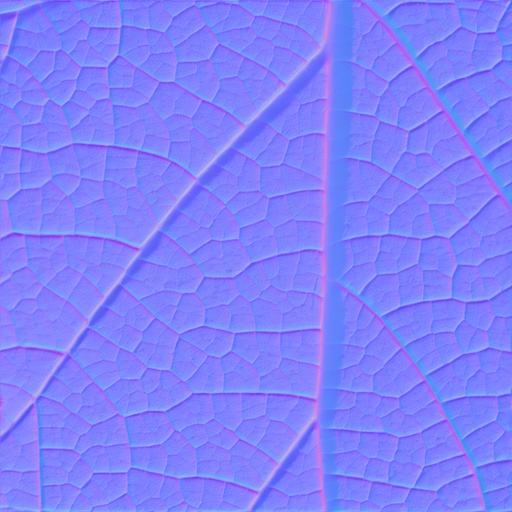} &
    \includegraphics[height=0.12\textwidth]{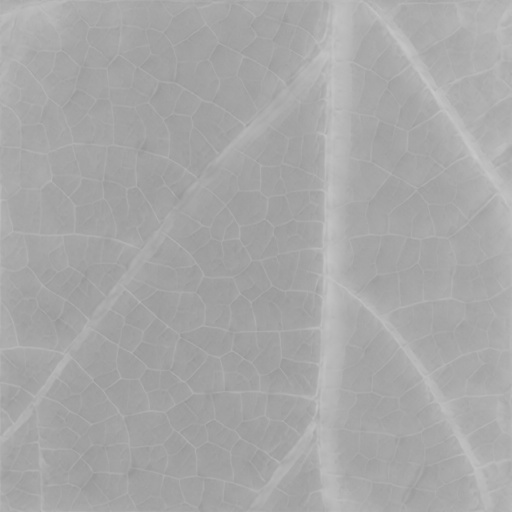} &
    \includegraphics[height=0.12\textwidth]{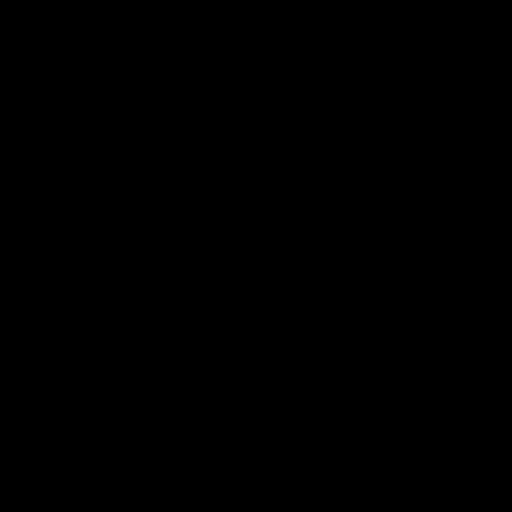} &
    \includegraphics[height=0.12\textwidth]{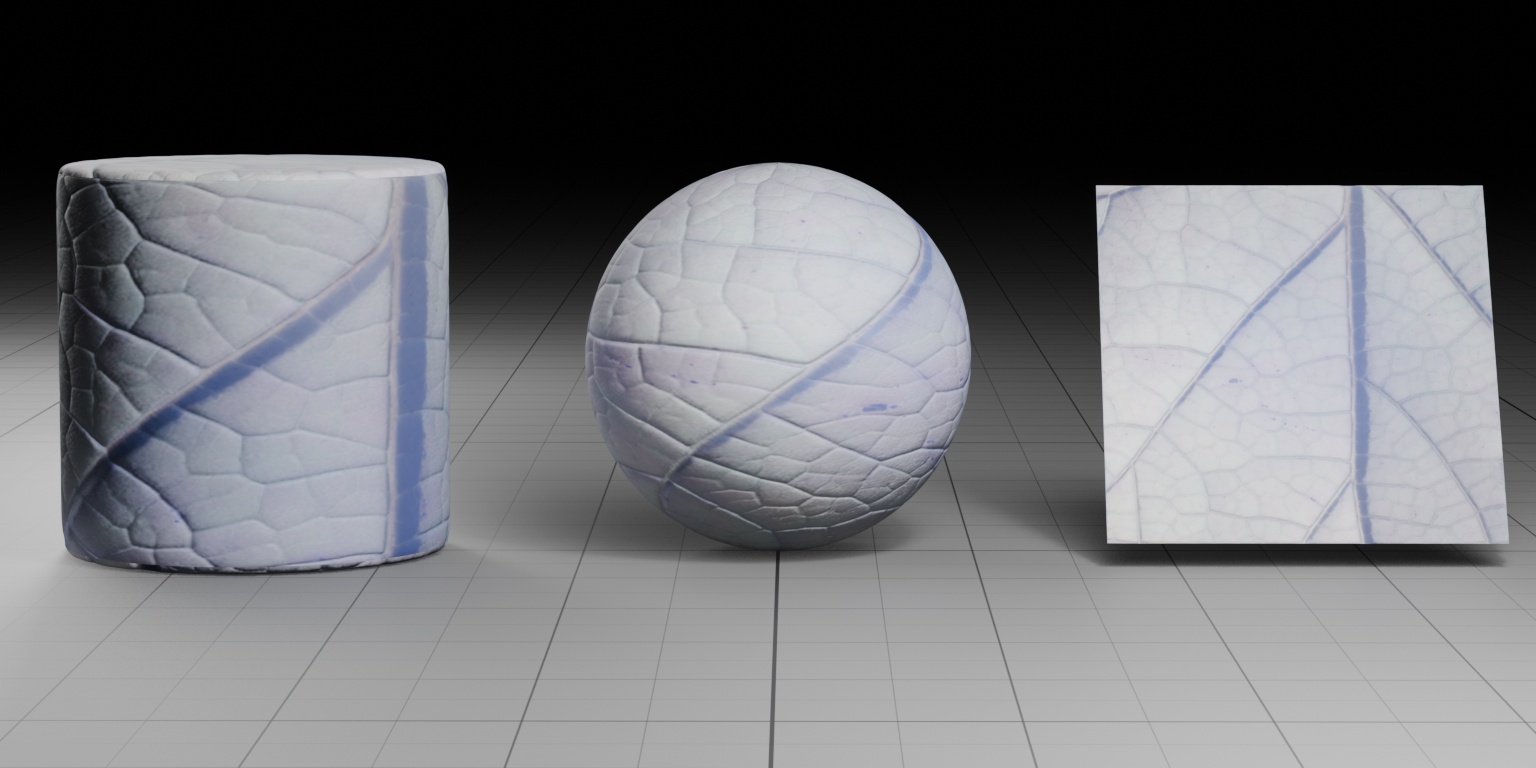}\\
    \end{tabular}
    \caption{\textbf{Material estimation from in-the-wild photographs.} We demonstrate the robustness of our Chord pipeline using real-world top-down photographs as inputs. Photo from Unsplash.}
    \label{fig:in_the_wild} 
      \vspace{-10pt}
\end{figure*}

%% file: figures/applications/applications.tex
\begin{figure*}[ht]
    \centering
    \begin{subfigure}{\textwidth}
    \centering
    \setlength\tabcolsep{1pt}
    \settoheight{\tempdima}{\includegraphics[width=0.12\textwidth]{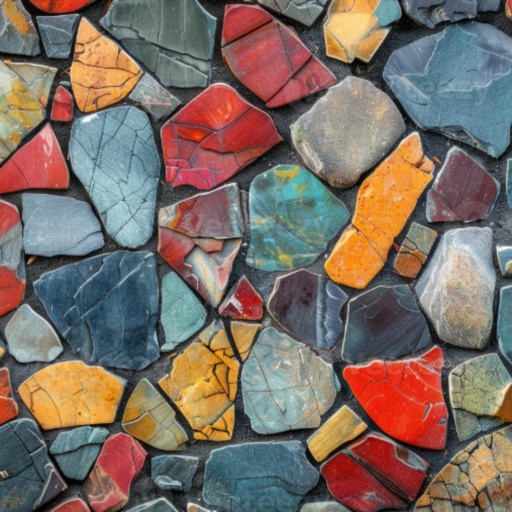}} 
    \begin{tabular}{@{} cc@{\hskip 0.03in}|@{\hskip 0.03in}cccccc @{}}
    Prompt & Reference & Texture RGB & Basecolor & Normal & Roughness & Metalness & Height \\
    \begin{minipage}[b]{0.12\linewidth}
    \centering
        \begin{prompt}
            \textbf{"texture of colorful stone wall, irregular shapes"}
        \end{prompt}
    \vspace{2\baselineskip}
    \end{minipage} &
    \includegraphics[height=0.12\textwidth]{figures/applications/from_reference/color_stone/ref_512.jpg} &
    \includegraphics[height=0.12\textwidth]{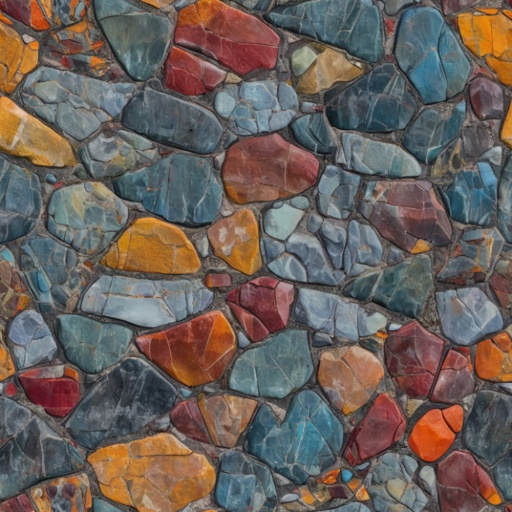} &
    \includegraphics[height=0.12\textwidth]{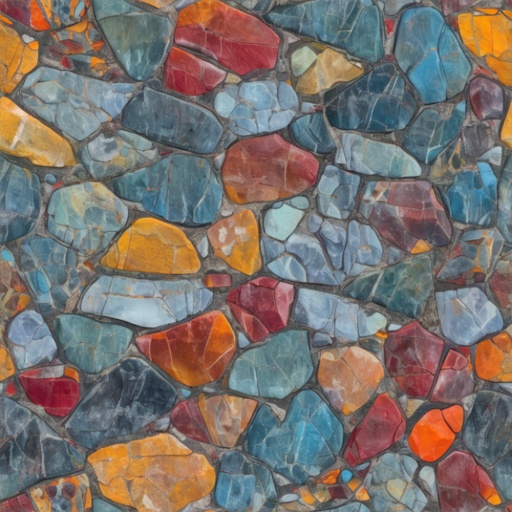} &
    \includegraphics[height=0.12\textwidth]{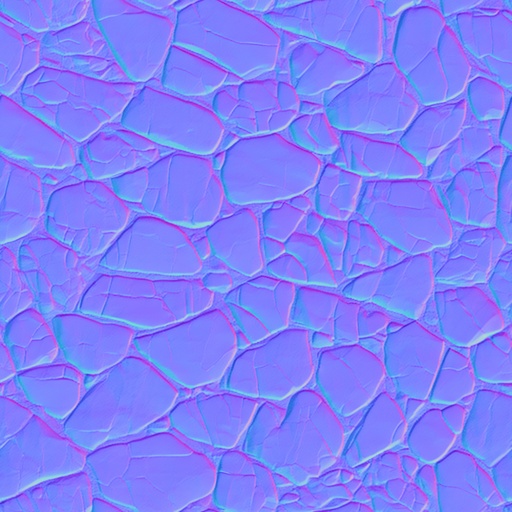} &
    \includegraphics[height=0.12\textwidth]{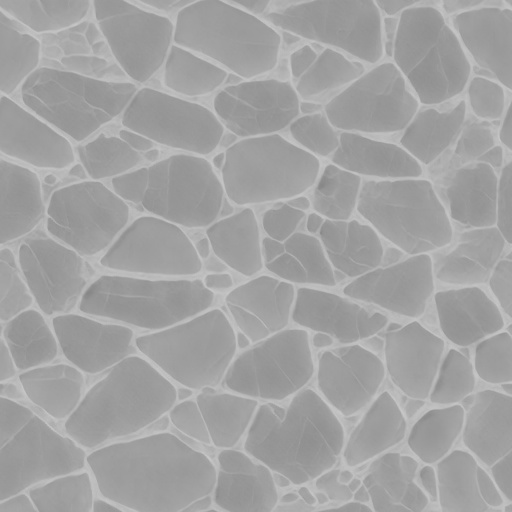} &
    \includegraphics[height=0.12\textwidth]{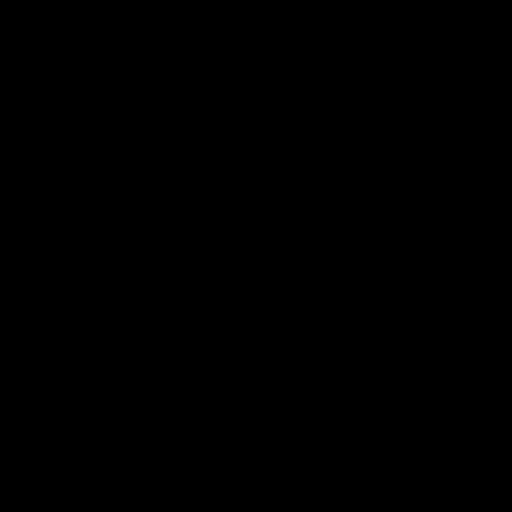} &
    \includegraphics[height=0.12\textwidth]{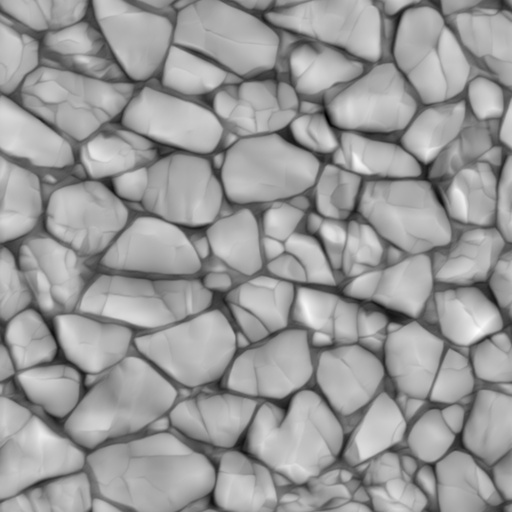} \\
    \begin{minipage}[b]{0.12\linewidth}
    \centering
        \begin{prompt}
            \textbf{"texture of iron surface, a few of \textcolor{ForestGreen}{green} rust"}
        \end{prompt}
    \vspace{2\baselineskip}
    \end{minipage} &
    \includegraphics[height=0.12\textwidth]{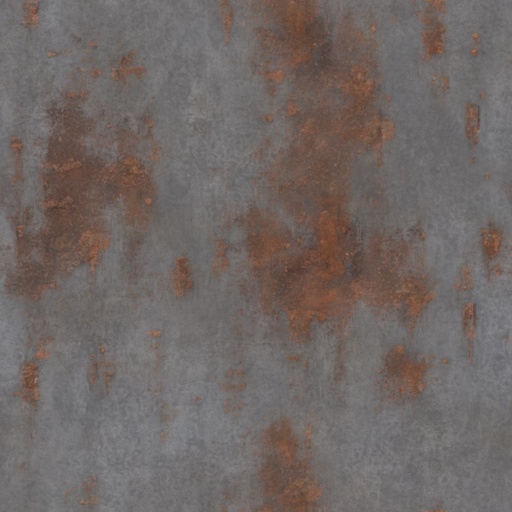} &
    \includegraphics[height=0.12\textwidth]{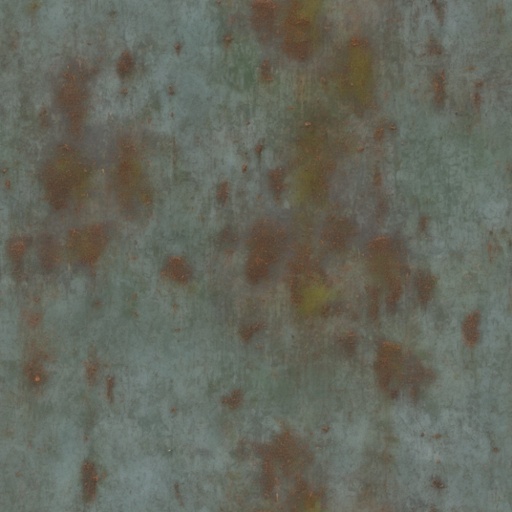} &
    \includegraphics[height=0.12\textwidth]{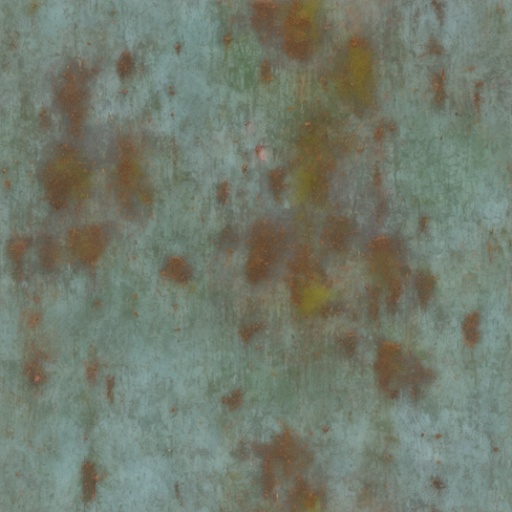} &
    \includegraphics[height=0.12\textwidth]{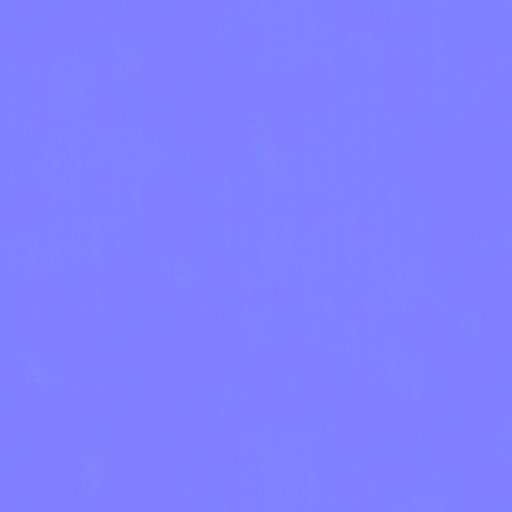} &
    \includegraphics[height=0.12\textwidth]{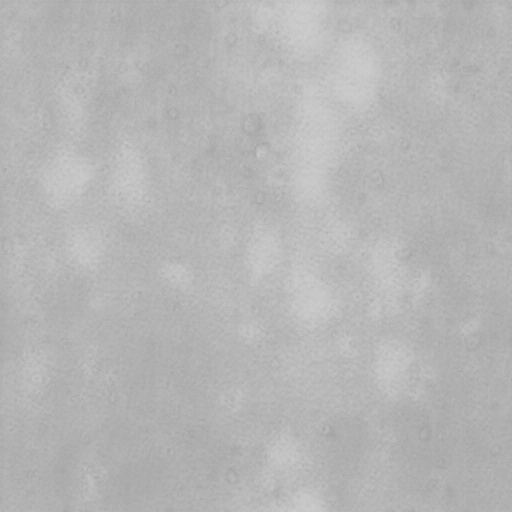} &
    \includegraphics[height=0.12\textwidth]{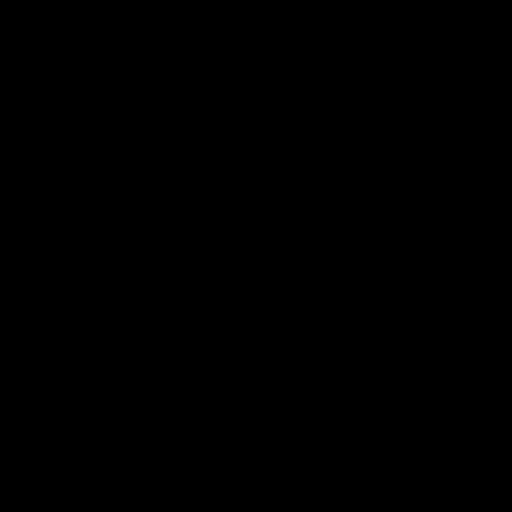}  &
    \includegraphics[height=0.12\textwidth]{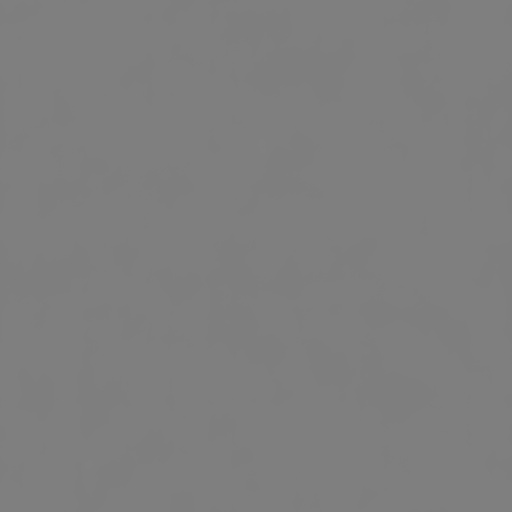}\\
    \end{tabular}
    \caption{\textbf{Image to Material.} Given a reference image, which may be non-tillable, our method generates tillable materials or material variations guided by the corresponding text prompt. Colorful stone reference image from Vecteezy}
    \end{subfigure}
    \begin{subfigure}{\textwidth}
    \centering
    \setlength\tabcolsep{1pt}
    \settoheight{\tempdima}{\includegraphics[width=0.12\textwidth]{figures/applications/from_reference/color_stone/ref_512.jpg}} 
    \begin{tabular}{cc@{\hskip 0.03in}|@{\hskip 0.03in}cccccc}
    Prompt & Condition & Texture RGB & Basecolor & Normal & Roughness & Metalness & Height \\
    \begin{minipage}[b]{0.12\linewidth}
    \centering
        \begin{prompt}
            \textbf{"texture of light gray sci-fi high tech steel"}
        \end{prompt}
    \vspace{2\baselineskip}
    \end{minipage} &
    \includegraphics[height=0.12\textwidth]{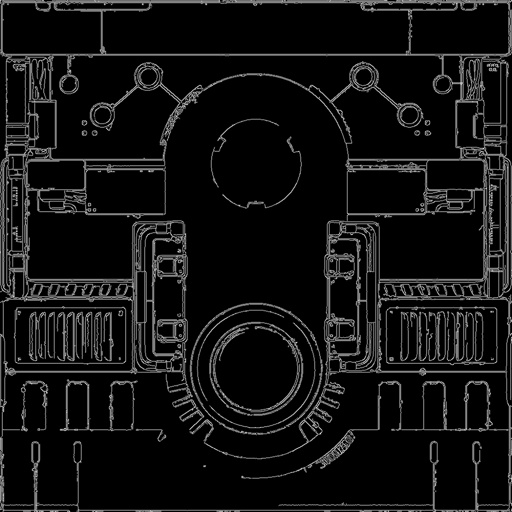} &
    \includegraphics[height=0.12\textwidth]{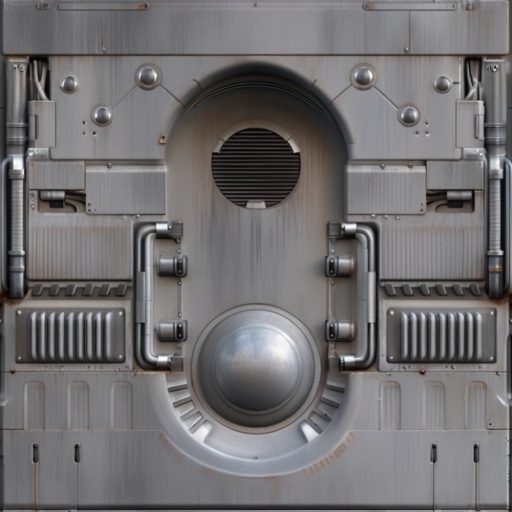} &
    \includegraphics[height=0.12\textwidth]{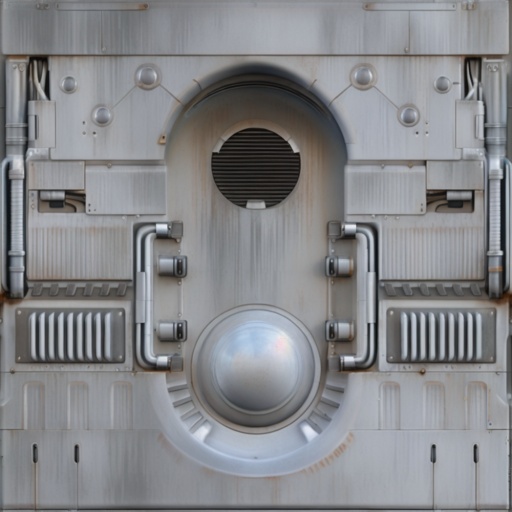} &
    \includegraphics[height=0.12\textwidth]{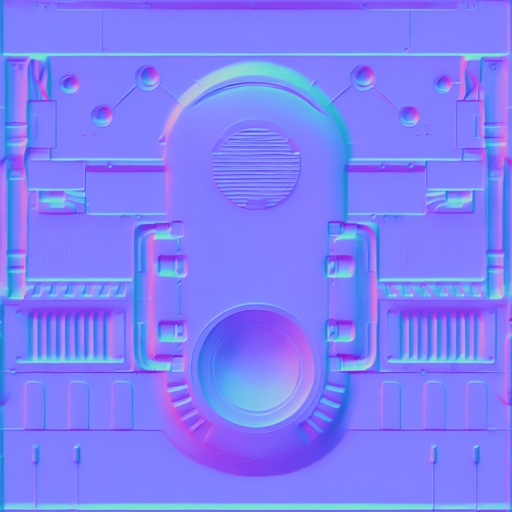} &
    \includegraphics[height=0.12\textwidth]{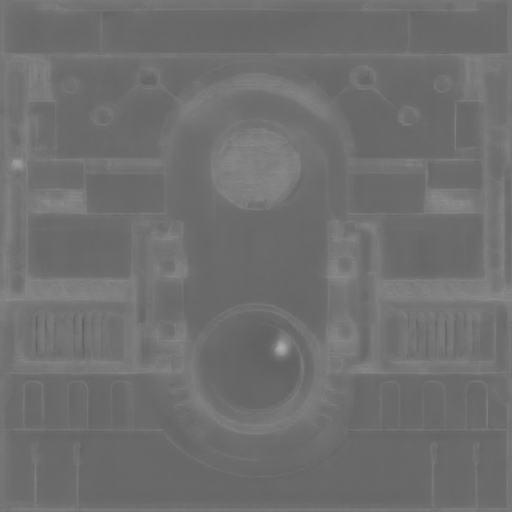} &
    \includegraphics[height=0.12\textwidth]{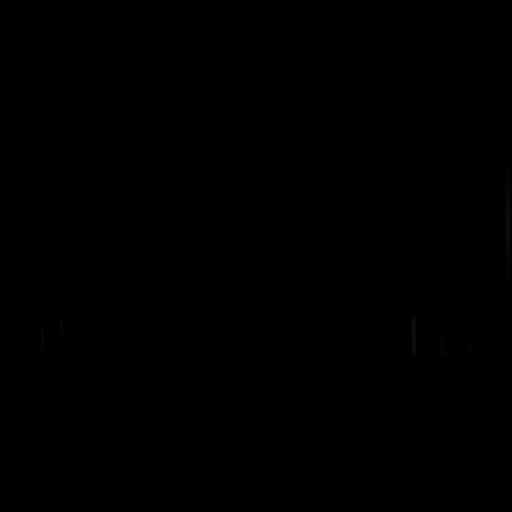} &
    \includegraphics[height=0.12\textwidth]{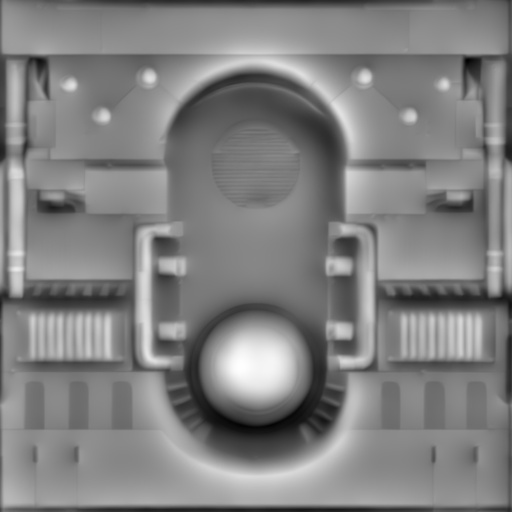}\\
    \begin{minipage}[b]{0.12\linewidth}
    \centering
        \begin{prompt}
            \textbf{"texture of magma pavement, with flowing lava surrounded, smooth surface, flames"}
        \end{prompt}
    \vspace{1\baselineskip}
    \end{minipage} &
    \includegraphics[height=0.12\textwidth]{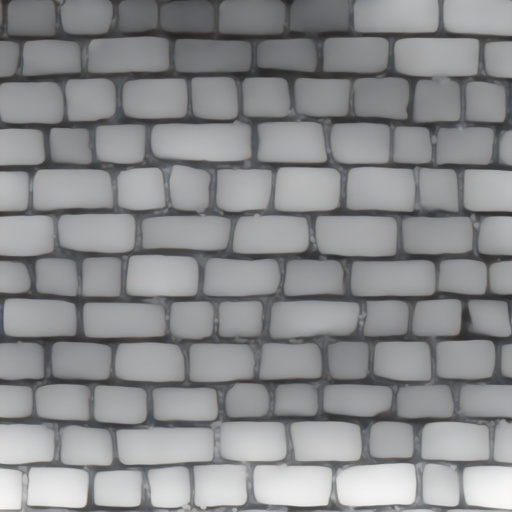} &
    \includegraphics[height=0.12\textwidth]{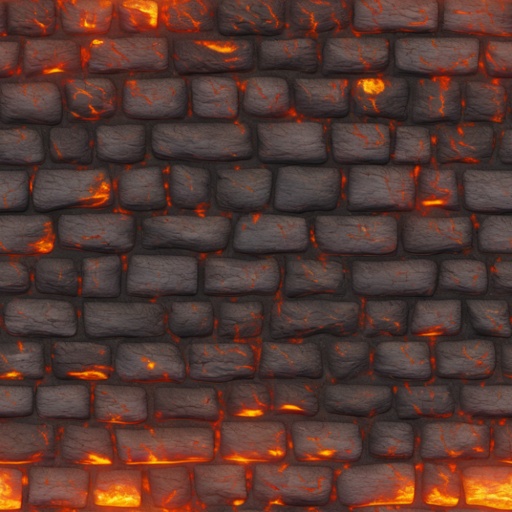} &
    \includegraphics[height=0.12\textwidth]{figures/applications/control_net/pavement/basecolor_512.jpg} &
    \includegraphics[height=0.12\textwidth]{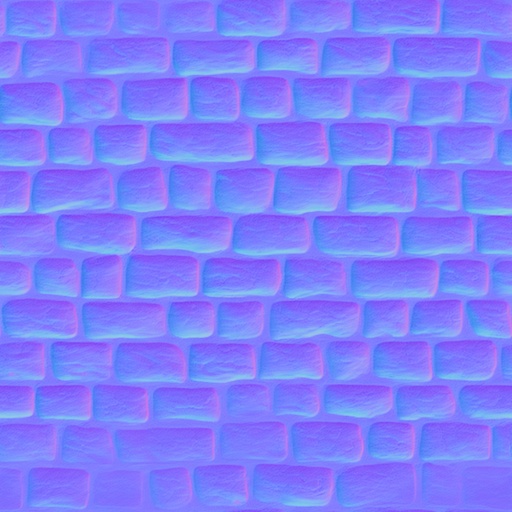} &
    \includegraphics[height=0.12\textwidth]{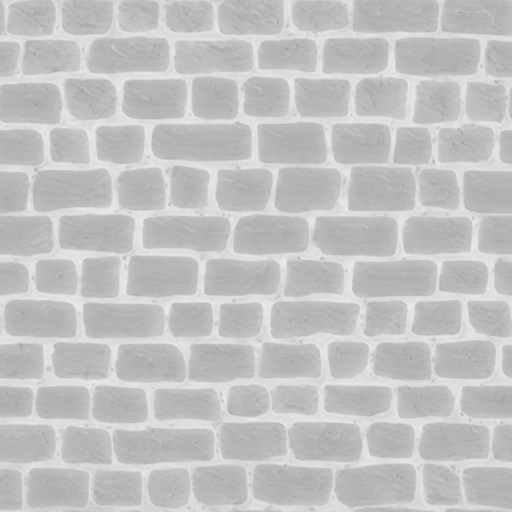} &
    \includegraphics[height=0.12\textwidth]{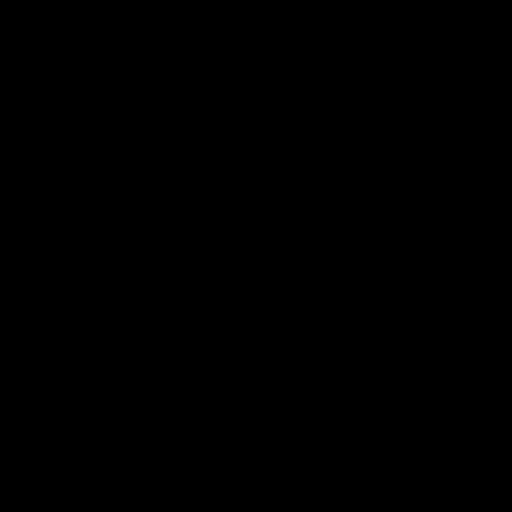} &
    \includegraphics[height=0.12\textwidth]{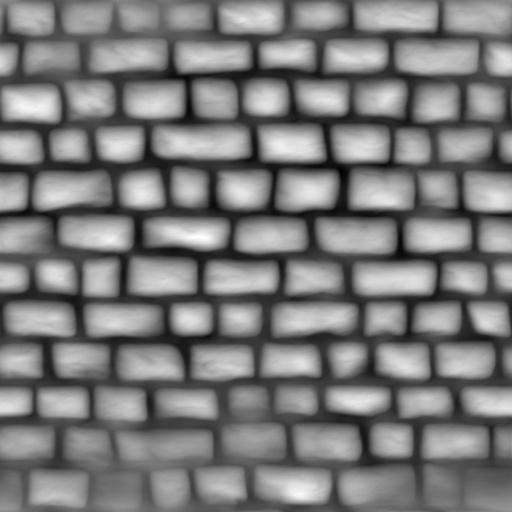}\\ 

    \end{tabular}
    \caption{\textbf{Structure-controlled Generation} We present two examples: one using line art as control and the other using depth map as control.}
    \end{subfigure}
    \begin{subfigure}{\textwidth}
    \centering
    \setlength\tabcolsep{1pt}
    \settoheight{\tempdima}{\includegraphics[width=0.12\textwidth]{figures/applications/from_reference/color_stone/ref_512.jpg}} 
    \begin{tabular}{cc@{\hskip 0.03in}|@{\hskip 0.03in}cccccc}
    Reference & Condition & Texture RGB & Basecolor & Normal & Roughness & Metalness & Height \\
    \includegraphics[height=0.12\textwidth]{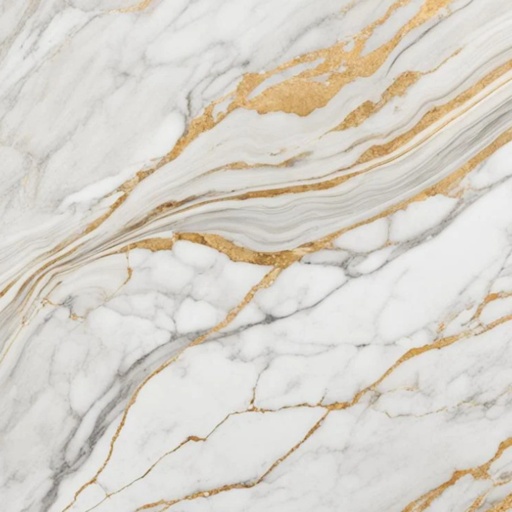} &
    \includegraphics[height=0.12\textwidth]{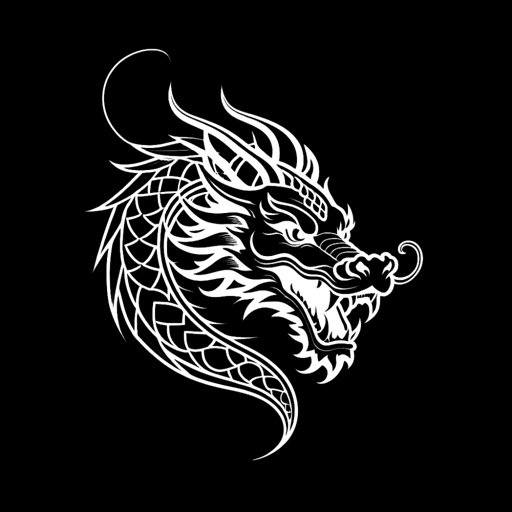} &
    \includegraphics[height=0.12\textwidth]{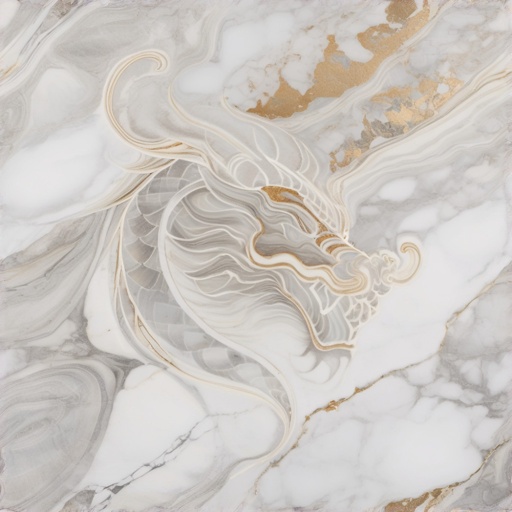} &
    \includegraphics[height=0.12\textwidth]{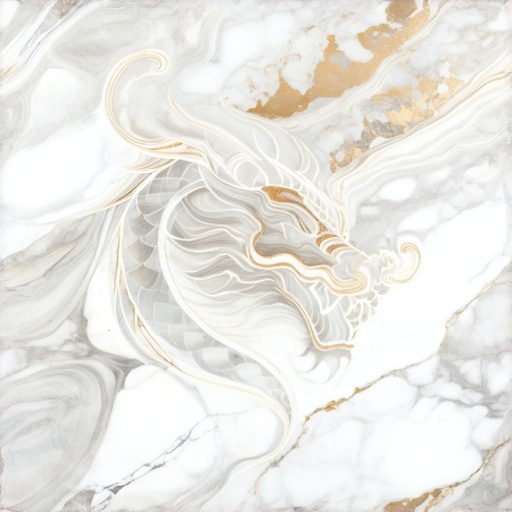} &
    \includegraphics[height=0.12\textwidth]{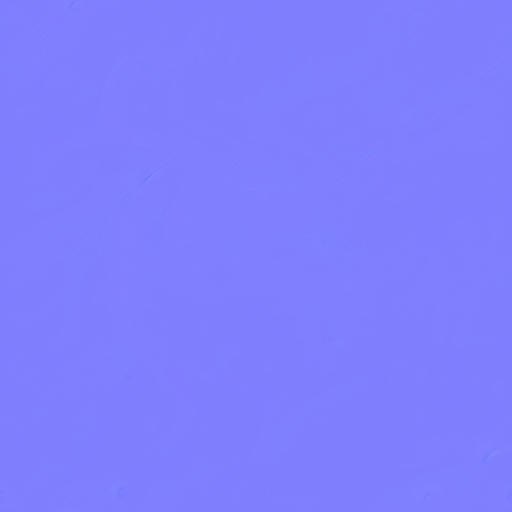} &
    \includegraphics[height=0.12\textwidth]{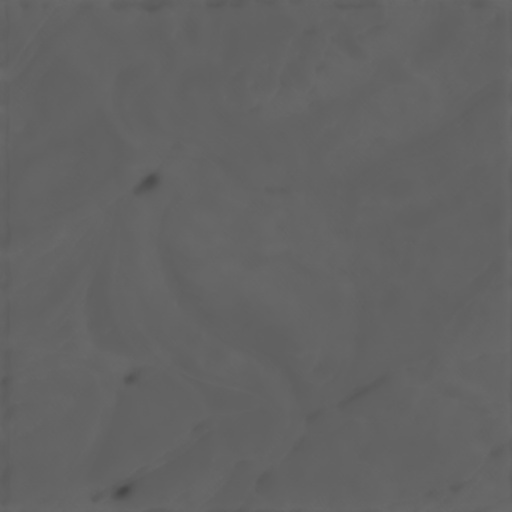} &
    \includegraphics[height=0.12\textwidth]{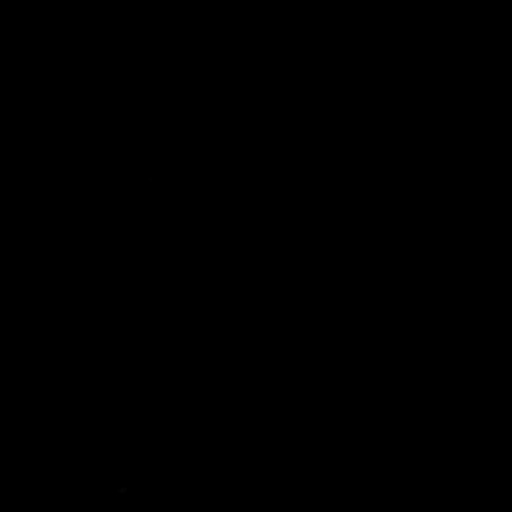} &
    \includegraphics[height=0.12\textwidth]{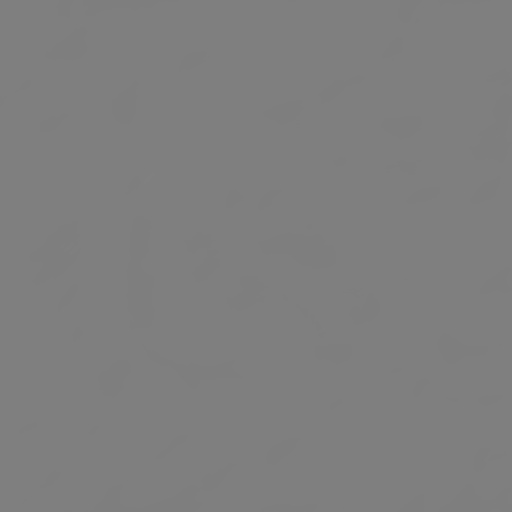} \\
    
    \end{tabular}
    \caption{This example illustrates how line art structure guidance is integrated with a reference image during generation.}
    \end{subfigure}
    \begin{subfigure}{\textwidth}
    \centering
    \setlength\tabcolsep{1pt}
    \settoheight{\tempdima}{\includegraphics[width=0.12\textwidth]{figures/applications/from_reference/color_stone/ref_512.jpg}} 
    \begin{tabular}{cc@{\hskip 0.03in}|@{\hskip 0.03in}cccccc}
    Prompt / Mask & Texture RGB & Relit & Basecolor & Normal & Roughness & Metalness & Height \\
    \begin{minipage}[b]{0.12\linewidth}
    \centering
        \begin{prompt}
            \textbf{"texture of ancient stone wall, green moss"}
        \end{prompt}
    \vspace{2\baselineskip}
    \end{minipage} &
    \includegraphics[height=0.12\textwidth]{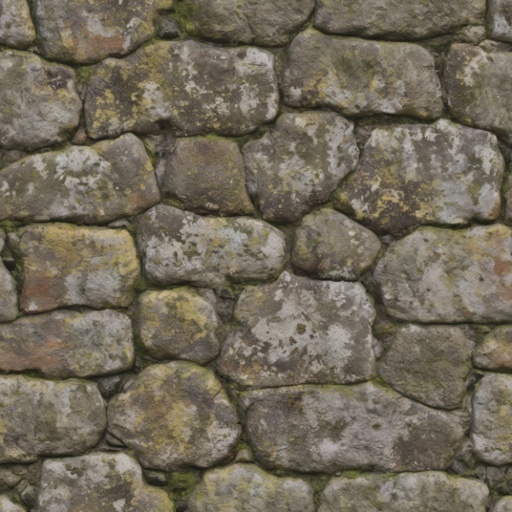} &
    \includegraphics[height=0.12\textwidth]{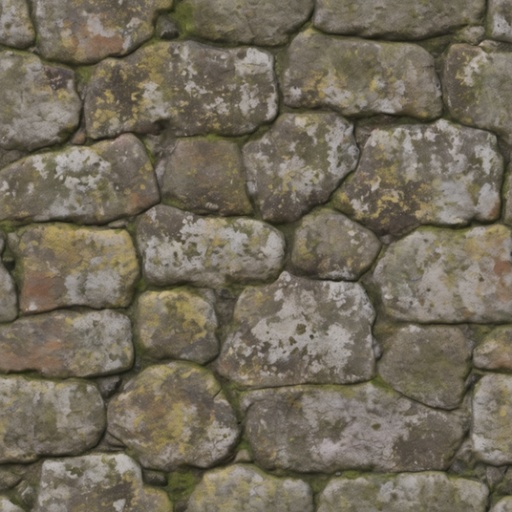} &
    \includegraphics[height=0.12\textwidth]{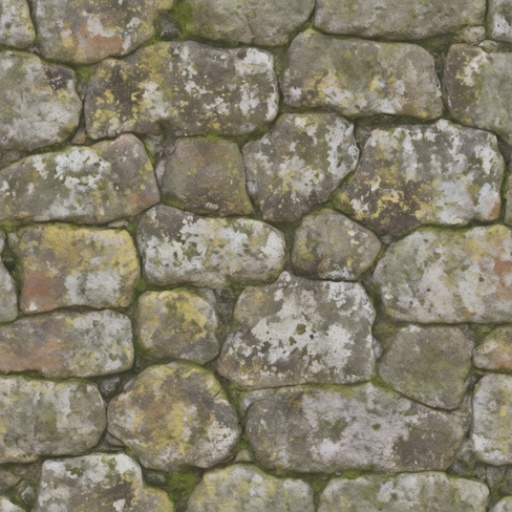} &
    \includegraphics[height=0.12\textwidth]{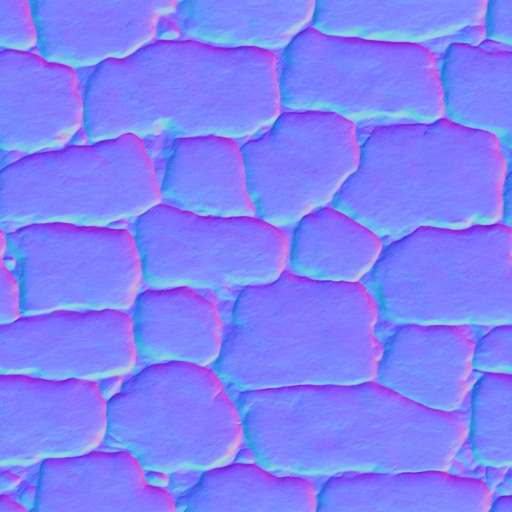} &
    \includegraphics[height=0.12\textwidth]{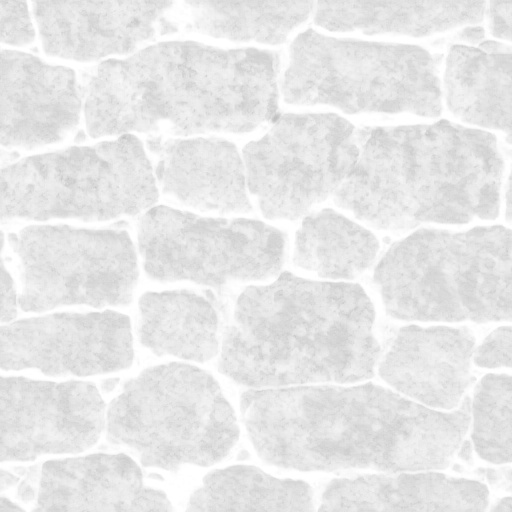} &
    \includegraphics[height=0.12\textwidth]{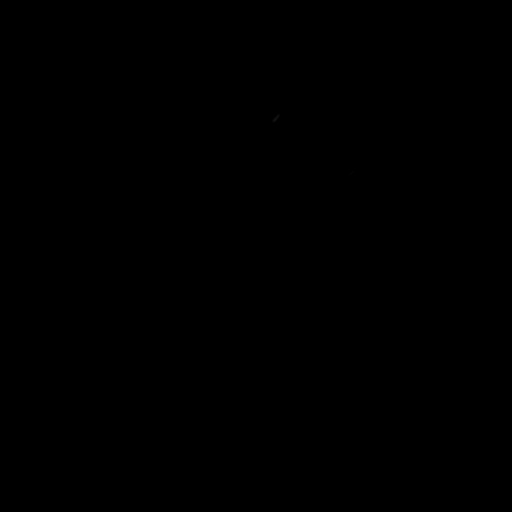} &
    \includegraphics[height=0.12\textwidth]{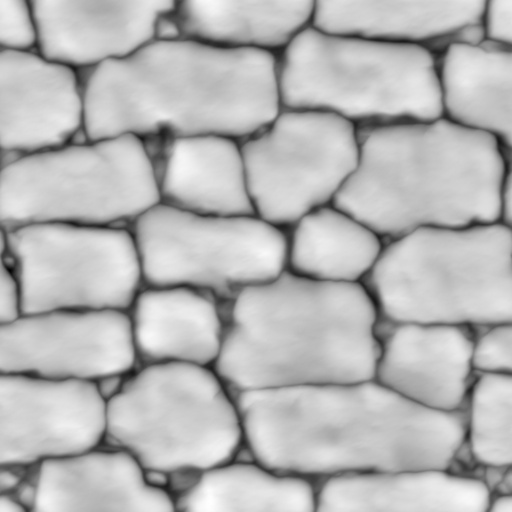}\\
    \includegraphics[height=0.12\textwidth]{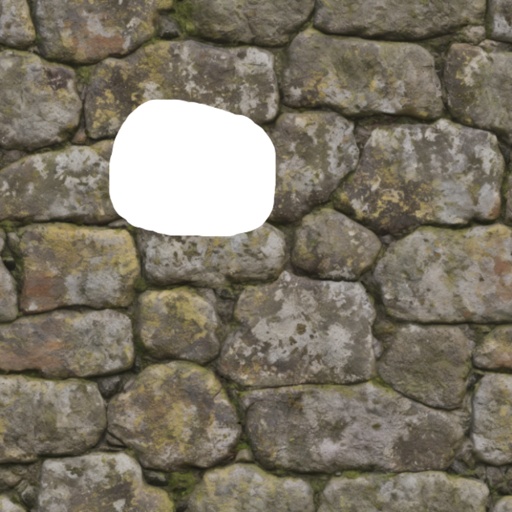} &
    \includegraphics[height=0.12\textwidth]{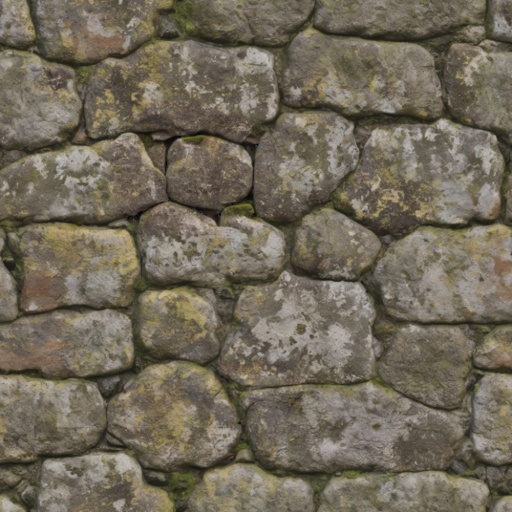} &
    \includegraphics[height=0.12\textwidth]{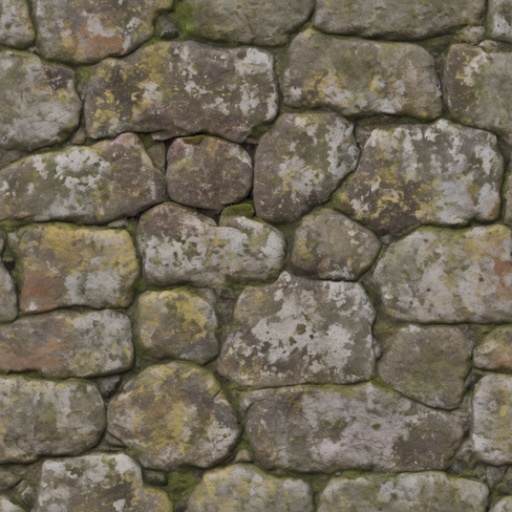} &
    \includegraphics[height=0.12\textwidth]{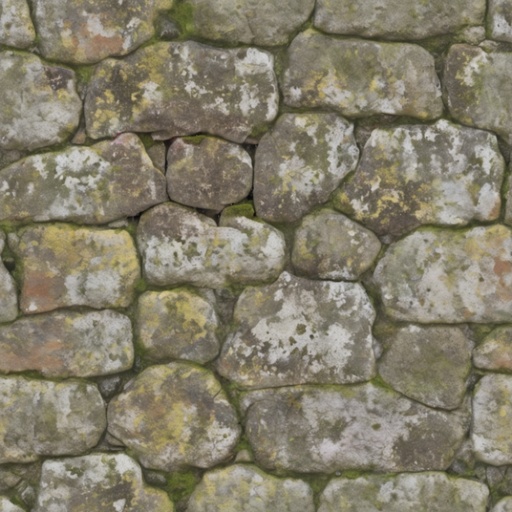} &
    \includegraphics[height=0.12\textwidth]{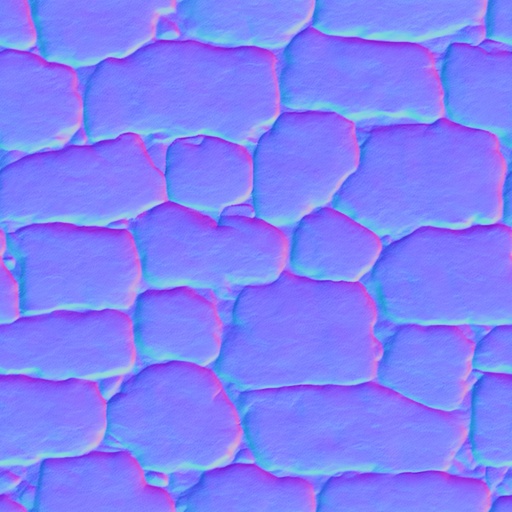} &
    \includegraphics[height=0.12\textwidth]{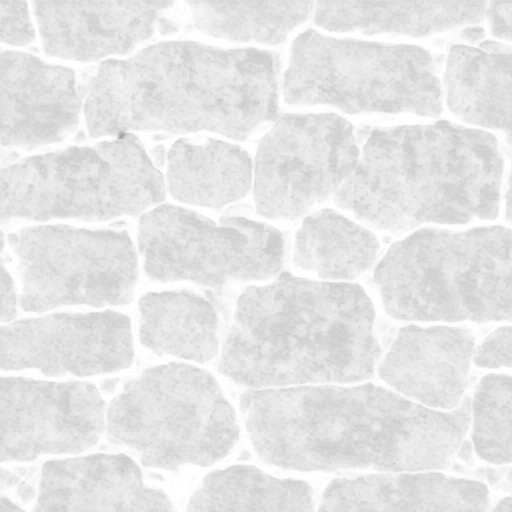} &
    \includegraphics[height=0.12\textwidth]{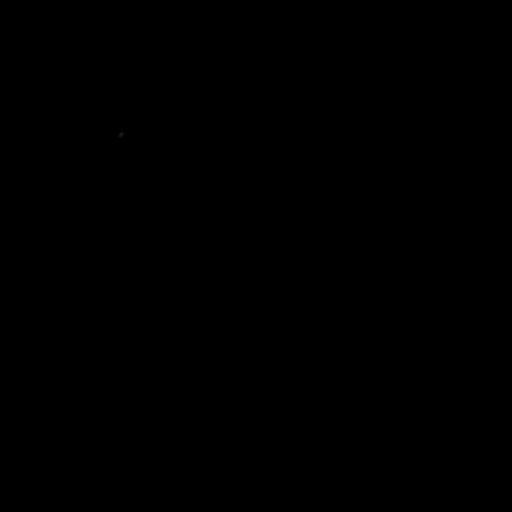} &
    \includegraphics[height=0.12\textwidth]{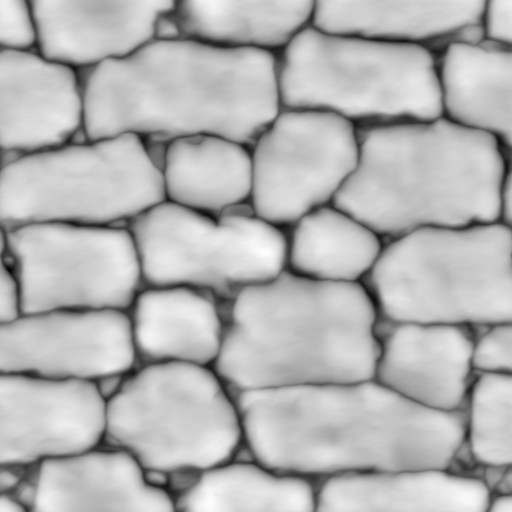}\\    
    \end{tabular}
    \caption{\textbf{Material Editing.} We demonstrate how in-painting can be used to regenerate a masked region in $I_\text{RGB}$, resulting in corresponding edits to the estimated material. The material estimation results outside the masked region remain unchanged.}
    \label{fig:app:editting} 
    \end{subfigure}
\caption{\textbf{More applications.}}
\label{fig:more_applications} 
\end{figure*}